\RequirePackage{xcolor}
\documentclass[12pt]{article}
\pdfoutput=1
\usepackage{jcapmodEFT}
\usepackage{shorthandEFT}
\usepackage[normalem]{ulem}
\usepackage{mathtools}
\usepackage{booktabs}
\usepackage[english]{babel}
\usepackage{amsmath,amssymb,amsbsy,amstext, amsthm, simplewick, amsfonts}
\usepackage{hyperref}
\usepackage{graphicx}
\usepackage[small]{caption}
\usepackage{slashed}
\usepackage{siunitx}
\usepackage{upgreek}
\usepackage{framed}
\usepackage{wrapfig}
\usepackage{multirow}
\usepackage{bbm}
\usepackage[utf8x]{inputenc}
\usepackage{selinput}
\usepackage{bm}
\usepackage{float}
\usepackage{geometry}
\usepackage{yfonts}
\usepackage{caption}
\usepackage{subcaption}
\usepackage{sidecap}
\usepackage{longtable}
\usepackage{anyfontsize}
\usepackage{dsfont}
\usepackage{tikz}
\usepackage{relsize}
\usepackage{tcolorbox}
\usepackage{braket}
\usepackage{mathtools}
\usepackage{tensor}

\renewcommand{\thefootnote}{\fnsymbol{footnote}}

\usepackage[framemethod=default]{mdframed}
\newmdenv[skipabove=7pt,
skipbelow=7pt,
rightline=false,
leftline=false,
topline=false,
bottomline=false,
backgroundcolor=gray!10,
linecolor=gray,
innerleftmargin=5pt,
innerrightmargin=5pt,
innertopmargin=5pt,
innerbottommargin=5pt,
leftmargin=0cm,
rightmargin=0cm,
linewidth=4pt]{eBox}
\newmdenv[skipabove=7pt,
skipbelow=7pt,
rightline=false,
leftline=false,
topline=false,
bottomline=false,
backgroundcolor=gray!10,
linecolor=gray,
innerleftmargin=5pt,
innerrightmargin=5pt,
innertopmargin=-5pt,
innerbottommargin=5pt,
leftmargin=0cm,
rightmargin=0cm,
linewidth=4pt]{eBox2}
\usepackage[framemethod=default]{mdframed}
\newmdenv[skipabove=7pt,
skipbelow=7pt,
rightline=true,
leftline=true,
topline=true,
bottomline=true,
backgroundcolor=gray!15,
linecolor=gray,
innerleftmargin=5pt,
innerrightmargin=5pt,
innertopmargin=5pt,
innerbottommargin=5pt,
leftmargin=0cm,
rightmargin=0cm,
linewidth=0.75pt]{eBox3}

\definecolor{pyblue}{RGB}{31, 119, 180}
\definecolor{pyorange}{RGB}{255, 127, 14}
\definecolor{pygreen}{RGB}{44, 160, 44}
\definecolor{pymiddle}{RGB}{105, 136, 79}
\definecolor{pyred}{RGB}{214, 39, 40}

\usepackage{colortbl}
\definecolor{lightgreen}{cmyk}{0.2, 0, 0.2, 0.2}
\definecolor{lightgray}{cmyk}{0.1,0.2,0,0.1}
\definecolor{lightgray2}{cmyk}{0.1,0.1,0,0.1}
\definecolor{greyish2}{rgb}{.96,.96,.96}

\makeatletter
\newlength{\apb@width}
\newcommand{\autoparbox}[2][c]{\settowidth{\apb@width}{#2}\parbox[#1]{\apb@width}{#2}}

\makeatother


\def\Mpl{M_{\text{Pl}}}
\def\d{\mathrm{d}}
\def\I {\mathcal{I}}

\def\kp{k_{\text{p}}}
\def\qp{q_{\text{p}}}

\def\H{\text{H}}
\def\Ms{\mathrm{s}}

\def\vpi {\varphi}
\def \bfk {\bm{k}}

\def\u{u}

\usetikzlibrary{arrows,calc,decorations.pathmorphing,decorations.markings,shapes.misc}
\pgfkeys{/tikz/.cd,
  circle color/.initial=black,
  circle color/.get=\circlecolor,
  circle color/.store in=\circlecolor,
}
\tikzset{/pgf/decoration/.cd,
    number of sines/.initial=10,
    angle step/.initial=20,
}
\newdimen\tmpdimen\pgfdeclaredecoration{complete sines}{initial}
{
    \state{initial}[
        width=+0pt,
        next state=move,
        persistent precomputation={
            \pgfmathparse{\pgfkeysvalueof{/pgf/decoration/angle step}}%
            \let\anglestep=\pgfmathresult%
            \let\currentangle=\pgfmathresult%
            \pgfmathsetlengthmacro{\pointsperanglestep}%
                {(\pgfdecoratedremainingdistance/\pgfkeysvalueof{/pgf/decoration/number of sines})/360*(\anglestep)}%
        }] {}
    \state{move}[width=+\pointsperanglestep, next state=draw]{
        \pgfpathmoveto{\pgfpointorigin}
    }
    \state{draw}[width=+\pointsperanglestep, switch if less than=1.25*\pointsperanglestep to final, 
        persistent postcomputation={
        \pgfmathparse{mod(\currentangle+\anglestep, 360)}%
        \let\currentangle=\pgfmathresult%
    }]{%
        \pgfmathsin{+\currentangle}%
        \tmpdimen=\pgfdecorationsegmentamplitude%
        \tmpdimen=\pgfmathresult\tmpdimen%
        \divide\tmpdimen by2\relax%
        \pgfpathlineto{\pgfqpoint{0pt}{\tmpdimen}}%
    }
    \state{final}{
        \ifdim\pgfdecoratedremainingdistance>0pt\relax
            \pgfpathlineto{\pgfpointdecoratedpathlast}
        \fi
   }
}

\allowdisplaybreaks[1]
\begin{document}


\newgeometry{top=3cm, bottom=2cm, left=3cm, right=3cm}

\begin{titlepage}
\setcounter{page}{1} \baselineskip=15.5pt 
\thispagestyle{empty}

\begin{center}
{\fontsize{18}{18} \bf Shapes of the Cosmological Low-Speed Collider}\\[14pt]
\end{center}

\vskip 20pt
\begin{center}
\noindent
{\fontsize{12}{18}\selectfont 
Sadra Jazayeri,\footnote[1]{\href{jazayeri@iap.fr}{jazayeri@iap.fr}}
S\'ebastien Renaux-Petel,\footnote[2]{\href{renaux@iap.fr}{renaux@iap.fr}}
and 
Denis Werth\footnote[4]{\href{werth@iap.fr}{werth@iap.fr}}\hskip 1pt}
\end{center}

\begin{center}
\textit{Sorbonne Universit\'e, CNRS, UMR 7095, Institut d'Astrophysique de Paris, 98 bis bd Arago, 75014 Paris, France}
\end{center}

\vspace{0.4cm}
\begin{center}{\bf Abstract}
\end{center}
\noindent

Massive particles produced during inflation leave specific signatures in soft limits of correlation functions of primordial fluctuations. When the Goldstone boson of broken time translations acquires a reduced speed of sound, implying that de Sitter boosts are strongly broken, we introduce a novel discovery channel to detect new physics during inflation, called the cosmological low-speed collider signal. This signal is characterised by a distinctive resonance lying in mildly-soft kinematic configurations of cosmological correlators, indicating the presence of a heavy particle, whose position enables to reconstruct its mass. We show that this resonance can be understood in terms of a non-local single-field effective field theory, in which the heavy field becomes effectively non-dynamical. This theory accurately describes the full dynamics of the Goldstone boson 
and captures all multi-field physical effects distinct from the non-perturbative particle production leading to the conventional cosmological collider signal. As such, this theory provides a systematic and tractable way to study the imprint of massive fields on cosmological correlators. We conduct a thorough study of the low-speed collider phenomenology in the scalar bispectrum, showing that
large non-Gaussianities with new shapes can be generated, in particular beyond weak mixing. We also provide a low-speed collider template for future cosmological surveys.

\end{titlepage}
\restoregeometry

\newpage
\setcounter{tocdepth}{3}
\setcounter{page}{2}

\linespread{1.6}
\tableofcontents
\linespread{1.1}

\renewcommand*{\thefootnote}{\arabic{footnote}}
\setcounter{footnote}{0}

\newpage
\section{Introduction}

The observed cosmological structures are believed to be of primordial origin. Thus, analysing correlation functions of density inhomogeneities can provide valuable insights into the physics of the early universe, possibly beyond the Standard Model. While the two-point correlation function of scalar fluctuations has been well measured, higher-point correlators are the main target of current and upcoming cosmological surveys, see e.g.~\cite{Achucarro:2022qrl} for a recent review and the references therein. 

\vskip 4pt
The discovery of primordial non-Gaussianities would represent a tremendous achievement both on theoretical and observational grounds, as they provide a probe of interactions during inflation.
Such detection may enable us to uncover new degrees of freedom around the Hubble scale, which can be as high as $10^{14}$ GeV. Additionally, this offers the prospect of identifying mass spectra and spins, characterise interactions, and infer dynamical properties such as sound speeds or, more generally, dispersion relations. All this information would help us establish a ``standard model of inflationary cosmology".

\vskip 4pt
The primordial fluctuations in any theory of inflation are at least of two types: the Goldstone boson associated with the spontaneous breaking of time translations $\pi$,\footnote{As for any spontaneously broken symmetry, there is a massless particle---the Goldstone boson---associated with this symmetry breaking. Its Lagrangian can be constructed without any knowledge of the specific mechanism that led to this breaking. This places inflaton fluctuations on the same footing as e.g.~pions of the chiral Lagrangian, phonons of solids, or longitudinal polarizations of the $W$ and $Z$ bosons. The Goldstone boson $\pi$ is related to the observed curvature perturbation fluctuation by $\zeta = -H \pi + \mathcal{O}(\pi^2)$.} and the transverse dynamical part of the metric $\gamma_{ij}$. The field $\pi$ being massless,\footnote{More precisely, the field $\pi$ is exactly massless in the decoupling limit, $M_{\text{pl}} \rightarrow \infty$ and $\dot{H} \rightarrow 0$ while keeping $M_{\text{pl}}^2\dot{H}$ fixed.} it is fully characterised at leading order in gradient expansion by its speed of sound $c_s$, that is, its dispersion relation reads $\omega = c_s \kp + \mathcal{O}(\kp^2)$, where $\kp=k/a$ is the physical momentum. As a consequence of $\pi$ non-linearly realising Lorentz symmetry, the size of non-linearities is correlated to the amount of symmetry breaking at the linear level. This generates well-known \textit{equilateral non-Gaussianities} whose amplitude is related to the speed of sound by symmetry, $f_{\text{NL}}^{\text{eq}} \sim 1/c_s^2$ \cite{Cheung:2007st} (see~\cite{Planck:2019kim} for current constraints). 
Possibly arising from a fundamental non-perturbative UV description of inflation \cite{Silverstein:2003hf, Alishahiha:2004eh}, a reduced sound speed can also be commonly generated by integrating out heavy fields with masses above the scale $\Lambda_\star$ beyond which primordial fluctuations become strongly coupled. Accordingly, $c_s$ is the primary parameter to be constrained by observations that encodes UV physics. The detection of sizeable equilateral non-Gaussianities then poses the question of what the subsequent steps in the investigation should entail.

\vskip 4pt
In addition to the observed massless fluctuations $\pi$, other degrees of freedom covering the entire mass spectrum are generically expected to be present during inflation. These fields can arise from UV completions of inflation, for example appearing in the Kaluza-Klein spectrum or as stringy states \cite{Baumann:2014nda}. Here we focus on fields with masses typically of the same order as the expansion rate. 
Their gravitational production and the subsequent decay of these species into inflaton fluctuations give rise to potentially observable signatures in the soft limits of cosmological correlators. These patterns, known as \textit{cosmological collider signals} \cite{Chen:2009zp, Noumi:2012vr, Arkani-Hamed:2015bza}, exhibit distinctive oscillatory behaviours whose frequencies are determined by the masses of heavy fields and whose amplitudes are Boltzmann-suppressed. The vast phenomenology of such heavy field signatures in cosmological correlators has been actively reported in e.g.~\cite{Chen:2009we,Chen:2009zp,Baumann:2011nk, Noumi:2012vr, Arkani-Hamed:2015bza, Chen:2015lza, Chen:2016nrs, Lee:2016vti, Chen:2016uwp, An:2017hlx, Iyer:2017qzw, Chen:2018xck, McAneny:2019epy, 
Alexander:2019vtb,Lu:2019tjj,Liu:2019fag, Wang:2019gbi,  
Wang:2019gok, Wang:2020ioa, Sou:2021juh, Lu:2021wxu,   Wang:2021qez, Pinol:2021aun,  Cui:2021iie, Tong:2022cdz, Reece:2022soh, Pimentel:2022fsc, Qin:2022lva, Jazayeri:2022kjy, Qin:2022fbv, Xianyu:2022jwk, Qin:2023ejc, Werth:2023pfl}. Although prospects for detecting these signatures in both future galaxy and 21cm surveys are rather optimistic \cite{Alvarez:2014vva, Munoz:2015eqa, Meerburg:2016zdz, MoradinezhadDizgah:2017szk, MoradinezhadDizgah:2018ssw},
the experimental challenge to observe them is enormous as it requires probing a large hierarchy of scales with high precision to resolve the oscillations.

\begin{figure}[t!]
    \centering
    \hspace*{-1.8cm}
    \subfloat{\includegraphics[width=1.2\textwidth]{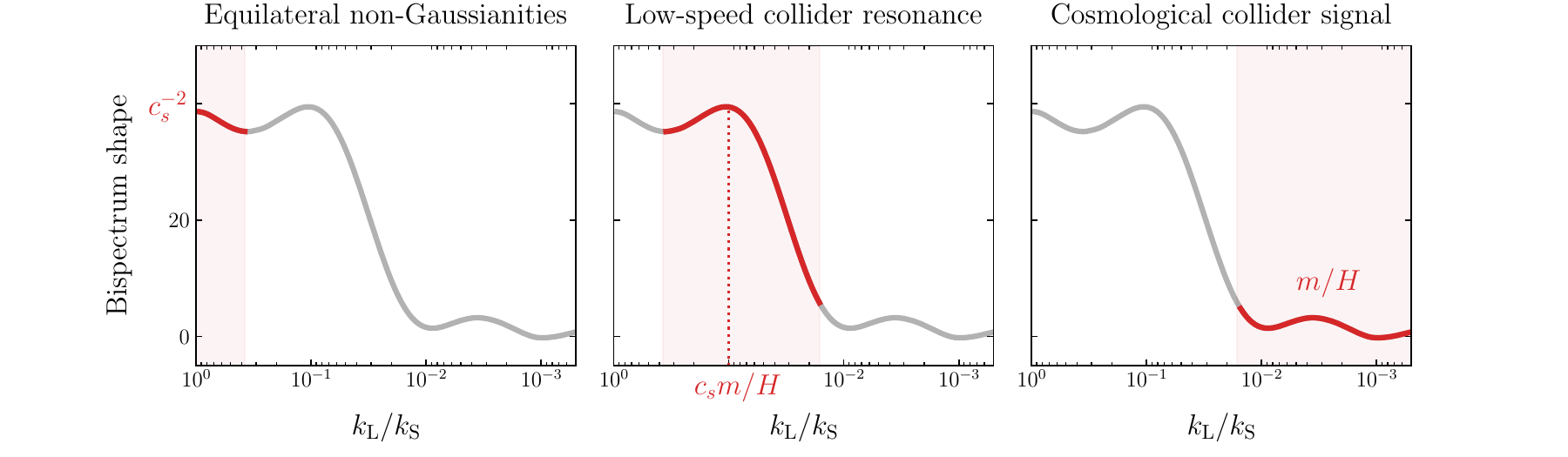}}
    \hfill
    \caption{Schematic illustration of the bispectrum shape in the isosceles kinematic configuration. A strong breaking of de Sitter boosts during inflation leads to large equilateral-type non-Gaussianities, whose size in equilateral configurations, $k_{\text{L}}/k_{\text{S}} \sim 1$ (\textit{left}), fixes the speed of sound $c_s$ of the Goldstone bosons $\pi$. The presence of an extra heavy particle is imprinted in the low-speed collider signal in mildly-squeezed configurations, $k_{\text{L}}/k_{\text{S}} \sim c_s m/H$ (\textit{middle}). Together with the detection of equilateral non-Gaussianities, this distinctive resonance provides a clean detection channel to pinpoint the mass of additional particles. In ultra-squeezed configurations, $k_{\text{L}}/k_{\text{S}} \ll c_s m/H$ (\textit{right}), the spontaneous production of these particles by the expanding spacetime is imprinted in the conventional cosmological collider signal, i.e.~characteristic oscillatory patterns whose frequency is set by $m/H$ and amplitude is suppressed by $e^{-\pi m/2H}$. As we delve into softer configurations of cosmological correlators, one requires an increasing level of precision. Therefore, it is highly probable that we will first detect the low-speed collider resonance before detecting any indication of a cosmological collider signal.}
    \label{fig: intro}
\end{figure}

\vskip 4pt
In this paper, building on the previous work \cite{Jazayeri:2022kjy}, we present a new discovery channel to detect additional heavy particles during inflation when the Goldstone boson has a reduced speed of sound $c_s \ll 1$, which implies that de Sitter boosts are strongly broken. These so-called \textit{cosmological low-speed collider signals} are characterised by a distinct \textit{resonance} in mildly-soft (internal) kinematic configurations of cosmological correlators, signalling the presence of a massive particle lighter than $H/c_s$.
This regime covers the entire parameter space if the additional field is lighter than the strong coupling scale $\Lambda_\star$.
In analogy with terrestrial collider experiments in which massive particles leave resonant signatures in the cross section of lighter particles' collisions when the center of mass energy coincides with the particle masses $\sqrt{s} \sim m$, the low-speed collider resonance is located at $k_{\text{L}}/k_{\text{S}} \sim c_s m/H$, where $k_{\text{L}}$ and $k_{\text{S}}$ are the long and short modes. After the detection of equilateral-type non-Gaussianities and hence the measurement of $c_s$, we would enter an era of precision measurements, analogous to electroweak precision physics in colliders. Measuring the low-speed collider peak would reveal the existence of an additional field coupled to the Goldstone boson and would give indications about its mass, which could be later confirmed by the measurement of the conventional cosmological collider signal in ultra-soft limits, see Fig.~\ref{fig: intro}.

\vskip 4pt
As we will see, a strong breaking of de Sitter boosts uncovers a concealed elegance within the analytical structure of cosmological correlators. In de Sitter spacetime, dealing with massive fields even in perturbation theory is notoriously challenging. Indeed, as a consequence of the inflationary background altering the free propagation of fields, their mode functions take the form of complicated special functions rather than simple plane waves. Consequently, except for certain soft limits of correlators, fully analytical predictions are very rare in the literature (see e.g.~\cite{Arkani-Hamed:2018kmz, Sleight:2019mgd, Sleight:2019hfp, Jazayeri:2022kjy, Pimentel:2022fsc, Qin:2022fbv, Xianyu:2022jwk, Qin:2023ejc} for recent heroic computations). However, the picture drastically simplifies when $\pi$ has a reduced speed of sound $c_s \ll 1$.
To better appreciate the underlying physics behind this simplicity, let us imagine the two-to-two scattering of Goldstone bosons at the energy scale $\omega$ and consider a particle with mass $m\lesssim \omega/c_s$ mediating a force between them. In this exchange process, the energy and the momentum of the off-shell particle are of order $\omega$ and $\omega/c_s$, respectively, implied by energy and momentum conservation. As a result, similar to the low-speed limit of the Yukawa theory, the massive field propagator can be simplified as 
\begin{equation}
    \frac{1}{\omega^2 - \bm{k}^2 - m^2} \sim \frac{-1}{\bm{k}^2 + m^2}\,.
\end{equation}
This approximation corresponds to replacing the real-space Feynman propagator of the massive field $D_{\text{F}}$ by a Yukawa potential 
\begin{equation}
    D_{\text{F}}(\bm{x}, t; \bm{y}, t') \rightarrow \delta(t-t')\, \frac{e^{-m |\bm{x} - \bm{y}|}}{4\pi |\bm{x} - \bm{y}|}\,,
\end{equation}
reflecting the instantaneous propagation of the massive field.\footnote{Notice that the \textit{local} EFT limit is recovered from the non-local one when the massive field is heavier than the transferred momentum $m\gtrsim \omega/c_s$. In this limit, the real-space propagator becomes $D_{\text{F}}(\bm{x}, t; \bm{y}, t') \rightarrow m^{-2}\delta(t-t')\,\delta^{(3)}(\bm{x} - \bm{y})$.} Similarly, for the computation of cosmological correlators, we show that the relativistic heavy field can be integrated out, albeit in a non-local manner in space, resulting in a \textit{non-local single-field EFT} for the Goldstone boson. This non-local EFT comes extremely handy when computing cosmological correlators for mainly two reasons. First, the only dynamical field in this EFT is the massless field $\pi$ which has a simple mode function, at least when the effect of the mixing with the additional field is weak. Second, exchange diagrams reduce to contact ones,
which drastically simplify computations. Corrections to the leading non-local EFT can also be systematically captured by adding extra higher-order time-derivative terms. Eventually, note that as a consequence of integrating over the dynamics of the heavy field, the non-perturbative particle production leading to the conventional cosmological collider signal cannot be captured in the realm of the non-local EFT. However, this simple theory, by essence, accurately describes the physics of the cosmological low-speed collider.

\paragraph{Outline.} In this paper, we present an exhaustive study of the low-speed collider signals in the three-point correlation function of $\pi$. In Sec.~\ref{sec: Inflationary Fluctuations}, we start by constructing the effective action (\ref{eq: full theory}) that couples the massless Goldstone boson $\pi$ to an additional massive scalar field $\sigma$. In Sec.~\ref{sec: Non-local EFT}, we show how a non-local single-field EFT (\ref{eq: non-local theory}) arises when $\pi$ has a reduced speed of sound $c_s \ll 1$. We then characterise this theory by inspecting the dispersion relation, showing that it encompasses several previously-known regimes, and discuss the regimes of validity of this non-local EFT. In Sec.~\ref{sec: Cosmological Correlators}, we divide our computation of cosmological correlators into two parts. When the massive field at the linear level is weakly mixed with the Goldstone boson, we analytically compute the two- and three-point correlators induced by the non-local EFT. We show that all correlators, including the (would-be) double- and triple-exchange diagrams, are obtained by acting with weight-shifting operators on a set of seed correlators. At strong mixing, we use the recently developed \textit{cosmological flow} approach \cite{Werth:2023pfl, CosmoFlow} to numerically compute these correlators. In Sec.~\ref{sec: pheno}, we present an in-depth analysis of the low-speed collider phenonenology, paying particular attention to its strong mixing regime.
We determine the size of non-Gaussianities and show that we can accomodate large non-Gaussianities while remaining under perturbative control. We then make an extensive analysis of the bispectrum shapes and give a separable template (\ref{eq: low speed collider template}) that accurately describes the low-speed collider resonance. Additional appendices expose technical details and derivations. We present our conclusions in Sec.~\ref{sec: conclusions}.

\paragraph{Notation and Conventions.} We use the mostly-plus signature for the metric $(-, +, +, +)$. Spatial three-dimensional vectors are written in boldface $\bm{k}$. We use Greek letters $(\mu, \nu, \dots)$ for space-time indices, and Latin letters $(i, j, \dots)$ for spatial indices as usual. Overdots and primes denote derivatives with respect to cosmic (physical) time $t$ and conformal time $\tau$ defined by $\mathrm{d}\tau = \mathrm{d}t/a$, respectively. A prime on a correlator is defined to mean that we drop the momentum conserving delta function, i.e.~$\braket{\mathcal{O}_{\bm{k}_1} \ldots \mathcal{O}_{\bm{k}_n}} = \delta^{(3)}(\bm{k}_1 + \ldots + \bm{k}_n) \braket{\mathcal{O}_{\bm{k}_1} \ldots \mathcal{O}_{\bm{k}_n}}'$.

\newpage
\section{Inflationary Fluctuations}
\label{sec: Inflationary Fluctuations}

Cosmology is about observing inhomogeneities and studying their statistical properties. According to the inflationary paradigm, these perturbations emerge from quantum fluctuations of the inflationary fields, which ultimately provide the seeds for small density fluctuations in the late universe. 
Remarkably, the simplest explanation of cosmological data involves weakly coupled inflationary fluctuations.
It means that in practice they can be described by an action that takes the form of a series expansion in powers of fluctuations and their derivatives. At every order, the terms are fixed by assumed symmetries up to a finite number of free coefficients. Such description is valid up to a UV cutoff scale, beyond which the UV physics become important. However, at energies below the UV cutoff, higher-order terms in the series expansion become irrelevant as they are suppressed by increasing powers of the UV cutoff.
In the end, only a finite number of terms is sufficient to accurately describe the physics of these fluctuations.
This is the very idea of \textit{effective field theories of inflationary fluctuations}, which provides a model-independent framework to study the physics of degrees of freedom that are directly linked to observations, making manifest the implications of symmetries.

\vskip 4pt
In this section, we identify the relevant degrees of freedom to describe inflationary fluctuations and construct a theory describing them. More specifically, we start by reviewing the effective action for the Goldstone boson of broken time translations interacting with an additional massive scalar field. Next, we provide bounds on the couplings that give the maximal size of the mixing interactions while keeping the effective theory under control.

\subsection{Goldstone Description}
\label{subsection: Inflationary fluctuations}

The effective field theory (EFT) of inflationary fluctuations (see the original papers \cite{Cheung:2007st, Senatore:2010wk} or review \cite{Piazza:2013coa} for more details) is based on the need of a physical ``clock", i.e. a preferred time slicing of spacetime, that breaks time-translation symmetry in order for inflation to end. This pattern of symmetry breaking gives rise to a Goldstone boson $\pi$ that transforms as $\pi \rightarrow \pi - \xi(\bm{x}, t)$ under a time shift $t \rightarrow t + \xi(\bm{x}, t)$. Together with fluctuations of the metric, it is guaranteed to be active during inflation and hence is the minimal relevant degree of freedom describing inflationary fluctuations.

\paragraph{Goldstone action.} In the unitary gauge, in which the fluctuations of the ``clock" are absorbed by the metric, the most general action is constructed out of all operators that are invariant under time-dependent spatial diffeomorphisms $x^i \rightarrow x^i + \xi^i(\bm{x}, t)$. At leading order in derivatives, the action can be written
\begin{equation}
\label{eq: unitary gauge action}
    \begin{aligned}
    S_\pi = \int \d^4x \sqrt{-g} \left[
    \frac{1}{2}M_{\text{pl}}^2 R + M_{\text{pl}}^2 \dot{H} g^{00} - M_{\text{pl}}^2(3H^2 + \dot{H}) + \sum_{n=2}^{\infty} \frac{M_n^4(t)}{n!} \left(\delta g^{00}\right)^n + \ldots
    \right]\,,
    \end{aligned}
\end{equation}
where $\delta g^{00} = g^{00}+1$ and $M_n(t)$ are general time-dependent mass scales. One recovers slow-roll inflation in the limit $M_n\rightarrow 0$. The coefficients of the first two operators have been fixed to remove all tadpoles so that the action starts quadratic in the fluctuations.

\vskip 4pt
The breaking of time diffeomorphisms during inflation is analogous to the breaking of the gauge group by the mass term in massive Yang-Mills theory, where the dynamical degrees of freedom are the longitudinal modes $\pi^a$ of the vector fields $A^a_\mu$. There, one can make them explicit via the St{\"u}ckelberg trick. In the context of the EFT of inflationary fluctuations, this trick boils down to performing a spacetime-dependent time reparameterization $t\rightarrow t+\pi(\bm{x}, t)$ in Eq.~(\ref{eq: unitary gauge action}). In general, the resulting action is rather complicated because it mixes the Goldstone mode to metric fluctuations. However, being interested just in effects that are not dominated by the mixing with gravity, one can neglect the metric fluctuations by taking the so-called decoupling limit $\Mpl \rightarrow \infty$ and $\dot{H}\rightarrow 0$ while keeping the product $\Mpl^2\dot{H}$ fixed. In this regime, the transformation reduces to $\delta g^{00} \rightarrow -2 \dot{\pi} - \dot{\pi}^2 + (\partial_i \pi)^2/a^2$, and the Goldstone boson $\pi$ is associated with the longitudinal scalar mode of the metric, which in the end is related to the curvature perturbation by $\zeta = -H \pi$ on super-Hubble scales. The Goldstone action becomes
\begin{equation}
\label{eq: Goldstone action}
    S_\pi = \int \d^4x \sqrt{-g}\, \frac{M_{\text{pl}}^2|\dot{H}|}{c_s^2}\left[\dot{\pi}^2 - c_s^2(\tilde{\partial}_i\pi)^2 + (1 - c_s^2)\left(\dot{\pi}^3 - \dot{\pi}(\tilde{\partial}_i\pi)^2\right) - \frac{4}{3}M_3^4 \frac{c_s^2}{M_{\text{pl}}^2|\dot{H}|}\dot{\pi}^3\right]\,,
\end{equation}
where $\tilde{\partial}_i = \partial_i/a$ and $c_s^{-2} = 1 - 2M_2^4/M_{\text{pl}}^2|\dot{H}|$ is the intrinsic speed of sound for the propagation of $\pi$. The non-linearly realised symmetry makes it explicit that a small value of $c_s$ generates an enhanced cubic interaction $\dot{\pi}(\partial_i \pi)^2$. The action (\ref{eq: Goldstone action}) accurately describes the dynamics of $\pi$ up to slow-roll corrections $\varepsilon = -\dot{H}/H^2$.

\paragraph{Additional sector.} We are interested in coupling the Goldstone boson associated with the breaking of time diffeomorphism to an additional sector composed of a single massive scalar field $\sigma$. We will exclusively consider the case where this scalar field is heavy with mass around the Hubble scale so that no additional internal symmetry is \textit{a priori} needed in order for this sector to be radiatively stable (see e.g.~\cite{Senatore:2010wk} where additional light fields are considered). The action for the $\sigma$-sector we consider is  
\begin{equation}
\label{eq: sigma sector Lagrangian}
    S_\sigma = - \int \d^4x \sqrt{-g}\left[\frac{1}{2}(\partial_\mu \sigma)^2 +  \frac{1}{2}m^2\sigma^2 + \mu \sigma^3 + \dots \right]\,,
\end{equation}
where the ellipses denote higher-order terms.\footnote{Up to cubic terms in the field, this is the most general Lagrangian one can write down up to dimension-4 operators because $\dot{\sigma}\sigma^2$ is a total derivative. Including dimension-5 operators demands to also consider the following operators: $\dot{\sigma}^2\sigma$ and $(\tilde{\partial}_i \sigma)^2\sigma$. These terms, being of higher-dimension, would be suppressed by an additional high-energy scale. Therefore, the cubic self-interaction $\sigma^3$ is responsible for the leading non-Gaussian signal in the $\sigma$-sector.} The non-linearities in this sector are entirely characterised by $\mu$.

\paragraph{Mixing sector.} We now describe the coupling of  $\sigma$ to the Goldstone boson $\pi$, see e.g.~\cite{Noumi:2012vr} for details. In the unitary gauge, we therefore construct the mixing sector out of the building blocks $\delta g^{00}$ and $\sigma$, invariant under time-dependent spatial diffeomorphisms. Among the terms allowed by symmetry, some give tadpoles that should add up to zero in order not to alter the background solution. The action then has to start at quadratic order in fluctuations. At leading order in derivatives, hence considering the leading deformation to the slow-roll action, the mixing action is\footnote{By construction, the Lagrangian is constructed out of building blocks that are invariant under time-dependent spatial diffeomorphisms. Neglecting operators involving the extrinsic curvature, terms with higher-order derivatives must therefore be contracted in a fully diffeomorphism-invariant way. At leading order in derivatives, this allows us to consider the operator $g^{0\mu}\partial_\mu\sigma$ that would generate the additional quadratic mixing $\dot{\pi}\dot{\sigma}$ and dimension-6 cubic operators. These terms have been fully classified in \cite{Baumann:2011nk}. In this work, we do not consider such interactions and only focus on the leading deformations of the slow-roll action.}
\begin{equation}
    S_{\pi-\sigma} = - \int \d^4x \sqrt{-g}\left[ \tilde{M}_1^3 \delta g^{00} \sigma + \tilde{M}_2^2 \delta g^{00} \sigma^2 + \tilde{M}_3^3 \left(\delta g^{00}\right)^2 \sigma\right] \,.
\label{eq:mixing-unitary-gauge}    
\end{equation}
This action simplifies when reintroducing the Goldstone boson and considering the decoupling limit in which couplings to metric fluctuations are negligible.
Up to cubic interactions, one obtains
\begin{equation}
    S_{\pi-\sigma} = \int \d^4x \sqrt{-g}\left[2\tilde{M}_1^3 \dot{\pi}\sigma + \tilde{M}_1^3(\partial_\mu \pi)^2\sigma + 2\tilde{M}_2^2\dot{\pi}\sigma^2 - 4\tilde{M}_3^3\dot{\pi}^2\sigma\right]\,.
\end{equation}
The parameter $\tilde{M}_1$ fully controls the size of the mixing $\dot{\pi} \sigma$ and the cubic interaction $(\tilde{\partial}_i \pi)^2 \sigma$. This is a consequence of the non-linearly realised symmetry i.e.~field operators are related at different orders in perturbation theory.

\paragraph{Full theory.}
Up to cubic order in fluctuations, collecting the various sectors gives the following full theory that we will consider

\begin{tcolorbox}[colframe=white,arc=0pt,colback=greyish2]
\begin{equation}
\label{eq: full theory}
    \begin{aligned}
    S = \int \d^4x &\sqrt{-g} \left( \frac{1}{2}\dot{\pi}_c^2 - \frac{c_s^2}{2} (\tilde{\partial}_i \pi_c)^2  - \frac{1}{2}(\partial_\mu \sigma)^2 - \frac{1}{2}m^2 \sigma^2 + \rho \dot{\pi}_c \sigma \right. \\
    &\left. - \lambda_1 \dot{\pi}_c (\tilde{\partial}_i \pi_c)^2 -  \lambda_2 \dot{\pi}_c^3 - \frac{1}{\Lambda_1} (\tilde{\partial}_i \pi_c)^2\sigma - \frac{1}{\Lambda_2} \dot{\pi}_c^2 \sigma - \lambda \dot{\pi}_c \sigma^2 - \mu \sigma^3\right)\,,
    \end{aligned}
\end{equation}
\end{tcolorbox}
\noindent where we have canonically normalised the Goldstone boson $\pi_c = c_s^{-3/2}f_\pi^2 \pi$ with $f_\pi^4 \equiv 2c_s \Mpl^2 |\dot{H}|$ being the symmetry breaking scale \cite{Baumann:2011su}. We have redefined the coupling constants as

\begin{equation}
    \begin{aligned}
    \rho &\equiv 2c_s^{3/2}\,\frac{\tilde{M}_1^3}{f_\pi^2}\,, \hspace*{1cm} \lambda_1 \equiv -c_s^{3/2}\,\frac{1-c_s^2}{2 f_\pi^2}\,, \hspace*{1.2cm} \lambda_2 \equiv -\frac{c_s^{3/2}}{2f_\pi^2} \left[1 - c_s^2 - \frac{8}{3}\frac{M_3^4}{f_\pi^4}c_s^3\right]\,,\\
    \Lambda_1^{-1} &\equiv -2c_s^3 \frac{\tilde{M}_1^3}{f_\pi^4}\,, \hspace*{0.8cm} \Lambda_2^{-1} \equiv -\frac{4c_s^3}{f_\pi^4}\left(\tilde{M}_1^3 + 4\tilde{M}_3^3\right)\,, \hspace*{0.34cm} \lambda \equiv -2c_s^{3/2}\frac{\tilde{M}_2^2}{f_\pi^2}\,.
    \end{aligned}
\end{equation}
Note that there is \textit{a priori} no model-building requirement on the size of the quadratic mixing $\rho/H$. In full generality, we therefore allow this coupling to exceed unity in Hubble unit. Importantly, because of the quadratic mixing, there is \textit{a priori} no one-to-one correspondence between fields and particle states. Instead, there is a mixing between the field content and the particle spectrum, and this mixing can be sizeable. In this paper, we will concentrate on the case of an additional heavy particle, which, as we will see, corresponds to $m$ or/and $\rho \gtrsim \mathcal{O}(H)$.

\vskip 4pt
The single-field dimensionless power spectrum, i.e. in the limit $\rho/H=0$, is given by
\begin{equation}
\label{eq: dimensionless power spectrum}
    \Delta_{\zeta, 0}^2 = \frac{k^3}{2\pi^2} \braket{\zeta_{\bm{k}} \zeta_{-\bm{k}}}' = \frac{1}{4\pi^2}\left(\frac{H}{f_\pi}\right)^4\,.
\end{equation}
The measured amplitude of the dimensionless power spectrum is $\Delta_\zeta^2 = 2.2\times10^{-9}$ \cite{Aghanim:2018eyx}. 

\subsection{Perturbativity Bounds on Couplings}
\label{subsection: Bounds on couplings}

The introduced coupling constants are free parameters of the theory. However, they must satisfy some bounds to ensure that the effective description of fluctuations is under theoretical control. Here, we give bounds on these couplings based on perturbativity, at weak and strong mixing. We will give additional details on these regimes in Sec.~\ref{sec: Non-local EFT}, and refer to \cite{CosmoFlow} for a detailed analysis (see also \cite{deRham:2017aoj, Grall:2020tqc} for methods to derive perturbativity bounds). We also comment on theoretically motivated values for the mass of the heavy field.

\paragraph{Weak mixing.} In the weak mixing regime, imposing that the interactions associated with relevant/marginal operators can be treated perturbatively at sound horizon crossing and that the strong coupling scales of irrelevant operators be larger than the Hubble scale gives 
\begin{equation}
\label{eq: perturbativity weak mixing}
    \rho/H \lesssim c_s^{-1/2}\,, \hspace*{0.5cm} H/\Lambda_2 \lesssim c_s\,, \hspace*{0.5cm} \lambda \lesssim c_s^{1/2}\,, \hspace*{0.5cm}\mu/H \lesssim 1\,.
\end{equation}
In the insert below, we show how to obtain these bounds from simple dimensional analysis arguments. For a sufficiently small mass $m/H\lesssim 1/c_s$, which is the case of interest for the low-speed collider regime, we will see that non-trivial physics also occurs outside the horizon, and it can be treated perturbatively only if one also imposes $\rho\lesssim m$, see the analysis in Sec.~\ref{sec:physics-LSC}.
This bound is more stringent and therefore defines the weak mixing regime.\footnote{Remarkably, the analysis in Sec.~\ref{sec:physics-LSC} will reveal that in the range $m \lesssim \rho \lesssim H/c_s$, the perturbative analysis still approximately holds, simply upon considering the effective mass $m^2 \to m^2+\rho^2$.} 

\begin{framed}
{\small \noindent {\it Perturbativity bounds.}---In this insert, we derive the perturbativity bounds (\ref{eq: perturbativity weak mixing}) by estimating the various terms in the Lagrangian (\ref{eq: full theory}). As the Goldstone boson propagates with a non-unity speed of sound, it is not possible to treat time and space on the same footing. Equivalently, we need to keep track of how the various terms in the Lagrangian scale with energy and momentum. To do this, let us fix the physical momentum $\kp$ as a reference. At weak mixing, the dispersion relation for the Goldstone boson is $\omega_\pi=c_s \kp$ and that of the massive mode reads $\omega_\sigma = \kp$, from which we get $\omega_\pi = c_s \omega_\sigma$. In the following, we will write all quantities in terms of $\omega\equiv \omega_\pi$. Note that placing ourselves deep in the UV, we neglect the mass contribution for simplicity. In the regime of interest for the low-speed collider $c_s m/H\ll 1$, one can indeed show that the mass term in the Lagrangian is negligible. We can directly read from the Lagrangian that $\pi_c \sim c_s^{-3/2}\omega$ and $\sigma \sim c_s^{-1}\omega$ by dimensional analysis. With these scalings, we deduce that the Goldstone boson kinetic term scales as $\dot{\pi}_c^2 \sim c_s^2(\tilde{\partial}_i \pi_c)^2 \sim c_s^{-3}\omega^4$ and that of the field $\sigma$ scales as $\dot{\sigma}^2\sim (\tilde{\partial}_i \sigma)^2 \sim c_s^{-4}\omega^4$. Note that the presence of a reduced sound speed suppresses the kinetic term of $\pi_c$ compared to that of $\sigma$. Similarly, the quadratic mixing scales as $\rho \dot{\pi}_c \sigma \sim c_s^{-5/2} \rho \,\omega^3$. Finally, requiring that this mixing be smaller than the most suppressed kinetic term, i.e.~$\dot{\pi}_c^2$, leads to $\rho/\omega \lesssim c_s^{-1/2}$, which after evaluating the expression at the sound-horizon crossing $\kp\sim H/c_s$---equivalent to $\omega\sim H$---gives the first bound in (\ref{eq: perturbativity weak mixing}). The same analysis leads to the perturbativity bounds for cubic interaction coupling constants. Note that the bound obtained from the cubic interaction $(\tilde{\partial}_i \pi_c)^2\sigma$ fixed by the quadratic mixing---$\rho/H \lesssim c_s^{3/2}(2 \pi \Delta_\zeta)^{-1}$---is less constraining than the one found after imposing perturbativity of the quadratic mixing. 
Finally, one can perform the same analysis in the strong mixing regime with a modified dispersion relation to obtain the perturbativity bounds (\ref{eq: perturbativity strong mixing}).}
\end{framed}

\paragraph{Strong mixing and modified dispersion relation.} We will see in Sec.~\ref{sec: Non-local EFT} that in the strong mixing regime, the quadratic theory is dominated by the mixing term and one can realise that (\ref{eq: full theory}) can be reduced to an effective single-field theory with non-linear dispersion relation for the Goldstone boson \cite{CosmoFlow, Baumann:2011su, Assassi:2013gxa}. It is then easy to obtain the following bounds 
\begin{equation}
\label{eq: perturbativity strong mixing}
\begin{aligned}
    \rho/H &\lesssim c_s\kappa^{1/2}\Delta_\zeta^{-1}\,, \hspace*{0.8cm} 
     H/\Lambda_2 \lesssim c_s^{5/4} \left(\rho/H\right)^{3/4} \,,\\ \lambda &\lesssim c_s^{1/4}\left(\rho/H\right)^{3/4}\,, \hspace*{0.5cm} \mu/H \lesssim c_s^{-3/4} \left(\rho/H\right)^{3/4}\,,
\end{aligned}
\end{equation}
where $\kappa = 2\Gamma(5/4)^2/\pi^3\approx 0.053$. These bounds allow non-Gaussian signals to be large without breaking perturbativity, see Sec.~\ref{sec:size}. 

\paragraph{Goldstone boson UV cutoff.} For the Goldstone boson sector, the UV cutoff scale $\Lambda_\star$, beyond which perturbative unitarity is violated, famously approaches the Hubble scale when the speed of sound is reduced. Indeed, it is given by \cite{Baumann:2014cja} 

\begin{equation}
 \Lambda_\star^4 = \frac{24}{5}\pi f_\pi^4 \, \frac{c_s^4}{1 - c_s^2}\,,
\end{equation}
which gives $\Lambda_\star \sim c_s f_\pi$ in the limit $c_s \ll 1$. Note that this strong coupling scale is associated with the operator $(\partial_i \pi_c)^4$, giving the more stringent UV cutoff. Imposing that $\Lambda_\star$ should exceed the Hubble scale at weak mixing gives a theoretical lower bound on the sound speed $c_s \geq 0.0087$. Heavy particles with masses above the strong coupling scale cannot be produced on-shell. However, they can be consistently coupled to the massless mode $\pi_c$ as long as the energy carried by the Goldstone boson does not exceed $\Lambda_\star$. This is analogous to the chiral Lagrangian where pions can be coupled to heavy baryons.\footnote{We thank Paolo Creminelli for helpful discussions on this topic.} In our situation, as we will see in Sec.~\ref{subsec: dispersion relation}, heavy particles with masses far above the Hubble scale can be integrated out in the usual local manner.
Instead, we will see in the following that the low-speed collider signature appears when $m \lesssim H/c_s$. This is automatically satisfied if the mass is smaller than the strong coupling scale, i.e.~$m \lesssim \Lambda_\star \approx 100 H c_s \lesssim  H/c_s$ as soon as $c_s \lesssim 0.1$.

\section{Non-local EFT of Inflation}
\label{sec: Non-local EFT}

We have identified the relevant degrees of freedom describing the theory of inflationary fluctuations that we consider, namely the (canonically normalised) Goldstone boson of broken time translations $\pi_c$ and an additional massive field $\sigma$. Our central interest is the low-speed collider signal which manifests itself as a resonance in squeezed configurations of the primordial three-point correlator, provided that the sound speed of $\pi_c$ is reduced $c_s<H/m$. This phenomenological interesting signal is of course well captured by the full theory (\ref{eq: full theory}). However, the multi-field nature of this theory makes analytical computations of correlators complex---if not intractable---and conceals the physics of the low-speed collider. 

\vskip 4pt
In this section, we derive an effective single-field theory that captures the physics of this signal in an uninvolved way. We discuss how one can obtain an accurate single-field, albeit non-local, effective description of the full dynamics, in the almost entire quadratic parameter space, by integrating out the heavy degree of freedom from the mass spectrum (see e.g.~\cite{Gwyn:2012mw} for other instances of non-local single-field EFTs).

\subsection{Effective Action}
\label{subsection: integrate out}

In a specific regime that we will determine later, we can integrate out the heavy field $\sigma$ to obtain an effective action for $\pi_c$. Formally, at the level of the partition function path integral,\footnote{In general, integrating out heavy degrees of freedom induces non-unitary effects---such as decoherence and dissipation---in the low-energy effective theory. When computing \textit{in-in} cosmological correlators, these effects arise from the interference between the two branches of the in-in contour (see e.g.~\cite{Hongo:2018ant} for more details). In addition, truncating the path integral at finite time leads to unusual boundary conditions for the fields that fix the non-vanishing homogenous solution of the heavy field equation of motion. In contrast, for flat-space \textit{in-out} scattering amplitudes, these boundary terms essentially vanish at the infinite past and future. An alternative approach to incorporate these subtle effects consists in working at the level of the wavefunctional path integral (see e.g.~\cite{Salcedo:2022aal} for recent developments). Given that the integrating out procedure in cosmology is not as well established as it is in particle physics, we will stick to the commonly used approach, and keep only the particular solution to \eqref{eq: formal classical sigma} below.} we can define a single-field effective action $S_{\text{eff}}$ by performing the path integral over the heavy state

\begin{equation}
    e^{i S_{\text{eff}}[\pi_c]} = \int \mathcal{D}\sigma\, e^{i S[\pi_c, \sigma]}\,.
\end{equation}
Generically and as we will see later, performing such path integral produces \textit{non-local interactions} in the Goldstone boson sector. A common procedure is then to perform an operator product expansion on the non-local interactions to produce local interactions in the effective theory. Note that in the usual $\hbar$ expansion of the effective action, hence accounting for loop corrections, this is often referred to as the process of ``matching". In this work, we only focus on the leading-order term in this expansion which describes tree-level processes. The \textit{tree-level} effective Lagrangian is particularly easy to determine as it can be obtained by performing a saddle point approximation of the functional integral over $\sigma$, i.e.

\begin{equation}
    e^{i S_{\text{eff}}[\pi_c]} = e^{iS[\pi_c, \sigma_{\text{cl}}]}\,,
\end{equation}
where $\sigma_{\text{cl}}$ is the classical solution of the equation of motion $\delta S[\pi_c, \sigma_{\text{cl}}]/\delta \sigma_{\text{cl}} = 0$.
The resulting effective Lagrangian $\mathcal{L}_{\text{eff}}$ is genuinely non-local and falls off for momenta larger than the mass of the heavy field. Varying the action (\ref{eq: full theory}) with respect to $\sigma_{\text{cl}}$ gives the following equation of motion

\begin{equation}
    (-\Box + m^2)\,\sigma_{\text{cl}} = \rho \dot{\pi}_c + \dots\,,
\end{equation}
where $\Box \equiv g^{\mu\nu}\nabla_\mu\nabla_\nu = -\partial_t^2 - 3H\partial_t + \tilde{\partial}_i^2$ denotes the d'Alembert operator on $\mathrm{dS}_4$, and the ellipses denote non-linear corrections. Note that it is sufficient to use the \textit{linear} equation of motion for $\sigma_{\text{cl}}$ and to plug it back in the original Lagrangian. Indeed, non-linear corrections to $\sigma_{\text{cl}}$ coming from cubic interactions identically vanish up to cubic order.\footnote{This can be shown in general terms as follows. The Lagrangian we are interested in schematically reads ${\cal L}=-\frac12 \sigma O \sigma+\sigma J(\pi)$ with the equation of motion $ O \sigma=J$ for the field $\sigma$. Replacing $\sigma$ by $O^{-1} J$ in the action gives ${\cal L}=\frac12 (O^{-1} J) J$, which, upon splitting $J=J_{1}+J_{2} +\ldots$ into terms linear in $\pi$, quadratic etc, gives the Lagrangian ${\cal L}=\frac12 (O^{-1} J_1)J_1+\frac12 (O^{-1} J_1 )J_2+\frac12 (O^{-1} J_2 )J_1$ up to cubic order. Instead, if one simply replaces $\sigma$ by the solution to its linear equation of motion $O^{-1} J_1$, one obtains  ${\cal L}=\frac12 (O^{-1} J_1)J_1+(O^{-1} J_1 )J_2$. It is easy to see that both actions agree in the regime of validity of the single-field EFT. For instance, in the local EFT where $O^{-1}$ is written as a series of $\Box^n/m^{2(n+1)}$ terms, the identification follows from $\int \d^4 x \sqrt{-g} J_2 \Box^n J_1=\int \d^4 x \sqrt{-g} J_1 \Box^n J_2$ upon integrations by parts. Analogous arguments hold in the non-local EFT regime when considering $\int \d^3 x J_2 {\cal D}^{-1} J_1=\int \d^3 x J_1 {\cal D}^{-1} J_2$, see the definition of ${\cal D}^{-1}$ after Eq.~\eqref{eq: non-local theory}, which is transparent in Fourier space.} Inverting the equation of motion for $\sigma_{\text{cl}}$ gives

\begin{equation}
\label{eq: formal classical sigma}
    \sigma_{\text{cl}} = \rho \, (-\Box + m^2)^{-1} \dot{\pi}_c\,.
\end{equation}
As long as the low-energy modes described by the effective action obey a non-relativistic dispersion relation such that $\omega^2 \ll k^2$, which is the case when the intrinsic speed of sound of the Goldstone boson is (significantly) reduced, the d'Alembert operator is dominated by spatial gradients. In \ref{subsec: regimes of validity}, we give the precise regime of validity of this approximation. Therefore, one can write the non-local operator in Eq.~(\ref{eq: formal classical sigma}) as a time derivative expansion\footnote{Note that $\tilde{\partial}_i = \partial_i/a$ bears a hidden (cosmic) time dependence in the scale factor $a(t)$. As a result, the order of the operators matters.}

\begin{equation}
\label{eq: time derivative expansion}
    \frac{1}{-\Box + m^2} = \frac{1}{-\tilde{\partial}_i^2 + m^2} \sum_{n=0}^{\infty} (-1)^n \left[(\partial_t^2 + 3H\partial_t)(-\tilde{\partial}_i^2 + m^2)^{-1}\right]^n\,.
\end{equation}
Plugging back Eq.~(\ref{eq: formal classical sigma}) in the action (\ref{eq: full theory}) and keeping only the leading order term in (\ref{eq: time derivative expansion}) leads to the following non-local single-field theory

\begin{tcolorbox}[colframe=white,arc=0pt,colback=greyish2]
\begin{equation}
\label{eq: non-local theory}
    \begin{aligned}
    S_{\text{eff}} &= \int \d^4x \sqrt{-g} \left( \frac{1}{2}\dot{\pi}_c\left[1 + \rho^2\,\mathcal{D}^{-1}\right]\dot{\pi}_c - \frac{c_s^2}{2} (\tilde{\partial}_i \pi_c)^2 - \lambda_1 \dot{\pi}_c (\tilde{\partial}_i \pi_c)^2 -  \lambda_2 \dot{\pi}_c^3 \right. \\
    &\left.-\frac{\rho}{\Lambda_1} (\tilde{\partial}_i \pi_c)^2 \, \mathcal{D}^{-1}\dot{\pi}_c- \frac{\rho}{\Lambda_2} \dot{\pi}_c^2 \, \mathcal{D}^{-1}\dot{\pi}_c - \lambda\rho^2 \dot{\pi}_c \left[\mathcal{D}^{-1}\dot{\pi}_c\right]^2 - \mu\rho^3 \left[\mathcal{D}^{-1}\dot{\pi}_c\right]^3\right)\,,
    \end{aligned}
\end{equation}
\end{tcolorbox}
\noindent where we have introduced the \textit{non-local} differential operator $\mathcal{D}^{-1} = (-\tilde{\partial}_i^2 + m^2)^{-1}$. This effective theory should be understood as the leading order contribution in the time derivative expansion (\ref{eq: time derivative expansion}) where higher-order terms are encoded in an infinite number of operators. As we will see in the next section, the non-local theory (\ref{eq: non-local theory}) accurately describes the dynamics of the Goldstone boson in the almost entire parameter space i.e.~for almost all ranges of $\rho/H$ and $m/H$. However, because this theory is intrinsically single-field, it does not capture the non-perturbative particle production, hence does not describe the conventional cosmological collider signal, visible in ultra-soft limits of cosmological correlators. Indeed, the effects of the massive field are encoded in (non-local) contact interactions which reflect the fact that it propagates instantaneously. As such, the theory (\ref{eq: non-local theory}) does not describe long-range interactions coming from the exchange of an additional massive field. 

\vskip 4pt
Let us stress that the non-local operator $\mathcal{D}^{-1}$ should be considered as a building block on its own, and cannot in general be expressed in terms of a series of simpler operators. When the field is sufficiently heavy, one can write $\mathcal{D}^{-1}=1/m^2 \sum_n (\tilde{\partial}_i^2/m^2)^n$ as a sum of local operators, with the caveat that this only provides an asymptotic expansion. However, the regime of parameters of interest for the low-speed collider rather corresponds to $m^2$ being negligible compared to $\tilde{\partial}_i^2$ around the critical time of sound horizon crossing, where this expansion is clearly not applicable. Conversely, one can wonder whether a tentative fully non-local expansion $\mathcal{D}^{-1} = \tilde{\partial}_i^{-2} \sum_n (m^2 \tilde{\partial}_i^{-2})^n$ as a series of inverse Laplacian operators would be relevant in that situation, but in fact it is not. A simple way to see this is that this expansion is analytic in $m$, hence it can never reproduce the dependence of correlators in $\log(m)$ that is found when considering the full $\mathcal{D}^{-1}$ \cite{Jazayeri:2022kjy}. Note also that for the interactions in \eqref{eq: non-local theory}, each term in this tentative expansion would be infrared divergent. Physically, this obstruction comes from the fact that interactions are not only localized around sound-horizon crossing, as usual for local theories, but rather \textit{spans over a range of time}, from sound horizon crossing to mass-shell crossing where $\tilde{\partial}_i^2 \sim m^2$, and beyond. Hence, any expansion centred around the former event (or any event) is doomed to fail. In the same way as propagators $(\Box-m^2)^{-1}$ in amplitudes cannot be expanded around resonances, the resonance of the cosmological low-speed collider signal is intrinsically non-perturbative. These physical explanations will find their counterparts below in the mathematical properties of the seed function $\I_1$ in \eqref{eq: EFT seed integrals}, see the discussion ``Properties of the EFT seed integrals'' in Sec.~\ref{sec:seed-integrals}.

\subsection{Dispersion Relation}
\label{subsec: dispersion relation}

The quadratic part of the non-local action (\ref{eq: non-local theory}) describes various phenomenologically interesting regimes for the Goldstone boson dynamics, that can be deciphered through its dispersion relation. Here, we clarify the dynamics of the Goldstone boson by looking at extreme regimes. We explicitly show that the non-local theory interpolates between the \textit{reduced speed of sound} regime and the \textit{modified dispersion relation} regime, analogously to the analysis in \cite{Baumann:2011su} for $c_s=1$.

\vskip 4pt
When the energy of the Goldstone boson is above the Hubble scale $\omega \gg H$, its dynamics is well described by plane-wave solutions. This motivated approximation can be extended up to horizon crossing where Hubble friction kicks in and $\pi_c$ starts to freeze. The quadratic part of the non-local action (\ref{eq: non-local theory}) leads to the following dispersion relation

\begin{tcolorbox}[colframe=white,arc=0pt,colback=greyish2]
\begin{equation}
\label{eq: full dispersion relation}
    \omega^2 = \frac{c_s^2 \kp^2}{1 + \frac{\rho^2}{\kp^2+m^2}}\,,
\end{equation}
\end{tcolorbox}
\noindent where $\kp=k/a$ is the physical momentum carried by the Goldstone boson. Let us now take the large mass and the large quadratic mixing limits.

\paragraph{Reduced effective speed of sound.} When the mass of the heavy field becomes large, it dominates the gradient term in the non-local differential operator $\mathcal{D}^{-1} \approx 1/m^2$. The heavy field therefore leaves its imprint in the dynamics of the Goldstone boson in the form of an induced speed of sound

\begin{equation}
\label{eq: induced sound speed}
    \tilde{c}_s^{-2} = c_s^{-2}\left(1 + \frac{\rho^2}{m^2}\right)\,.
\end{equation}
The massless mode then freezes at the \textit{effective} sound horizon $|\tilde{c}_s k\tau| \sim 1$ while having a linear dispersion relation $\omega \approx \tilde{c}_s \kp$. Clearly, this limiting regime is valid as long as the physical momentum $k_{\text{p}}$ of the Goldstone boson verifies $k_{\text{p}}\ll m$. Using the dispersion relation, one can see that this regime is an accurate description of the quadratic dynamics if $\omega\ll \omega_{\text{new}}=\tilde{c}_s m$, where $\omega_{\text{new}}$ marks the energy scale at which higher-order terms in the derivative $-\tilde{\partial}_i^2/m^2$ expansion should be taken into account. In order to be able to initialise the mode functions in the proper vacuum, one must require that $\omega_{\text{new}}$ be well above the Hubble scale i.e. $\omega_{\text{new}} \gg H$. Therefore, the reduced effective speed of sound regime is valid if

\begin{equation}
\label{eq: reduced eff speed of sound validity}
    \frac{H}{\tilde{c}_s} \ll m\,.
\end{equation}
It should be noted that the region with a significantly reduced effective speed of sound compared to the ``bare'' one $\tilde{c}_s\ll c_s$ is rather limited as it requires $\rho\gg m$ while maintaining (\ref{eq: reduced eff speed of sound validity}). This region of parameter space does not give rise to the low-speed collider and is not the one of interest in this paper. In the two-field picture of this regime, we show in the insert below that the energy gap between the heavy and the light modes increases due to the quadratic mixing.

\paragraph{Modified dispersion relation.} In the regime where $\kp \gg m$, the gradient term dominates the mass in the non-local differential operator $\mathcal{D}^{-1} \approx -\tilde{\partial}_i^{-2}$. Additionally, for $\rho \gg \kp$, the second term containing $\mathcal{D}^{-1}$ in the quadratic part of the action (\ref{eq: non-local theory}) dominates over the usual $\dot{\pi}_c^2$ term. This regime is similar but still different from ghost inflation \cite{ArkaniHamed:2003uz}, and describes the dynamics of the Goldstone boson having a modified dispersion relation $\omega \approx c_s \kp^2/\rho$. Evaluating this dispersion relation at the Hubble scale $\omega = H$, one can see that the Goldstone boson freezes at the time scale set by $c_s^{1/2}|k\tau| \sim \sqrt{\rho/H}$, before the usual horizon crossing $|k\tau|\sim 1$.\footnote{In this regime, the quadratic theory can be fully solved analytically. The equation of motion is found to be

\begin{equation}
    \ddot{\pi}_c + 5H \dot{\pi}_c + \frac{c_s^2}{\rho^2} \,\kp^4\,\pi_c = 0\,.
\end{equation}
After quantizing the theory and imposing Bunch-Davis initial conditions, the mode functions are completely determined in term of the Hankel function (see \cite{Baumann:2011su, CosmoFlow} for more details) 

\begin{equation}
\label{eq: MDR mode function}
    \pi_{c, k}(\tau) = \sqrt{\frac{\pi}{4}} \frac{H}{\rho} \frac{H}{\sqrt{2k^3}} (-k\tau)^{5/2} \H^{(1)}_{5/4} \left(\frac{c_s}{2}\frac{H}{\rho}(k\tau)^2\right)\,.
\end{equation}
}

\vskip 4pt
Let us now derive the precise regime of validity of this regime. Using the dispersion relation, the first condition $\kp \gg m$ evaluated at the Hubble scale gives $c_s m^2/\rho \ll H$. This defines a new energy scale $M \equiv c_s m^2/\rho$. When $M$ approaches the Hubble scale, the dynamics becomes sensitive to higher-order powers of $m^2/\tilde{\partial}_i^2$, i.e.~these terms are no longer irrelevant and the effective description of the Goldstone boson having a non-linear dispersion relation breaks down.
The second condition $\rho/\kp\gg 1$, upon using the dispersion relation at $\omega=H$, gives $\rho/H\gg 1/c_s$. To sum up, the modified dispersion relation regime is valid as long as
\begin{equation}
\label{eq: MDR regime of validity}
    M\equiv \frac{ c_s m^2}{\rho} \ll H \,, \hspace*{0.5cm} \text{and} \hspace*{0.5cm} \frac{\rho}{H} \gg \frac{1}{c_s}\,.
\end{equation}

\vskip 4pt
From the two asymptotic regimes above, it is clear that the dispersion relation (\ref{eq: full dispersion relation}) gives a unified description in the $(c_s, m, \rho)$ parameter space of the full quadratic dynamics, except in the regime where $m<H$ and $\rho<H$, where the non-local EFT is not valid.
Indeed, if the additional field is light and weakly mixed to $\pi_c$, the Goldstone boson is continuously sourced by $\sigma$ on super-Hubble scales. The dynamics is then sensitive to the amount of $e$-folds elapsed from the moment at which the mode crosses the horizon to the end of inflation (see \cite{CosmoFlow} for an exhaustive analysis of the full parameter space). 
In the next section, we derive the precise regime of validity of the non-local single-field theory (\ref{eq: non-local theory}).

\begin{framed}
{\small \noindent {\it Reduced effective sound speed energy gap.}---In this insert, we examine the energy gap between the heavy and the massless modes in the reduced effective speed of sound regime when $c_s \ll 1$. In flat-space, the quadratic two-field dynamics can be solved analytically. The exact dispersion relations are given in Eq.~(\ref{eq: exact dispersion relations}). The leading-order solution at weak mixing, corresponding to the completely decoupled case $\rho/m \rightarrow 0$, is given by $\omega_-^2 = c_s^2 \kp^2$ and $\omega_+^2 = \kp^2 + m^2$. Let us now examine the perturbative limit $\rho\lesssim m$. Taking this limit in the exact dispersion relations (\ref{eq: exact dispersion relations}) at next-to-leading order leads to 
\begin{equation}
    \omega_-^2 \approx c_s^2 \kp^2 - \rho^2\, \frac{c_s^2 \kp^2}{(1-c_s^2)\kp^2 + m^2}\,,\hspace*{0.5cm}\text{and}\hspace*{0.5cm} \omega_+^2 \approx \kp^2 + m^2 + \rho^2 \frac{\kp^2+m^2}{(1-c_s^2)\kp^2 + m^2}\,.
\end{equation}
At reduced sound speed $c_s\ll 1$, the heavy mode dispersion relation reduces to $\omega_+^2 \approx \kp^2 + m_{\text{eff}}^2$ where $m_{\text{eff}}^2 = m^2 + \rho^2$. For the massless mode, assuming $(1-c_s^2)\kp^2\ll m^2$, the dispersion relation boils down to $\omega_-^2\approx c_s^2\,(1-\rho^2/m^2)\kp^2$, recovering the induced sound speed $\tilde{c}_s^2 = c_s^2\,(1+\rho^2/m^2)^{-1} \approx c_s^2\,(1 - \rho^2/m^2)$ in the limit $\rho\lesssim m$, see Eq.~(\ref{eq: induced sound speed}). Note that the Goldstone boson cannot acquire an effective mass from the quadratic mixing because it is protected by shift symmetry. We summarise the effect of the quadratic coupling at weak mixing in the following energy-level diagram 

\begin{center}
\hspace*{-2cm}
\begin{tikzpicture}[line width=1. pt, scale=2]
		
		\draw[thick, ->] (-0.2, -0.5) -- (-0.2, 1.7);
		\node at (-0.2, 1.9) {$\omega^2(k)$};
		
		
		\draw[very thick, -, pyblue] (0.3, 0.9) -- (1.2, 0.9);
		\node at (0.75, 1.1) {\textcolor{pyblue}{$k^2+m^2$}};
		
		\draw[very thick, -, pyred] (0.3, 0.4) -- (1.2, 0.4);
		\node at (0.8, 0.2) {\textcolor{pyred}{$c_s^2k^2$}};
		
		
		\draw[very thick, -, pyblue] (1.9, 1.4) -- (2.8, 1.4);
		\node at (2.35, 1.6) {\textcolor{pyblue}{$k^2+m_{\text{eff}}^2$}};
		
		\draw[very thick, -, pyred] (1.9, -0.1) -- (2.8, -0.1);
		\node at (2.35, -0.3) {\textcolor{pyred}{$\tilde{c}_s^2k^2$}};
		
		
		\draw[thick, -, pygreen] (-0.2, 0.9) -- (-0.2, 1.4);
		\draw[thick, -, pygreen] (-0.3, 0.9) -- (-0.1, 0.9);
		\draw[thick, -, pygreen] (-0.3, 1.4) -- (-0.1, 1.4);
		\node at (-0.9, 1.25) {\textcolor{pygreen}{effective mass}};
		\node at (-0.9, 1.05) {\textcolor{pygreen}{effect}};

		\draw[thick, -, pygreen] (-0.2, -0.1) -- (-0.2, 0.4);
		\draw[thick, -, pygreen] (-0.3, -0.1) -- (-0.1, -0.1);
		\draw[thick, -, pygreen] (-0.3, 0.4) -- (-0.1, 0.4);
		\node at (-0.9, 0.25) {\textcolor{pygreen}{induced sound}};
		\node at (-0.9, 0.05) {\textcolor{pygreen}{speed effect}};

		
		\node at (0.8, -0.7) {uncoupled limit};
		\node at (0.8, -0.9) {$\rho/m\rightarrow 0$};
		
		\node at (2.4, -0.7) {$\rho/m$ correction};
\end{tikzpicture} 
\end{center}
}
\end{framed}

\subsection{Regimes of Validity}
\label{subsec: regimes of validity}

It is important to determine the precise regime of validity of the non-local theory (\ref{eq: non-local theory}), as the leading-order term in the derivative expansion (\ref{eq: time derivative expansion}). Indeed, we wish to treat higher-order terms as perturbative corrections to the effective action.

\paragraph{Flat-space intuition.} Before considering cosmological correlators, for pedagogical reasons, it is instructive to analyse the dynamics in the flat-space limit $H\rightarrow 0$. Let us come back to the full two-field picture (\ref{eq: full theory}) and consider the four-point amplitude of Goldstone bosons due to the exchange of a $\sigma$ field, with interactions $g\pi_c^2\sigma$ for definiteness. This amplitude is given by
\begin{equation}
    \mathcal{A} =  \mathcal{A}_s + \mathcal{A}_t + \mathcal{A}_u\,,   
\end{equation}
where the three contributions are the $s, t$ and $u$ channels respectively. We denote the particle four-momenta as $p_a^\mu = (\omega_a, \bm{p}_a)$ where $a\in \{1, \ldots, 4\}$ labels the momentum of each particle. Importantly, since boost invariance is broken, the scattering amplitude is no longer a function of the Lorentz-invariant momentum inner products. Instead, the $s$-channel amplitude is given by
\begin{equation}
    \mathcal{A}_s = \frac{-g^2}{(\omega_1+\omega_2)^2 - (\bm{p}_1 + \bm{p}_2)^2 - m^2}\,.
\end{equation}
where we use all incoming particles convention so that $\sum_a \omega_a = 0$ and $\sum_a \bm{p}_a = 0$. When writing the non-local effective theory (\ref{eq: non-local theory}), we have precisely considered that the term $(\omega_1+\omega_2)^2$ is negligible compared to the term $(\bm{p}_1 + \bm{p}_2)^2$, which is equivalent to considering that the time derivative is negligible compared to the gradient term in the equation of motion of $\sigma$. Because the Goldstone boson propagates with a reduced sound speed, we have $(\omega_1+\omega_2)^2 = c_s^2(p_1+p_2)^2 \approx 4c_s^2 k^2$, where $k$ is the characteristic momentum carried by incoming particles.

\vskip 4pt
Let us now consider two situations. The first one is when the transfer momentum carried by the exchanged particle, denoted $q=|\bm{q}|$, is of the same magnitude as the incoming particle, that is $q^2\equiv (\bm{p}_1 + \bm{p}_2)^2 \approx 4k^2$. If the sound speed of the Goldstone boson is sufficiently reduced $c_s \ll 1$, then we always have $(\omega_1+\omega_2)^2 \ll (\bm{p}_1 + \bm{p}_2)^2$ and the non-local theory correctly describes the dynamics for this kinematic configuration. The second case is that of the transfer momentum being soft. In this case, $k$ is the hard momentum. We immediately see that in the ultra-soft limit---i.e. when $q/k \ll 2c_s$---, the term $(\omega_1+\omega_2)^2$ is no longer negligible compared to $(\bm{p}_1 + \bm{p}_2)^2$. In other words, the propagator is no longer dominated by the gradient term and the non-local theory (\ref{eq: non-local theory}) breaks down. Crucially, this means that the regime of validity of the non-local effective theory not only depends on the parameters of the theory but also on the \textit{kinematics}, and fails to correctly describe ultra-soft limits of scattering amplitudes or---as we will see in the next paragraph---cosmological correlators.

\paragraph{Soft limits of cosmological correlators.} We now extend this picture to cosmological correlators. The time dependence of the background and of the physical momenta makes the analysis more intricate. However, when the energy of the Goldstone boson is above the Hubble scale $\omega\gtrsim H$, we still can use the notion of dispersion relation to derive the precise regime of validity of the non-local effective theory. It should be noted that when $\omega \lesssim H$, corrections due to Hubble friction become important. Although these corrections can be accounted for systematically by working at the level of the mode functions, we do not follow this path for simplicity of presentation (see \cite{Jazayeri:2022kjy} for more details).

\vskip 4pt
In the full two-field picture, let us consider a generic cubic interaction describing the decay of a massive field $\sigma$ into two Goldstone bosons appearing on the external legs of a correlator. We are voluntarily being agnostic on the precise form of this interaction because it is unimportant for the following argument. Such process is described by $(-\Box +m^2)\sigma = J(\pi_c)$ where the sourcing $J$ is quadratic in the field $\pi_c$. Depending on the precise channel that we consider, the massive field $\sigma$ can carry a hard (i.e.~short) or a soft (i.e.~long) momentum. 
In the following, we examine the most constraining situation, where the Goldstone bosons are the short modes $\kp = k_{\text{S}}/a$ and the massive field is the long mode $\qp = k_{\text{L}}/a$. Considering only the leading order term in the time derivative expansion (\ref{eq: time derivative expansion}) enforces the following inequality to hold for the effective action to be correct:

\begin{equation}
\label{eq: regime of validity inequality}
    \omega^2(\kp) \lesssim \qp^2 + m^2\,.
\end{equation}
Here, $\omega^2(\kp)$ should be understood as the sum of the energies of both Goldstone bosons. Of course, we stress that the stronger the inequality is, the better the approximation of working with the effective action will be. The complication when dealing with cosmological correlators is that (\ref{eq: regime of validity inequality}) depends on time, which is hidden in the physical momenta. 
In order for the non-local EFT to give an accurate description of cosmological correlators, we do not need this inequality to hold at all times (the fields are effectively decoupled deep inside the horizon $\omega\gg H$ where the system asymptotically reaches the vacuum state), but it should be valid at the critical time of horizon crossing of the Goldstone boson $\omega\sim H$.\footnote{A crucial aspect of the low-speed collider is that non-trivial interactions also occur \textit{after} horizon crossing, where the plane-wave analysis ceases to be valid. The full analysis done is \cite{Jazayeri:2022kjy} shows that limiting the reasoning to $\omega \sim H$ is sufficient for our purpose here.} Let us now examine the cases of weak and strong mixing separately. 

 \begin{figure}[t!]
     \centering
     \includegraphics[scale=1]{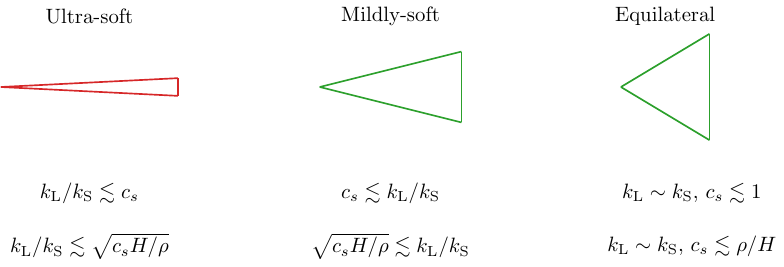}
     \caption{Summary of the regimes of validity of the effective action (\ref{eq: non-local theory}) for cosmological correlators, in the low-speed collider regime $\rho \lesssim H/c_s$ (\textit{top}) and in the modified dispersion relation regime (\textit{bottom}). We have coloured in \textcolor{pyred}{red} the kinematic configuration that is not accurately described by the non-local effective theory and in \textcolor{pygreen}{green} those that are. When the massive field is heavy $m\gg H$, the effective description is valid in all configurations. For purpose of illustration, we have displayed the regimes of validity for the three-point correlators as they are our main interest in this work, but the same conditions should be verified for higher-point correlators.}
     \label{fig: regimes of validity}
 \end{figure}

\vskip 4pt
At weak mixing, the Goldstone boson dynamics is described by a linear dispersion relation $\omega^2(\kp) = c_s^2 \kp^2$. Evaluating this expression for energies of order the Hubble scale gives $c_sk_{\text{S}} = aH$. Plugging back this condition in (\ref{eq: regime of validity inequality}) gives 

\begin{equation}
\label{eq: regime of validity weak mixing}
    1 \lesssim \frac{1}{c_s^2} \left(\frac{k_{\text{L}}}{k_{\text{S}}}\right)^2 + \left(\frac{m}{H}\right)^2\,.
\end{equation}
Two situations appear. If the field $\sigma$ is heavy $m\gtrsim H$, then (\ref{eq: regime of validity weak mixing}) is always valid. If the mass of the field $\sigma$ is of order the Hubble scale $m\sim H$, then one needs to ensure that the first term on the right-hand side of the inequality is greater than unity, i.e. $k_{\text{L}}/k_{\text{S}} \gtrsim c_s$. As a result, the momentum carried by $\sigma$ cannot be arbitrarily soft.\footnote{We now see that if we had chosen one of the Goldstone boson to be the long mode and the other one to be the short mode, then the massive field would have been a short mode. Therefore, $\omega$ on the left-hand side of (\ref{eq: regime of validity inequality}) would have been the sum of the energies of a long and a short mode, therefore completely dominated by the short mode, ultimately creating no hierarchy of scales.} Notice that within the regime of the low-speed collider, i.e. for masses such that $m\lesssim H/c_s$, one can neglect the mass term for $\omega(\kp) \sim H$, as the non-local dynamics is dominated by the gradients. However, as explained in Sec.~\ref{subsection: integrate out}, it is important to keep in mind that it can not be thrown away as it plays the role of an infrared regulator for the non-local theory. Also, from (\ref{eq: regime of validity weak mixing}), it is clear that close to equilateral configurations $k_{\text{L}}\sim k_{\text{S}}$, the effective action is valid as long as the sound speed is reduced. As we will see later in Sec.~\ref{sec:physics-LSC}, the weak mixing condition (\ref{eq: regime of validity weak mixing}) also extends to the interesting strongly-mixed low-speed collider regime $m/H \lesssim \rho/H \lesssim 1/c_s$, see Fig.~\ref{fig: weak&strong low-speed collider pheno}.

\vskip 4pt
At strong enough mixing, the Goldstone boson dynamics is described by the non-linear dispersion relation $\omega^2(\kp) = c_s^2 \kp^4/\rho^2$. For energies around the Hubble scale, this gives $c_s k_{\text{S}}^2 = a^2 H \rho$. Substituting this in (\ref{eq: regime of validity inequality}) yields

\begin{equation}
\label{eq: regime of validity strong mixing}
    1 \lesssim \frac{1}{c_s} \left(\frac{k_{\text{L}}}{k_{\text{S}}}\right)^2 \frac{\rho}{H} + \left(\frac{m}{H}\right)^2\,.
\end{equation}
If $m\gtrsim H$, this inequality is always satisfied. Likewise the weak mixing case, if $m\sim H$, one needs to require $k_{\text{L}}/k_{\text{S}} \gtrsim  \sqrt{c_s H/\rho}$. Certainly, as we have illustrated in the flat-space picture, ultra-soft limits of cosmological correlators are not accurately described by the non-local effective theory. For the reader's convenience, we have collected the established regimes of validity in Fig.~\ref{fig: regimes of validity}. 

\section{Cosmological Correlators}
\label{sec: Cosmological Correlators}

We will now compute the effects of the additional massive field $\sigma$ on correlators of the Goldstone boson. However, the quadratic mixing---which can be strong---leads to complications. In this section, we treat the cases of weak and strong mixings separately, using two different approaches.

\vskip 4pt
In the weak mixing regime, we will treat the quadratic coupling perturbatively. The various cubic interactions in the full theory (\ref{eq: full theory}) therefore gives rise to single-, double-, and triple-exchange diagrams for the bispectrum. 
Not being interested in the cosmological collider signal here, but rather in the more striking low-speed collider signature, we will extract it by using the effective single-field theory (\ref{eq: non-local theory}), within which all diagrams collapse into simple (non-local) contact ones.

\vskip 4pt
In the strong mixing regime, we will exploit the cosmological flow approach \cite{Werth:2023pfl, CosmoFlow} to numerically compute the desired correlators. This way, we will confirm the validity of our analytical, albeit necessarily approximate, reasonings, see Sec.~\ref{sec:physics-LSC}. 

\subsection{Weak Mixing}

Before specifying our analysis to the non-local EFT, it is worth taking a detour and coming back to the full theory (\ref{eq: full theory}). Following the bootstrap approach used in \cite{Jazayeri:2022kjy} (see \cite{Baumann:2022jpr} for a recent review), we will first trade the Goldstone boson for an auxiliary field $\varphi$ with mass $m_\varphi^2 = 2H^2$ that propagates with a speed of sound $c_s$, and identify a set of seed correlators that serve as building blocks for all desired correlators. 
We then introduce well-chosen weight-shifting operators that map correlators of $\varphi$ to correlators of Goldstone bosons as external fields, which is the case of interest in this work. This strategy enables one to obtain a complete set of correlators, regardless of the precise interactions.

\vskip 4pt
In a second phase, we will see that the seed correlators collapse into simple contact diagrams when the exchanged massive field is integrated out in a non-local manner. Eventually, within the framework of the non-local EFT (\ref{eq: non-local theory}), we will give explicit solutions for single-, double- and triple-exchange correlators of massless external fields.

\subsubsection{Weight-Shifting Operators}
\label{subsubsec : weight-shifting operators}

Let us introduce an auxiliary scalar field $\varphi(\bm{x}, t)$ with mass $m_\varphi^2 = 2H^2$ that propagates with a speed of sound $c_s$.\footnote{For $c_s=1$, $\varphi$ reduces to a conformally coupled field in $\text{dS}_4$.}
The associated mode function is determined by solving its free equation of motion and imposing the Bunch-Davies state as initial condition. It is given by

\begin{equation}
    \varphi_k(\tau) = -\frac{H}{\sqrt{2c_s k}}\, \tau e^{-ic_s k \tau}\,,
\end{equation}
where $\tau$ is the conformal time. The fundamental objects that generate \textit{all} correlators---regardless of the precise form of the interactions---are the equal-time four- and six-point correlation functions of $\varphi$ generated by the interactions 

\begin{equation}
\mathcal{L}/a^3 = - g_{1}\varphi^2 \sigma - g_{2}\varphi^2 \sigma^2 - g_{3}\sigma^3\,,
\end{equation}
where $g_i$ are arbitrary coupling constants. Let us stress that by considering this proxy theory---which may seem over-complicated at first sight---we are able to easily obtain all correlators generated by arbitrary interactions, even beyond those considered in (\ref{eq: non-local theory}), see App.~\ref{app: WS Operators} for more details. Using scaling symmetry, we define the seed functions $\textcolor{pyred}{F_{1, 2, 3}}$ as the corresponding solutions of the following diagrams

\begin{subequations}
\label{vpicordef}
\begin{align}
    \braket{\varphi_{\bm{k}_1} \ldots \varphi_{\bm{k}_4}}_{(1)}' &=
    \hspace*{0.5cm}
    \raisebox{-\height/3}{
\begin{tikzpicture}[line width=1. pt, scale=2.5]
\draw[black] (0.2, 0) -- (0.4, -0.6);
\draw[black] (0.6, 0) -- (0.4, -0.6);
\draw[black] (1, 0) -- (1.2, -0.6);
\draw[black] (1.4, 0) -- (1.2, -0.6);
\draw[pyblue] (0.4, -0.6) -- (1.2, -0.6);
\draw[fill=black] (0.4, -0.6) circle (.03cm);
\draw[fill=black] (1.2, -0.6) circle (.03cm);
\draw[lightgray2, line width=0.8mm] (0, 0) -- (1.6, 0);
\node at (0.2, 0.15) {\footnotesize{$\bfk_1$}};
\node at (0.6, 0.15) {\footnotesize{$\bfk_2$}};
\node at (1, 0.15) {\footnotesize{$\bfk_3$}};
\node at (1.4, 0.15) {\footnotesize{$\bfk_4$}};
\end{tikzpicture} 
} \hspace*{0.5cm} = g_1^2 \left(\prod_{i=1}^{4} \dfrac{\tau_0}{c_s k_i}\right)\, \\
& \hspace*{5.8cm} \times \dfrac{1}{\Ms_{12}} \textcolor{pyred}{F_1\left(\dfrac{c_s k_{12}}{\Ms_{12}},\dfrac{c_s k_{34}}{\Ms_{12}}\right)}\,, \nonumber \\
\braket{\varphi_{\bm{k}_1} \ldots \varphi_{\bm{k}_6}}_{(2)}' &= 
\raisebox{-\height/3}{
\begin{tikzpicture}[line width=1. pt, scale=2.5]
\node at (0.2, 0.15) {\footnotesize{$\bfk_1$}};
\node at (0.6, 0.15) {\footnotesize{$\bfk_2$}};
\node at (0.85, 0.15) {\footnotesize{$\bfk_3$}};
\node at (1.2, 0.15) {\footnotesize{$\bfk_4$}};
\node at (1.42, 0.15) {\footnotesize{$\bfk_5$}};
\node at (1.85, 0.15) {\footnotesize{$\bfk_6$}};
\draw[black] (0.2, 0) -- (0.4, -0.6);
\draw[black] (0.6, 0) -- (0.4, -0.6);
\draw[black] (0.8, 0) -- (1, -0.6);
\draw[black] (1.2, 0) -- (1, -0.6);
\draw[black] (1.4, 0) -- (1.6, -0.6);
\draw[black] (1.8, 0) -- (1.6, -0.6);
\draw[pyblue] (0.4, -0.6) -- (1, -0.6);
\draw[pyblue] (1, -0.6) -- (1.6, -0.6);
\draw[fill=black] (0.4, -0.6) circle (.03cm);
\draw[fill=black] (1, -0.6) circle (.03cm);
\draw[fill=black] (1.6, -0.6) circle (.03cm);
\draw[lightgray2, line width=0.8mm] (0, 0) -- (2, 0);
\end{tikzpicture} 
} =g_1^2g_2 \left(\prod_{i=1}^{6} \dfrac{\tau_0}{c_s k_i}\right)\, \\
& \hspace*{5.8cm} \times \dfrac{1}{\Ms^3_{12}} \textcolor{pyred}{F_2\left(\dfrac{c_s k_{12}}{\Ms_{12}},\dfrac{c_s k_{34}}{\Ms_{12}},\dfrac{c_s k_{56}}{\Ms_{12}},\dfrac{\Ms_{56}}{\Ms_{12}}\right)}\,, \nonumber \\
\braket{\varphi_{\bm{k}_1} \ldots \varphi_{\bm{k}_6}}_{(3)}' &=
\raisebox{-\height/3}{
\begin{tikzpicture}[line width=1. pt, scale=2.5]
\draw[black] (0.2, 0) -- (0.4, -0.6);
\draw[black] (0.6, 0) -- (0.4, -0.6);
\draw[black] (0.8, 0) -- (1, -0.3);
\draw[black] (1.2, 0) -- (1, -0.3);
\draw[black] (1.4, 0) -- (1.6, -0.6);
\draw[black] (1.8, 0) -- (1.6, -0.6);
\draw[pyblue] (1, -0.3) -- (1, -0.6);
\draw[pyblue] (0.4, -0.6) -- (1, -0.6);
\draw[pyblue] (1, -0.6) -- (1.6, -0.6);
\draw[fill=black] (0.4, -0.6) circle (.03cm);
\draw[fill=black] (1, -0.6) circle (.03cm);
\draw[fill=black] (1, -0.3) circle (.03cm);
\draw[fill=black] (1.6, -0.6) circle (.03cm);
\draw[lightgray2, line width=0.8mm] (0, 0) -- (2, 0);
\node at (0.2, 0.15) {\footnotesize{$\bfk_1$}};
\node at (0.55, 0.15) {\footnotesize{$\bfk_2$}};
\node at (0.83, 0.15) {\footnotesize{$\bfk_3$}};
\node at (1.15, 0.15) {\footnotesize{$\bfk_4$}};
\node at (1.43, 0.15) {\footnotesize{$\bfk_5$}};
\node at (1.8, 0.15) {\footnotesize{$\bfk_6$}};
\end{tikzpicture} 
} =g_1^3 g_3 \left(\prod_{i=1}^{6} \dfrac{\tau_0}{c_s k_i}\right)\, \\ 
& \hspace*{5.8cm}\times \dfrac{1}{\Ms^3_{12}} \textcolor{pyred}{F_3\left(\dfrac{c_s k_{12}}{\Ms_{12}},\dfrac{c_s k_{34}}{\Ms_{12}},\dfrac{c_s k_{56}}{\Ms_{12}},\dfrac{\Ms_{34}}{\Ms_{12}},\dfrac{\Ms_{56}}{\Ms_{12}}\right)}\,, \nonumber
\end{align}
\end{subequations}
where $\Ms_{ij} = |\bm{k}_i + \bm{k}_j|$, $k_{ij} = k_i + k_j$, and $\tau_0$ marks the end of inflation surface. Having 
 defined these seed functions, we now derive correlators of the Goldstone boson by introducing a set of weight-shifting operators that map diagrams with external fields $\pi_c$ to those with external fields $\varphi$. The reason why this procedure is universal, regardless of the form of the considered interactions, is because the mode functions of both fields can be easily related by simple differential operators. Importantly, as the weight-shifting operators only act on external fields of a specific diagram, they are valid for any bulk theory, i.e.~be it the full two-field picture or within the effective single-field theory. To avoid clutter, we leave their precise derivation in Appendix \ref{app: WS Operators}.

\paragraph{Power spectrum correction.} The effect of the quadratic mixing $\dot{\pi}_c\sigma$ can be obtained from the cubic interaction $\varphi^2\sigma$ after taking a suitable soft limit on the external field.
From this, one obtains the leading correction to the two-point correlator of $\pi_c$ as

\begin{equation}
\raisebox{-\height/2}{
\begin{tikzpicture}[line width=1. pt, scale=2.5]
\draw[red] (0.2, 0) -- (0.4, -0.6);
\draw[pyred] (1.4, 0) -- (1.2, -0.6);
\draw[pyblue] (0.4, -0.6) -- (1.2, -0.6);
\draw[fill=black] (0.4, -0.6) circle (.03cm);
\draw[fill=black] (1.2, -0.6) circle (.03cm);
\draw[lightgray2, line width=0.8mm] (0, 0) -- (1.6, 0);
\node at (0.4, -0.8) {\footnotesize{$\dot{\pi}_c\sigma$}};
\node at (1.2, -0.8) {\footnotesize{$\dot{\pi}_c\sigma$}};
\node at (0.2, 0.15) {\footnotesize{$\bfk_1$}};
\node at (1.4, 0.15) {\footnotesize{$\bfk_2$}};
\end{tikzpicture} 
} = \dfrac{\rho^2}{2c_s^2}\dfrac{1}{k_1^3} \, F_1(c_s,c_s)\,.
\end{equation}
The corresponding weight-shifting operator simply consists in taking the soft limit of two external legs $k_2$ and $k_3$ while keeping the diagram connected.

\paragraph{Single-exchange diagrams.} Both single-exchange diagrams arising from the interactions $\dot{\pi}_c^2\sigma$ and $(\partial_i \pi_c)^2\sigma$ can be obtained from the same seed function $F_1$ \cite{Jazayeri:2022kjy}. They read

\begin{subequations}
\label{wsform}
    \begin{align}
    \raisebox{-\height/2}{
\begin{tikzpicture}[line width=1. pt, scale=2.5]
\draw[pyred] (0.2, 0) -- (0.4, -0.6);
\draw[pyred] (0.6, 0) -- (0.4, -0.6);
\draw[pyred] (1.4, 0) -- (1.2, -0.6);
\draw[pyblue] (0.4, -0.6) -- (1.2, -0.6);
\draw[fill=black] (0.4, -0.6) circle (.03cm);
\draw[fill=black] (1.2, -0.6) circle (.03cm);
\draw[lightgray2, line width=0.8mm] (0, 0) -- (1.6, 0);
\node at (0.4, -0.8) {\footnotesize{$(\partial_i \pi_c)^2\sigma$}};
\node at (1.2, -0.8) {\footnotesize{$\dot{\pi}_c\sigma$}};
\node at (0.2, 0.15) {\footnotesize{$\bfk_1$}};
\node at (0.6, 0.15) {\footnotesize{$\bfk_2$}};
\node at (1.4, 0.15) {\footnotesize{$\bfk_3$}};
\end{tikzpicture} 
} &= -\dfrac{\rho}{2c_s^3}\dfrac{H}{\Lambda_1}\dfrac{1}{(k_1 k_2 k_3)^2}\dfrac{\bfk_1\cdot\bfk_2}{k_1 k_2}\\
&\times \left(1-k_{12}\dfrac{\partial}{\partial k_{12}}+k_1 k_2 \dfrac{\partial^2}{\partial k_{12}^2}\right) F_1\left(\dfrac{c_s k_{12}}{k_3},c_s\right)\,, \nonumber \\
\raisebox{-\height/2}{
\begin{tikzpicture}[line width=1. pt, scale=2.5]
\draw[pyred] (0.2, 0) -- (0.4, -0.6);
\draw[pyred] (0.6, 0) -- (0.4, -0.6);
\draw[pyred] (1.4, 0) -- (1.2, -0.6);
\draw[pyblue] (0.4, -0.6) -- (1.2, -0.6);
\draw[fill=black] (0.4, -0.6) circle (.03cm);
\draw[fill=black] (1.2, -0.6) circle (.03cm);
\draw[lightgray2, line width=0.8mm] (0, 0) -- (1.6, 0);
\node at (0.4, -0.8) {\footnotesize{$\dot{\pi}_c^2\sigma$}};
\node at (1.2, -0.8) {\footnotesize{$\dot{\pi}_c\sigma$}};
\node at (0.2, 0.15) {\footnotesize{$\bfk_1$}};
\node at (0.6, 0.15) {\footnotesize{$\bfk_2$}};
\node at (1.4, 0.15) {\footnotesize{$\bfk_3$}};
\end{tikzpicture} 
} &= -\dfrac{\rho}{4 c_s^5}\frac{H}{\Lambda_2}\dfrac{k_1 k_2}{(k_1 k_2 k_3)^2} \dfrac{\partial^2}{\partial k_{12}^2} F_1\left(\dfrac{c_s k_{12}}{k_3},c_s\right)\,.
    \end{align}
\end{subequations}

\paragraph{Double-exchange diagram.} The double-exchange diagram is related to the double-exchange six-point function $F_2$ by the relation

\begin{equation}
\label{wsform2}
\raisebox{-\height/2}{
\begin{tikzpicture}[line width=1. pt, scale=2.5]
\draw[pyred] (0.2, 0) -- (0.4, -0.6);
\draw[pyred] (1.8, 0) -- (1.6, -0.6);
\draw[pyred] (1, 0) -- (1, -0.6);
\draw[pyblue] (0.4, -0.6) -- (1, -0.6);
\draw[pyblue] (1, -0.6) -- (1.6, -0.6);
\draw[fill=black] (0.4, -0.6) circle (.03cm);
\draw[fill=black] (1, -0.6) circle (.03cm);
\draw[fill=black] (1.6, -0.6) circle (.03cm);
\draw[lightgray2, line width=0.8mm] (0, 0) -- (2, 0);
\node at (0.4, -0.8) {\footnotesize{$\dot{\pi}_c\sigma$}};
\node at (1.6, -0.8) {\footnotesize{$\dot{\pi}_c\sigma$}};
\node at (1, -0.8) {\footnotesize{$\dot{\pi}_c\sigma^2$}};
\node at (0.2, 0.15) {\footnotesize{$\bfk_1$}};
\node at (1, 0.15) {\footnotesize{$\bfk_2$}};
\node at (1.8, 0.15) {\footnotesize{$\bfk_3$}};
\end{tikzpicture} 
} = 
- \lambda\,\dfrac{\rho^2}{c_s^3} \dfrac{1}{k_1^4 k_2 k_3} F_2\left(c_s,\dfrac{c_s k_2}{k_1},\dfrac{c_s k_3}{k_1},\dfrac{k_3}{k_1}\right)\,.
\end{equation}

\paragraph{Triple-exchange diagram.} The triple-exchange diagram is related to the seed function $F_3$ after taking suitable soft limits. It is given by

\begin{equation}
\label{wsform3}
    \raisebox{-\height/2}{
\begin{tikzpicture}[line width=1. pt, scale=2.5]
\draw[pyred] (0.2, 0) -- (0.4, -0.6);
\draw[pyred] (1.8, 0) -- (1.6, -0.6);
\draw[pyred] (1, 0) -- (1, -0.3);
\draw[pyblue] (1, -0.3) -- (1, -0.6);
\draw[pyblue] (0.4, -0.6) -- (1, -0.6);
\draw[pyblue] (1, -0.6) -- (1.6, -0.6);
\draw[fill=black] (0.4, -0.6) circle (.03cm);
\draw[fill=black] (1, -0.6) circle (.03cm);
\draw[fill=black] (1.6, -0.6) circle (.03cm);
\draw[fill=black] (1, -0.3) circle (.03cm);
\draw[lightgray2, line width=0.8mm] (0, 0) -- (2, 0);
\node at (0.4, -0.8) {\footnotesize{$\dot{\pi}_c\sigma$}};
\node at (1.6, -0.8) {\footnotesize{$\dot{\pi}_c\sigma$}};
\node at (1, -0.8) {\footnotesize{$\sigma^3$}};
\node at (1.2, -0.3) {\footnotesize{$\dot{\pi}_c\sigma$}};
\node at (0.2, 0.15) {\footnotesize{$\bfk_1$}};
\node at (1, 0.15) {\footnotesize{$\bfk_2$}};
\node at (1.8, 0.15) {\footnotesize{$\bfk_3$}};
\end{tikzpicture} 
} 
= -\mu\,\dfrac{\rho^3}{c_s^3}\dfrac{1}{k_1^4 k_2 k_3} F_3\left(c_s,\dfrac{c_s k_2}{k_1},\dfrac{c_s k_3}{k_1},\dfrac{k_2}{k_1},\dfrac{k_3}{k_1}\right)\,.
\end{equation}
The seed function $F_1$ with $c_s=1$, corresponding to the de Sitter-invariant four-point function, was first analytically computed in \cite{Arkani-Hamed:2018kmz}, where the solution is given in terms of a power series representation. Its extension to the case of a reduced sound speed was obtained in \cite{Jazayeri:2022kjy}, which allowed to bootstrap correlators arising from interactions that strongly break boosts. As a result, more general correlators can be obtained after deformation of the four-point function, see \cite{Pimentel:2022fsc} for an alternative approach. Determining a closed-form solution for $F_{2, 3}$ in the full two-field theory requires more sophisticated tools and is beyond the scope of this paper. However, when the heavy field is integrated out, the seed functions $F_{2, 3}$ collapse into contact interactions that can be easily computed.

\subsubsection{EFT Seed Integrals and Correlators}
\label{sec:seed-integrals}

The non-local single-field EFT at leading-order in the time-derivative expansion (\ref{eq: non-local theory}) has a remarkable property: the two- and three-point correlators can be derived out of a single \textit{EFT seed integral} by applying a set of bespoke weight-shifting operators. Exploiting this feature, in this section, we give closed form expressions for the power spectrum correction and all three-point correlators.

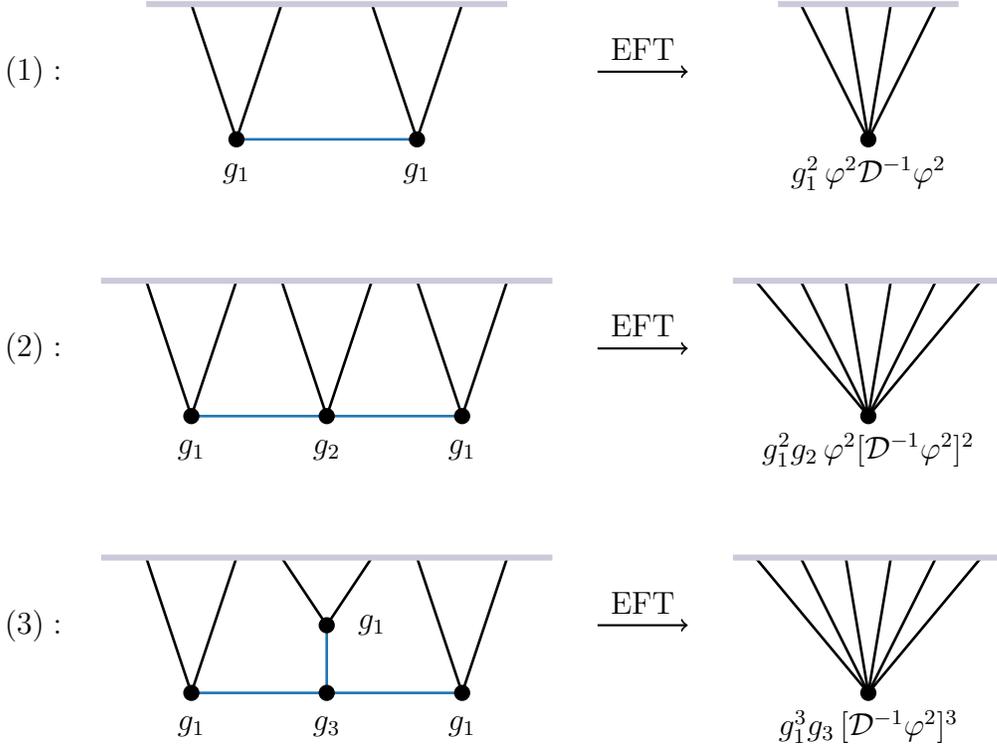
\begin{figure}[t!]
    \hspace*{0.5cm}
\begin{tikzpicture}[line width=1. pt, scale=3]
\node at (-0.3, -0.3) {$(1):$};
\draw[black] (0.4, 0) -- (0.6, -0.6);
\draw[black] (0.8, 0) -- (0.6, -0.6);
\draw[black] (1.2, 0) -- (1.4, -0.6);
\draw[black] (1.6, 0) -- (1.4, -0.6);
\draw[pyblue] (0.6, -0.6) -- (1.4, -0.6);
\draw[fill=black] (0.6, -0.6) circle (.03cm);
\node at (0.6, -0.75) {$g_1$};
\draw[fill=black] (1.4, -0.6) circle (.03cm);
\node at (1.4, -0.75) {$g_1$};
\draw[lightgray2, line width=0.8mm] (0.2, 0) -- (1.8, 0);
\draw[->, thick] (2.2, -0.3) -- (2.6, -0.3);
\node at (2.4, -0.2) {EFT};
\draw[black] (3.1, 0) -- (3.4, -0.6);
\draw[black] (3.3, 0) -- (3.4, -0.6);
\draw[black] (3.5, 0) -- (3.4, -0.6);
\draw[black] (3.7, 0) -- (3.4, -0.6);
\draw[fill=black] (3.4, -0.6) circle (.03cm);
\draw[lightgray2, line width=0.8mm] (3, 0) -- (3.8, 0);
\node at (3.4, -0.75) {$g_1^2\, \varphi^2\mathcal{D}^{-1}\varphi^2$};
\end{tikzpicture} 

\vspace*{1cm}
\hspace*{0.5cm}
\begin{tikzpicture}[line width=1. pt, scale=3]
\node at (-0.3, -0.3) {$(2):$};
\draw[black] (0.2, 0) -- (0.4, -0.6);
\draw[black] (0.6, 0) -- (0.4, -0.6);
\draw[black] (0.8, 0) -- (1, -0.6);
\draw[black] (1.2, 0) -- (1, -0.6);
\draw[black] (1.4, 0) -- (1.6, -0.6);
\draw[black] (1.8, 0) -- (1.6, -0.6);
\draw[pyblue] (0.4, -0.6) -- (1, -0.6);
\draw[pyblue] (1, -0.6) -- (1.6, -0.6);
\draw[fill=black] (0.4, -0.6) circle (.03cm);
\node at (0.4, -0.75) {$g_1$};
\draw[fill=black] (1, -0.6) circle (.03cm);
\node at (1, -0.75) {$g_2$};
\draw[fill=black] (1.6, -0.6) circle (.03cm);
\node at (1.6, -0.75) {$g_1$};
\draw[lightgray2, line width=0.8mm] (0, 0) -- (2, 0);
\draw[->, thick] (2.2, -0.3) -- (2.6, -0.3);
\node at (2.4, -0.2) {EFT};
\draw[black] (2.9, 0) -- (3.4, -0.6);
\draw[black] (3.1, 0) -- (3.4, -0.6);
\draw[black] (3.3, 0) -- (3.4, -0.6);
\draw[black] (3.5, 0) -- (3.4, -0.6);
\draw[black] (3.7, 0) -- (3.4, -0.6);
\draw[black] (3.9, 0) -- (3.4, -0.6);
\draw[fill=black] (3.4, -0.6) circle (.03cm);
\draw[lightgray2, line width=0.8mm] (2.8, 0) -- (4, 0);
\node at (3.4, -0.75) {$g_1^2g_2\, \varphi^2[\mathcal{D}^{-1}\varphi^2]^2$};
\end{tikzpicture} 

\vspace*{1cm}
\hspace*{0.5cm}
\begin{tikzpicture}[line width=1. pt, scale=3]
\node at (-0.3, -0.3) {$(3):$};
\draw[black] (0.2, 0) -- (0.4, -0.6);
\draw[black] (0.6, 0) -- (0.4, -0.6);
\draw[black] (0.8, 0) -- (1, -0.3);
\draw[black] (1.2, 0) -- (1, -0.3);
\draw[black] (1.4, 0) -- (1.6, -0.6);
\draw[black] (1.8, 0) -- (1.6, -0.6);
\draw[pyblue] (1, -0.3) -- (1, -0.6);
\draw[pyblue] (0.4, -0.6) -- (1, -0.6);
\draw[pyblue] (1, -0.6) -- (1.6, -0.6);
\draw[fill=black] (0.4, -0.6) circle (.03cm);
\node at (0.4, -0.75) {$g_1$};
\draw[fill=black] (1, -0.6) circle (.03cm);
\node at (1.2, -0.3) {$g_1$};
\draw[fill=black] (1, -0.3) circle (.03cm);
\node at (1, -0.75) {$g_3$};
\draw[fill=black] (1.6, -0.6) circle (.03cm);
\node at (1.6, -0.75) {$g_1$};
\draw[lightgray2, line width=0.8mm] (0, 0) -- (2, 0);
\draw[->, thick] (2.2, -0.3) -- (2.6, -0.3);
\node at (2.4, -0.2) {EFT};
\draw[black] (2.9, 0) -- (3.4, -0.6);
\draw[black] (3.1, 0) -- (3.4, -0.6);
\draw[black] (3.3, 0) -- (3.4, -0.6);
\draw[black] (3.5, 0) -- (3.4, -0.6);
\draw[black] (3.7, 0) -- (3.4, -0.6);
\draw[black] (3.9, 0) -- (3.4, -0.6);
\draw[fill=black] (3.4, -0.6) circle (.03cm);
\draw[lightgray2, line width=0.8mm] (2.8, 0) -- (4, 0);
\node at (3.4, -0.75) {$g_1^3g_3\, [\mathcal{D}^{-1}\varphi^2]^3$};

\end{tikzpicture} 
\caption{Schematic representation of the single-, double-, and triple-exchange correlators of $\varphi$ [\textcolor{black}{black}] in the full fundamental theory that we consider (\textit{left}), and the resulting (non-local) contact correlators of $\varphi$ after integrating out the field $\sigma$ [\textcolor{pyblue}{blue}] (\textit{right}). The diagrams (1), (2) and (3) provide the seeds for the complete set of correlation functions of the Goldstone boson, that are derived by acting on the seed diagrams with weight-shifting operators.}
     \label{fig: EFT diagrams}
\end{figure}

\vskip 4pt
As illustrated in Fig.~\ref{fig: EFT diagrams}, within the non-local single-field EFT framework, interactions that mix $\varphi$ and $\sigma$ become effective contact interactions $\mathcal{L}_{\text{eff}}/a^3 = - g_{1}^2\varphi^2 \mathcal{D}^{-1} \varphi^2 - g_{1}^2 g_{2}\varphi^2 [\mathcal{D}^{-1} \varphi^2]^2 - g_{1}^3 g_{3}[\mathcal{D}^{-1} \varphi^2]^3$. The seed functions $F_{1, 2, 3}$ in Eqs.~\eqref{vpicordef} then take the following simple forms

\begin{subequations}
\label{eq: seed Fs non-local EFT}
    \begin{align}
    F_1 &= \dfrac{c_s k_t}{\Ms_{12}}\,{\cal I}_1\left(\dfrac{c_s k_t}{\Ms_{12}}\dfrac{m}{H}\right)\,, \\
    F_2 &= \dfrac{c_s k_T \Ms_{12}}{2\Ms_{56}^2} \,{\cal I}_2\left(\dfrac{c_s k_T}{\Ms_{12}}\dfrac{m}{H},\dfrac{c_s k_T}{\Ms_{56}}\dfrac{m}{H} \right)\,,\\
    F_3 &= \dfrac{3c_s^3 k_T^3 \Ms_{12}}{2\Ms_{34}^2 \Ms_{56}^2} \,{\cal I}_3\left(\dfrac{c_s k_T}{\Ms_{12}}\dfrac{m}{H},\dfrac{c_s k_T}{\Ms_{34}}\dfrac{m}{H},\dfrac{c_s k_T}{\Ms_{56}}\dfrac{m}{H}\right)\,,
    \end{align}
\end{subequations}
with $k_t = \sum_{i=1}^4 k_i, k_{\rm{T}} = \sum_{i=1}^6 k_i$, and where we have defined the \textit{EFT seed integrals} as

\begin{equation}
\begin{aligned}
\label{eq: EFT seed integrals}
    \I_1(x) &= \int_{-\infty}^0 \d \u \,\frac{ \sin(\u)}{\u^2 + x^2}\,, \\
    \I_2(x, y) &= \int_{-\infty}^0 \d \u \, \frac{\u^2\sin(u)}{(\u^2 + x^2)(\u^2 + y^2)}\,,\\
    \I_3(x, y, z) &=\int_{-\infty}^0 \d \u \, \frac{u^2\sin(\u)}{(\u^2 + x^2)(\u^2 + y^2)(\u^2 + z^2)}\,.
\end{aligned}
\end{equation}
\noindent All correlators can therefore be derived by acting with weight-shifting operators on these building-block integrals.

\paragraph{Properties of the EFT seed integrals.} Anticipating the following development, it is useful to give some properties about the seed integrals. First, the integral $\I_1$ can be readily integrated analytically, giving

\begin{equation}
\label{eq: I1 EFT integral}
    \I_1(x) = \frac{e^x \text{Ei}(-x) - e^{-x}\text{Ei}(x)}{2x}\,,
\end{equation}
where $\text{Ei}$ is the exponential integral. Note that one uses $\text{Ei}(-x)=\frac12[\text{Ei}(-x+\epsilon)+\text{Ei}(-x-\epsilon)]$ due to the branch cut on the negative real axis. Remarkably, the integrals $\I_2$ and $\I_3$ can be written in terms of $\I_1$, which therefore constitutes the building block of all correlators. Indeed, decomposing the rational functions of $u^2$ in the integrands as sums of simple poles, one obtains

\begin{equation}
\begin{aligned}
    \I_2(x, y) &= \frac{x^2\, \I_1(x) - y^2\, \I_1(y)}{x^2 - y^2}\,,\\
    \I_3(x, y, z) &= \frac{x^2\, \I_1(x)}{(x^2 - z^2)(y^2 - x^2)} + \frac{y^2\, \I_1(y)}{(y^2 - z^2)(x^2 - y^2)} + \frac{z^2\, \I_1(z)}{(z^2 - y^2)(x^2 - z^2)}\,.
\end{aligned}
\end{equation}
It is straightforward to see that the apparent singularities when some arguments coincide are artificial, and to deduce the expressions

\begin{equation}
\begin{aligned}
    \I_2(x, x) &= \I_1(x) + \frac{x}{2}\frac{\partial \I_1}{\partial x}(x)\,,\\
    \I_3(x, x, x) &= -\frac{3}{8x}\frac{\partial \I_1}{\partial x}(x) - \frac{1}{8}\frac{\partial^2 \I_1}{\partial x^2}(x)\,,\\
    \I_3(x, x, y) &= \frac{x}{2(x^2 - y^2)} \frac{\partial \I_1}{\partial x}(x) + y^2\, \frac{\I_1(x) - \I_1(y)}{(x^2 - y^2)^2}\,.
\end{aligned}
\end{equation}
These forms of the integrals, as we have seen, all related to $\I_1$, appear in the correlators of the Goldstone boson.

\vskip 4pt
We have explained in Sec.~\ref{subsection: integrate out} that the differential operator ${\cal D}^{-1}$ should be considered as a building block on its own, and that any expansion either has pitfalls (in the large mass regime) or is doomed to fail (in the low-speed collider regime). Let us now explain how these statements translate into properties of the seed integral $\I_1$, respectively for $x \gg 1$, and $x \ll 1$ (see also Sec.~6.2 of \cite{Jazayeri:2022kjy}). As the rapid oscillations of the $\sin(\u)$ term in the integrand of \eqref{eq: EFT seed integrals} implies that most of the contribution to the integral comes from $|\u| \lesssim 1$, it is natural, for $x \gg 1$, to simply Taylor expand the denominator $(u^2+x^2)^{-1}$ for large $x$. After proper Wick rotations of the different frequencies in $\sin(u)$, one then obtains the asymptotic expansion $\I_1=-\sum_{n=0}^{\infty}(2n)!/x^{2(n+1)}$, corresponding to the local EFT expansion in the large mass regime. Naturally, what this computation misses is the contribution from $|u| \geq x$, and hence the poles of the Wick-rotated integrand at the Euclidean times $u=\pm i x$.
As for $x \ll 1$, the naive expansion of the denominator $(u^2+x^2)^{-1}=\sum_{n=0}^{\infty} (-1)^n x^{2n}/u^{2(n+1)}$, valid for $x <|\u|$, cannot work simply because the integrand gets most of its contribution precisely from $|\u| \sim x$.\footnote{Note that the corresponding integrals, divergent near $u=0$, could be regularized by an IR cutoff, but this would not change the fact that the corresponding expansion is anyway not valid for $|u|\leq x$, and hence cannot provide a reliable estimate of $\I_1$.} In the end, only the full function can reproduce the logarithmic behaviour $\I_1=\log(x)+\gamma-1+\ldots$ for small $x$.

\paragraph{Correlators.} Using Eqs.~(\ref{eq: seed Fs non-local EFT}) as the corresponding forms of the seed functions within the single-field EFT and using the weight-shifting operators defined in \ref{subsubsec : weight-shifting operators}, we now explicitly give analytical expressions of the two- and three-point correlators generated by (\ref{eq: non-local theory}). The curvature perturbation dimensionless power spectrum is given by
\begin{equation}
\Delta_\zeta^2 = \Delta_{\zeta, 0}^2\left[1 + 2c_s^2 \left(\frac{\rho}{H}\right)^2 \I_1\left(2c_s\frac{m}{H}\right) \right]\,,
\label{correction-power-spectrum}
\end{equation}
where $\Delta_{\zeta, 0}^2$ is defined in Eq.~(\ref{eq: dimensionless power spectrum}).  The three-point correlator of the Goldstone boson associated with the interaction $(\tilde{\partial}_i \pi_c)^2\mathcal{D}^{-1}\dot{\pi}_c$ is

\begin{equation}
    \begin{aligned}
    \frac{\braket{\pi_{c, \bm{k}_1} \pi_{c, \bm{k}_2} \pi_{c, \bm{k}_3}}'}{H^3} &= \frac{1}{2c_s^6}\frac{\rho}{\Lambda_1} \frac{k_1 k_2}{(k_1k_2k_3)^3} \frac{\bm{k}_1\cdot\bm{k}_2}{K} \\
    &
    \times\left[1 +
    \left(\frac{K^2}{k_1k_2} - \frac{K(k_1+k_2)}{k_1k_2} + c_s\frac{m}{H}\frac{K}{k_3}\right)\mathcal{I}_1\left(c_s\frac{m}{H}\frac{K}{k_3}\right) \label{universal-weak-mixing} \right.\\
    &\left.\hspace*{1cm}- c_s\frac{m}{H}\frac{K}{k_3} \frac{K(k_1+k_2)}{k_1k_2} \mathcal{I}_1'\left(c_s\frac{m}{H}\frac{K}{k_3}\right)\right] +\text{ 2 perms}\,,
    \end{aligned}
\end{equation}
where we have defined $K = k_1 + k_2 + k_3$, and that associated with $\dot{\pi}_c^2 \mathcal{D}^{-1}\dot{\pi}_c$ reads

\begin{equation}
    \frac{\braket{\pi_{c, \bm{k}_1} \pi_{c, \bm{k}_2} \pi_{c, \bm{k}_3}}'}{H^3} = \frac{1}{4c_s^4}\frac{\rho}{\Lambda_2} \frac{1}{(k_1 k_2 k_3)^2} \frac{k_1 k_2}{K k_3}\left[1 + \left(c_s \frac{m}{H} \frac{K}{k_3}\right)^2 \mathcal{I}_1\left(c_s \frac{m}{H} \frac{K}{k_3}\right)\right] + \text{ 2 perms}\,.
    \label{3point}
\end{equation}
The expressions \eqref{correction-power-spectrum}-\eqref{3point} agree with the ones in \cite{Jazayeri:2022kjy}. The double-exchange correlator of Goldstone bosons is

\begin{equation}
    \frac{\braket{\pi_{c, \bm{k}_1} \pi_{c, \bm{k}_2} \pi_{c, \bm{k}_3}}'}{H^3} = -\frac{\lambda}{2} \left(\frac{\rho}{H}\right)^2 \frac{k_3^2K}{c_s^2(k_1 k_2 k_3)^3} \,\mathcal{I}_2\left(c_s\frac{m}{H}\frac{K}{k_1}, c_s\frac{m}{H}\frac{K}{k_2}\right) + \text{ 2 perms}\,,
    \label{result-double-exchange}
\end{equation}
while the triple-exchange correlator of Goldstone bosons is

\begin{equation}
    \frac{\braket{\pi_{c, \bm{k}_1} \pi_{c, \bm{k}_2} \pi_{c, \bm{k}_3}}'}{H^3} = -\frac{3}{2} \frac{\mu}{H}\left(\frac{\rho}{H}\right)^3 \frac{K^3}{(k_1 k_2 k_3)^3} \,\mathcal{I}_3\left(c_s\frac{m}{H}\frac{K}{k_1}, c_s\frac{m}{H}\frac{K}{k_2}, c_s\frac{m}{H}\frac{K}{k_3}\right)\,.
    \label{triplediag} 
\end{equation}
With these expressions at hand, using $\zeta = -H c_s^{3/2} f_\pi^{-2} \pi_c$, one can obtain the corresponding correlators of $\zeta$.

\paragraph{Non-locality.} Before moving on, let us make some comments on (non)-locality. Loosely speaking, locality is related to the effects of a field decreasing with distance. For instance, in the non-relativistic limit, a massive scalar field with mass $m$ generates the well-known Yukawa potential $V(r) = \frac{\e^{-m r}}{4\pi r} $ when sourced by a point-like charge. This action at distance is non-local, but its effect is confined in a region $r\lesssim 1/m$, i.e.~it is a mild form of non-locality. In contrast, an inverse gradient term of the form $\frac{m^2}{-\tilde{\partial}_i^2} \sim m^2 r^2$ becomes increasingly important for $r\gtrsim 1/m$ and represents a severe violation of locality. Our single-field effective theory is non-local, as can be seen from the presence of the non-local differential operator $\mathcal{D}^{-1} = (-\tilde{\partial}_i^2 + m^2)^{-1}$ in cubic and quadratic interactions. 
The action in real space indeed contains terms evaluated at two distinct positions, i.e.~$S \sim \int \mathrm{d}t \int\mathrm{d}^3x \,\mathcal{D}^{-1} J(\bm{x}) = - \int \mathrm{d}t \int\mathrm{d}^3x \int\mathrm{d}^3y \,\Pi(\bm{x}, \bm{y}) J(\bm{y})$ where $\Pi(\bm{x}, \bm{y}) = \frac{e^{-m |\bm{x} - \bm{y}|}}{4\pi |\bm{x} - \bm{y}|}$ and $J(\bm{x})$ is a general source term. However, being of Yukawa-type, we see that this represents a mild form of non-locality.

\vskip 4pt
Recently, the Manifestly Local Test (MLT) was proposed as a check of correlators arising from a manifestly-local theory \cite{Jazayeri:2021fvk}. It is stated at the level of wavefunction coefficients \cite{Anninos:2014lwa, Goon:2018fyu}, and reads in our case

\begin{equation}
    \left.\frac{\partial}{\partial k_i} \psi_3(k_1, k_2, k_3) \right\vert_{k_i = 0} = 0\,.
\end{equation}
Its derivation is based on a simple property satisfied by the mode function of a massless field in de Sitter. We have explicitly checked that our correlators satisfy the MLT for $\text{Re} \,\psi_3$, and extends to the full wavefunction coefficient $\psi_3$ by inspecting the MLT derivation from a bulk perspective. The reason is that the non-local differential operator $\mathcal{D}^{-1}$ appearing in the diagram vertices cannot be singular as the momentum is taken to be soft due to the non-zero mass, which is naturally related to the mild non-locality that we have just discussed. To put it short, a manifestly local theory satisfies the MLT, but satisfying the MLT does not offer a diagnosis of the underlying theory being manifestly local.

\subsection{Strong Mixing}
\label{subsec: strong mixing}

In the strong mixing regime, the quadratic theory cannot be solved analytically. Therefore, it is necessary to numerically compute the correlators. Recently, the cosmological flow approach has been developed to systematically compute the two- and three-point correlators in any theory of inflationary fluctuations \cite{Werth:2023pfl, CosmoFlow}. Here, we briefly summarise this method.

\paragraph{Flow equations.} 
The cosmological flow is based on solving differential equations in time satisfied by the correlators, and is completely equivalent to the in-in formalism \cite{Mulryne:2013uka}. Essentially, this emerges from: $i$) the evolution equation
$\frac{\d}{\d t}\langle \mathcal{O} \rangle =i \langle [H,\mathcal{O}] \rangle $ for expectation values of operators $\mathcal{O}$ without explicit time-dependence (Ehrenfest theorem), and $ii$) the fact that the commutator between $\mathcal{O}$ and the Hamiltonian $H$ can be systematically computed in perturbation theory using the canonical commutation relations and the fact that $H$ takes a  polynomial form in fields and momenta. Working at tree-level, one deduces that the time evolution of two- and three-point correlators in any theory takes the form

\begin{subequations}
\begin{align}
    \frac{\d}{\d t} \langle \bm{X}^{\sf{a}} &\bm{X}^{\sf{b}} \rangle = \tensor{u}{^{\sf{a}}}{_{\sf{c}}} \langle \bm{X}^{\sf{c}} \bm{X}^{\sf{b}} \rangle + \tensor{u}{^{\sf{b}}}{_{\sf{c}}} \langle \bm{X}^{\sf{a}} \bm{X}^{\sf{c}} \rangle\,,\label{eq: 2pt time evolution}\\
    \begin{split}
    \frac{\d}{\d t} \langle \bm{X}^{\sf{a}} \bm{X}^{\sf{b}}&\bm{X}^{\sf{c}} \rangle = \tensor{u}{^{\sf{a}}}{_{\sf{d}}} \langle \bm{X}^{\sf{d}} \bm{X}^{\sf{b}}\bm{X}^{\sf{c}} \rangle 
    + \tensor{u}{^{\sf{a}}}{_{\sf{de}}}\langle \bm{X}^{\sf{b}} \bm{X}^{\sf{d}} \rangle\langle \bm{X}^{\sf{c}} \bm{X}^{\sf{e}} \rangle + (2\text{ perms})\,,\label{eq: 3pt time evolution}
    \end{split}
\end{align}
\end{subequations}
where $\bm{X}^{\sf{a}}$ encompasses all phase-space variables in Fourier space (the index $\sf{a}$ thus refers to a Fourier wavevector as well as to a field or momenta, $(\pi_c, \sigma, p_{\pi}, p_\sigma)$ in our situation of interest), and the entire theory dependence is encoded in the precise form of the coefficients $\tensor{u}{^{\sf{a}}}{_{\sf{b}}}$ and $\tensor{u}{^{\sf{a}}}{_{\sf{bc}}}$, which can be systematically deduced from the action.

\vskip 4pt
Solving Eq.~(\ref{eq: 2pt time evolution}) is completely equivalent to fully solving the quadratic theory. De facto, solving these equations, which couple all two-point correlators of fields and momenta, is equivalent to dressing the two-point correlators with an infinite number of quadratic mixing insertions---hence performing a resummation---, when using an interaction scheme where quadratic mixings are treated perturbatively. 
Similarly, Eq.~\eqref{eq: 3pt time evolution} couples all types of three-point correlators. In contrast, it is noteworthy that each kinematic configuration follows its flow independently of the others: Eqs.~\eqref{eq: 2pt time evolution} and \eqref{eq: 3pt time evolution} can be conveniently solved triangle by triangle. Eventually, considering the Bunch-Davies vacuum, the coefficients $\tensor{u}{^{\sf{a}}}{_{\sf{b}}}$ and $\tensor{u}{^{\sf{a}}}{_{\sf{bc}}}$ fully determine the initial conditions for the correlators $\langle \bm{X}^{\sf{a}} \bm{X}^{\sf{b}} \rangle$ and $\langle \bm{X}^{\sf{a}} \bm{X}^{\sf{b}}\bm{X}^{\sf{c}} \rangle$.

\section{Phenomenology}
\label{sec: pheno}

The presence of a reduced sound speed for the propagation of the Goldstone boson leads to a striking signature of heavy fields---referred to as the \textit{low-speed collider signal}---that is captured by the single-field non-local EFT (\ref{eq: non-local theory}). For heavy particles lighter than $H/c_s$, this signal manifests itself as a resonance in the squeezed limit of the bispectrum. In this section, we discuss in details both the size and the shape dependence of this signal, both in the weak and the strong mixing regimes.

\subsection{The Physics of the Low-Speed Collider}
\label{sec:physics-LSC}

Before diving in the phenomenology, it is worth highlighting that the low-speed collider signal has a simple physical explanation, related to the presence of two characteristic time scales in the dynamics, see~\cite{Jazayeri:2022kjy} for more details. Because modes of different wavelength freeze or decay at different times, these time scales are encoded in the momentum dependence of primordial correlators. 
Indeed, due to a reduced sound speed $c_s$, a given short Goldstone boson mode freezes when it exits the sound horizon $c_s k_{\text{S}}/a(t_1) \sim H$. Likewise, a long mode of the massive field starts to decay when it crosses the mass-shell horizon $k_{\text{L}}/a(t_2) \sim m$. When $m\ll H/c_s$, the time $t_1$ occurs before $t_2$ and the Goldstone boson interacts with the massive field during the period of time $t_1<t<t_2$ as if the latter was massless. This effect leads to the amplification of the bispectrum and mimics local non-Gaussianities close to equilateral configurations for $c_s m/H\lesssim k_{\text{L}}/k_{\text{S}}$, equivalent to the condition $t_1\lesssim t_2$. 
In the ultra-squeezed limit $k_{\text{L}}/k_{\text{S}} \lesssim c_s m/H$, equivalent to $t_2\lesssim t_1$, the bispectrum fades away due to the decay of the massive field, which is then encoded in the conventional cosmological collider signal.
When $k_{\text{L}}/k_{\text{S}} \sim c_s m/H$, equivalent to both time scales coinciding $t_1\sim t_2$, the bispectrum displays a characteristic resonance. 
Instead, for $m \gg H/c_s$, the time $t_1$ occurs after $t_2$ for all kinematic configurations, and the situation qualitatively resembles the one with $c_s=1$. Namely, the heavy field can be integrated out in the usual local manner, corresponding to the regime ``Reduced effective speed of sound" in Sec.~\ref{subsec: dispersion relation}, and the bispectrum shapes generated by all interactions are of the conventional equilateral type. Consequently, the low-speed collider phenomenology is governed by the dimensionless parameter
\begin{equation}
\label{eq: alpha parameter}
    \alpha \equiv c_s\, \frac{m}{H}\,,
\end{equation}
with $\alpha \gtrsim 1$ corresponding to usual equilateral shapes, and $\alpha \lesssim 1$ to the genuine low-speed collider regime, on which we focus from now on.

\begin{figure}[h!]
     \centering
     \hspace*{-0.7cm}
     \includegraphics[scale=1]{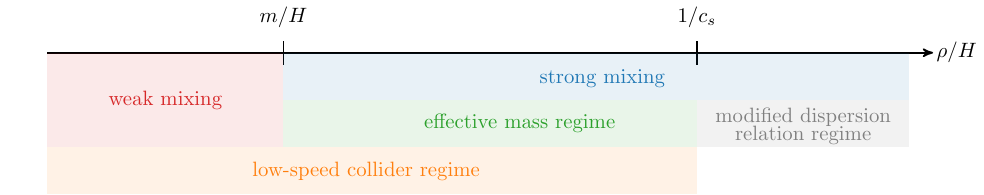}
     \caption{Schematic illustration of the different regimes for a reduced Goldstone boson intrinsic sound speed $c_s\lesssim 1$ and $m/H\lesssim 1/c_s$, varying the mixing strength $\rho/H$. Of course, the boundaries of these regions are not meant to be taken as strict demarcations, but rather as faint delimitations.}
     \label{fig: weak&strong low-speed collider pheno}
 \end{figure}

\vskip 4pt
The physics of the low-speed collider explained above is valid in the weak mixing regime, i.e.~when the quadratic mixing $\rho \dot{\pi}_c \sigma$ can be treated perturbatively. This corresponds to $\rho \lesssim m$, as we show in the insert below. A natural question is how this description is changed at strong mixing. We show in the insert that the effect of the mixing beyond perturbative treatments is to induce an effective mass $m^2\rightarrow m_{\text{eff}}^2 = m^2 +\rho^2$ for the heavy mode, see also \cite{Castillo:2013sfa,An:2017hlx, Iyer:2017qzw,Werth:2023pfl} for $c_s=1$.
Moreover, we show that the mixing, despite having this important effect for the dynamics of the heavy mode, has a small effect on the dynamics of the Goldstone boson inside the horizon as long as $m_{\text{eff}} \lesssim H/c_s$, corresponding to the parameter 
\begin{equation}
\label{eq: alpha-eff parameter}
    \alpha_{\text{eff}} \equiv c_s\, \frac{m_{\text{eff}}}{H}\,,
\end{equation}
being less than unity. This entails that the physics of the low-speed collider valid at weak mixing $\rho \lesssim m$, also holds qualitatively in the regime $m \lesssim \rho \lesssim H/c_s$, simply upon considering the effective mass $m_{\text{eff}}$, i.e~ with $\alpha \to  \alpha_{\text{eff}}$. If the mixing is further increased to $\rho/H \gtrsim 1/c_s$, corresponding to $ \alpha_{\text{eff}} \gtrsim 1$, the dynamics of the system enters the regime where the Goldstone boson has a modified dispersion relation, see Sec.~\ref{subsec: dispersion relation} and the conditions \eqref{eq: MDR regime of validity},\footnote{Note that when $m/H\lesssim 1/c_s$ and $1/c_s\lesssim \rho/H$, the condition $m^2/\rho\lesssim H/c_s$ is automatically satisfied.} giving rise to standard equilateral-type shapes.
In Fig.~\ref{fig: weak&strong low-speed collider pheno}, we summarise the low-speed collider phenomenology at weak and strong mixing.

\vskip 4pt
We will see that the strong mixing regime of the low-speed collider, i.e.~the effective mass regime depicted in green in Fig.~\ref{fig: weak&strong low-speed collider pheno}, is particularly interesting. Indeed, the low-speed collider signal can be observationally large and is not dwarfed by the ever-present equilateral-type non-Gaussianities coming from self-interactions of the Goldstone boson. With this in mind, one aspect is worth highlighting: fluctuations in this regime can be strongly mixed without being strongly coupled. Strong mixing refers to the fact that the quadratic mixing $\rho \dot{\pi}_c \sigma$ cannot be treated in perturbation theory. At the same time, because of the non-linearly realized time diffeomorphism invariance, this interaction is necessarily accompanied by the dimension-5 cubic interaction $\propto \rho (\tilde{\partial}_i \pi)^2 \sigma$. Hence, one can wonder whether large values of $\rho$ in this regime are such that this interaction becomes strongly coupled at energies of order the Hubble scale, making the theory strongly coupled. This is not the case. As the arguments above show, all reasonings in the weak mixing regime approximately hold also in the effective mass regime upon considering the effective mass $m_{\text{eff}}$.\footnote{Note that the bound on $\rho$ in Eq.~\eqref{eq: perturbativity weak mixing}, coming from being able to treat the quadratic mixing perturbatively, is ineffective by definition in the strong mixing regime.} Thus, see the insert in Sec.~\ref{subsection: Bounds on couplings}, the bound on $\rho$ coming from avoiding strong coupling reads $\rho/H \lesssim c_s^{3/2}(2 \pi \Delta_\zeta)^{-1} \sim 100$ for the representative value $c_s \sim 0.1$, showing that the theory remains healthy in the strong mixing regime of the low-speed collider $\rho/H \lesssim 1/c_s$.

\begin{framed}
{\small \noindent {\it Dispersion relation analysis.}---In this insert, we examine the full two-field quadratic dynamics. First, we give the corresponding dispersion relations and we show that $\rho\lesssim m$ is required in order to treat the quadratic mixing perturbatively. Then, we explain that in the parameter space of interest $c_s m/H \lesssim 1$ for the low-speed collider, $m_{\text{eff}}$ plays the role of the mass of the heavy mode. Furthermore, going beyond weak mixing, we show that $\alpha_{\text{eff}}\equiv c_s m_{\text{eff}}/H \lesssim 1$ is the correct non-perturbative criterion for the existence of the low-speed collider, giving rise to the ``effective mass" regime of Fig.~\ref{fig: weak&strong low-speed collider pheno}. Remarkably, this regime is amenable to an approximate analytical understanding.

\vskip 4pt
Taking the flat-space limit $H\rightarrow 0$ of the full theory (\ref{eq: full theory}), the quadratic dynamics can be solved exactly after injecting plane-wave solutions. The dispersion relations for both degrees of freedom are given by
\begin{equation}
\label{eq: exact dispersion relations}
    \omega_{\pm}^2 = \frac{1+c_s^2}{2}\, \kp^2 + \frac{m_{\text{eff}}^2}{2} \pm \frac{1}{2}\sqrt{m_{\text{eff}}^4 + (1-c_s^2)^2\kp^4 + 2(1+c_s^2)\kp^2\left[\rho^2 + \frac{1-c_s^2}{1+c_s^2}\,m^2\right]}\,,
\end{equation}
where $m_{\text{eff}}^2 = m^2+\rho^2$ is simply a short-hand notation at this stage. We will look below at physically interesting limits in which this exact solution simplifies, but first, as a sanity check, we note that the uncoupled limit $\rho/H=0$ gives $\omega_-^2 = c_s^2 \kp^2$ and $\omega_+^2 = \kp^2 + m^2$. We recover that the spectrum is then composed of a massless mode propagating with a speed of sound $c_s$ and a relativistic heavy mode with mass $m$.

\vskip 4pt
We now explain the criterion $\rho\lesssim m$ as the perturbativity bound on the quadratic coupling, directly at the level of the non-local EFT (\ref{eq: non-local theory}). To do this, we need to estimate the propagator $\mathcal{D}^{-1}=(\kp^2+m^2)^{-1}$ at the relevant time scales of the dynamics, which, as we have stressed in Sec.~\ref{subsection: integrate out}, includes sound-horizon crossing but also mass-shell crossing and beyond. At the sound-horizon crossing of the massless mode $\kp\sim H/c_s$ and in the parameter space of interest with $c_s m/H\lesssim 1$, the propagator scales as $\mathcal{D}^{-1} \approx c_s^2/H^2$. Therefore, the correction to the quadratic Lagrangian scales as $c_s^2 (\rho/H)^2\lesssim c_s^2 (m/H)^2\lesssim 1$ in the would-be weak mixing regime, and is indeed negligible. At and after the mass-shell crossing of the heavy mode $\kp \lesssim m$, it scales as 
$\mathcal{D}^{-1} \approx m^{-2}$, giving a correction of order $(\rho/m)^2$, which is consistently negligible as long as $\rho\lesssim m$.
From the exact dispersion relations (\ref{eq: exact dispersion relations}), it is then interesting to consider the first correction to the decoupled dynamics of the heavy mode in this weak mixing regime
\begin{equation}
 \omega_+^2 \approx \kp^2 + m^2 + \rho^2 \frac{\kp^2+m^2}{(1-c_s^2)\kp^2 + m^2}\,.
\end{equation}
One can appreciate that for a low sound speed $c_s\ll 1$, the last term in the heavy mode dispersion relation $\omega_+^2$ is given by $\rho^2$, and therefore combines with the mass parameter to generate the combination $m_{\text{eff}}^2$, which indeed appears as the physical mass of the heavy excitations. This effect is small by definition in the weak mixing regime. However, remarkably, this picture holds beyond weak mixing for $\alpha\equiv c_s m/H\lesssim 1$. Indeed, in the regime of small $\alpha$, the exact dispersion relations boil down to
\begin{equation}
    \omega_-^2 \approx c_s^2 \kp^2\, \frac{\kp^2 + m^2}{\kp^2 + m_{\text{eff}}^2}\,,\hspace*{0.5cm}\text{and}\hspace*{0.5cm} \omega_+^2 \approx \kp^2 + m_{\text{eff}}^2 + c_s^2\rho^2 \frac{\kp^2}{\kp^2 + m_{\text{eff}}^2}\,.
    \label{dispersion-relations-low-speed-collider}
\end{equation}
These relations are non-perturbative in the quadratic mixing as they do not assume weak mixing. Notice that the last term, $c_s^2\rho^2 \frac{\kp^2}{\kp^2 + m_{\text{eff}}^2} \lesssim c_s^2\rho^2$, is negligible compared to the term $\rho^2$ appearing in $m_{\text{eff}}^2$. Therefore, in the regime $\alpha\lesssim 1$, the heavy mode dynamics is governed by $\omega_+^2 = \kp^2 + m_{\text{eff}}^2$.
This explains why the low-speed collider resonance occurs also at strong mixing as long as $\alpha_{\text{eff}} \equiv c_s\, m_{\text{eff}}/H \lesssim 1$. Indeed, in this regime, the mixing between the massless and the heavy mode is important enough to generate an effective mass different from the bare one, but it does not appreciably affect the dynamics of the Goldstone boson, and the heavy mode can still be considered as effectively massless
at sound horizon crossing for $\pi_c$. To see this, let us recall that the massless mode dispersion relation is only valid for $\omega_- \gtrsim H$ as Hubble friction becomes important at smaller energies. The requirement that the dispersion relation is not affected by the mixing even at this energy, corresponding to momenta $\kp\sim H/c_s$ (the two modes obviously decouple in the large $\kp$ limit), is equivalent to $\frac{1+\alpha^2}{1+\alpha_{\text{eff}}^2} \simeq 1$, which is satisfied for $\alpha_{\text{eff}} \ll 1$.
This condition also guarantees that the effective mass in $\omega_+^2 = \kp^2 + m_{\text{eff}}^2$ is negligible for $\kp\sim H/c_s$, giving rise to the peculiar local-like behaviour of the bispectrum for near equilateral configurations. Furthermore, one can check that in this effective mass regime, the corrections to the dispersion relation of the Goldstone mode around sound horizon crossing, $\omega_-^2/(c_s^2 \kp^2)-1\simeq -c_s^2 \rho^2/H^2$, takes exactly the same form as in the weak mixing regime. 
Eventually, as $m_{\text{eff}}$ also controls the dynamics at late times, i.e~the quasi-normal modes of the coupled system (a statement that does not depend on sound speeds), the decay of the heavy mode starts at effective mass crossing $\kp \sim m_{\text{eff}}$, hence giving rise to the low-speed collider resonance at the location $k_{\text{L}}/k_{\text{S}} \sim c_s m_{\text{eff}}/H$.
This analysis suggests that, to a good approximation, the analytical results of the low-speed collider in the weak mixing regime can be transposed to the effective mass regime $m/H \lesssim \rho/H \lesssim 1/c_s$, simply upon considering the ``dressed'' effective mass $m^2\rightarrow m_{\text{eff}}^2 = m^2 +\rho^2$, with any small discrepancy attributed to the intermediate regime $m_{\text{eff}} \lesssim \kp \lesssim H/c_s$.
}
\end{framed}

\vskip 4pt

In this work, we are interested in the low-speed collider signatures of effectively heavy fields, corresponding to $m_{\text{eff}} \gtrsim H$, a situation that is amenable to a description in terms of a single-field (non-local) description. But interesting signatures also arise in the complementary regime of a light field weakly mixed to the Goldstone boson. In this regime, notice that remarkably, the bootstrap analytical formulae for the single-exchange diagrams derived in \cite{Jazayeri:2022kjy} are also valid after proper substitution $\mu \rightarrow -i \nu$. The low-speed collider signal still manifests itself as a resonance in mildly-squeezed configurations, with the usual quasi-single field scalings taking over in the ultra-squeezed configuration, and with the resonance ultimately fading away as the field becomes massless. We leave the detailed analysis of this region of parameter space for an upcoming work.

\subsection{Size of non-Gaussianities}
\label{sec:size}

Let us estimate the size of the bispectrum generated by the non-linear interactions in (\ref{eq: full theory}). In the weak mixing regime, one can easily deduce the size of the bispectrum from the analytical results presented in the previous section. However, here, we will present a systematic procedure to quickly estimate the size of non-Gaussianities based on dimensional analysis, which is also valid at strong mixing.
Following standard conventions, we first define the momentum dependence of the three-point correlator of $\zeta$ through the dimensionless shape function $S$ such that

\begin{equation}
\label{eq: shape definition}
    \braket{\zeta_{\bm{k}_1} \zeta_{\bm{k}_2} \zeta_{\bm{k}_3}}' \equiv (2\pi)^4 \frac{S(k_1, k_2, k_3)}{(k_1 k_2 k_3)^2} \, \Delta_\zeta^4\,.
\end{equation}
We then use the standard measure of the size of the non-Gaussian signal by the following non-linearity parameter

\begin{equation}
    f_{\text{NL}} \equiv \frac{10}{9} S(k, k, k)\,,
\end{equation}
with the shape evaluated in the equilateral configuration $k_1=k_2=k_3=k$. Using the single-field non-local EFT (\ref{eq: non-local theory}), the size of the bispectrum can be easily recovered using purely dimensional analysis arguments because the theory contains only one dynamical degree of freedom. An estimate of the size of the bispectrum is then given by $f_{\text{NL}} \sim \frac{\mathcal{L}_3}{\mathcal{L}_2} \, \Delta_\zeta^{-1}$, where mass-dimension quantities in the Lagrangian are evaluated at the energy scale probed during inflation, i.e.~$\omega \sim H$. This method provides a fast estimate without setting up the full calculation, and ultimately gives a more intuitive physical explanation of the various regimes covered by the full theory (\ref{eq: full theory}).\footnote{Note that in the low-speed collider regime, this method cannot reproduce the peculiar logarithmic dependence of the bispectrum on $\alpha$, precisely as the latter comes from the super-Hubble regime $\omega \lesssim H$, see the discussions in Sec.~\ref{subsection: integrate out} and \ref{sec:seed-integrals}.} In practice, after inspecting the Lagrangian (\ref{eq: non-local theory}), we need to estimate the non-local differential operator $\mathcal{D}^{-1}$ and the typical size of the fluctuations $\pi_c/H$ in the various regimes of interest.

\vskip 4pt
In the weak mixing regime, and as we have seen in Sec.~\ref{sec:physics-LSC}, also approximately in the effective mass regime, the Goldstone boson freezes while having a linear dispersion relation $\omega = c_s \kp$, with $\kp=k/a$ being the physical momentum. Evaluating at $\omega\sim H$, the gradient term in the propagator $\mathcal{D}^{-1}$ dominates over the mass term for the parameter space of interest $m\lesssim H/c_s$.
In the end, we obtain $\mathcal{D}^{-1} \sim (c_s/H)^2$. Additionally, one deduces from the action \eqref{eq: non-local theory} that $\pi_c/H\sim c_s^{-3/2}$, in agreement with the power spectrum (\ref{eq: dimensionless power spectrum}).
In this regime, the quadratic part of the Lagrangian is dominated by the first term in (\ref{eq: non-local theory}), i.e.~$\mathcal{L}_2\sim \dot{\pi}_c^2\sim H^2 \pi_c^2$.

\vskip 4pt
Deep in the strong mixing regime $\rho/H \gtrsim 1/c_s$, the Goldstone boson freezes while having a modified dispersion relation $\omega=c_s \kp^2/\rho$. Evaluating the gradient term in the propagator---which also dominates over the mass term---at $\omega\sim H$ leads to $\mathcal{D}^{-1} \sim c_s/(\rho H)$. From the action \eqref{eq: non-local theory}, one deduces the typical size of the fluctuations $\pi_c/H\sim c_s^{-5/4} (\rho/H)^{1/4}$, in agreement with (\ref{eq: MDR mode function}). Note that this respects continuity between the weak and strong mixing regimes at the threshold value
$\rho/H\sim 1/c_s$. Naturally in this regime, the quadratic part of the Lagrangian is given by $\mathcal{L}_2 \sim \rho^2 \dot{\pi}_c\mathcal{D}^{-1}\dot{\pi}_c$. We now have the necessary ingredients to estimate the size of non-Gaussianities for all interactions in (\ref{eq: non-local theory}).

\vskip 4pt
Before giving the results, let us stress that we include for completeness the modified dispersion relation regime, but that shapes there are of conventional equilateral type and do not exhibit the characteristic resonances of the low-speed collider. The latter arise both for $\rho/H \lesssim m/H$ (the weak mixing regime) and for $m/H \lesssim \rho/H \lesssim 1/c_s$ (the effective mass regime), which make up the low-speed collider regime, see Fig.~\ref{fig: weak&strong low-speed collider pheno}. Remarkably, this regime can be described in a unified manner.

\begin{itemize}
    \item Let us first consider the interaction $(\tilde{\partial}_i \pi_c)^2\sigma$, suppressed by the scale $\Lambda_1 \sim c_s^{-3/2}f_\pi^2/\rho$. Using the estimates previously derived, the size of the bispectrum is given by
    \begin{equation}
    \label{eq: fNL size of S11}
        f_{\text{NL}} \sim \frac{\rho}{\Lambda_1} \frac{(\tilde{\partial}_i \pi_c)^2\mathcal{D}^{-1}\dot{\pi}_c}{\mathcal{L}_2 \, \Delta_\zeta} \sim
        \begin{cases}
            \left(\rho/H\right)^2 & \text{(low-speed collider)}\,,\\
           c_s^{-1} \rho/H   & \text{(modified dispersion relation)}\,.
    \end{cases}
    \end{equation}
    In the low-speed collider regime $\rho/H \lesssim 1/c_s$, the signal can be as large as $1/c_s^2$, parametrically of the same size as the signal from self-interactions, see Sec.~\ref{sec:self-interaction}. Note that the signal grows with the strength of the quadratic mixing in the strong mixing regime, in which  the perturbativity bound on the quadratic coupling given in Sec.~\ref{subsection: Bounds on couplings} yields $f_{\text{NL}}< \Delta_\zeta^{-1}$.
 
    \item We now consider the interaction $\dot{\pi}_c^2\sigma$, whose coupling constant is not fixed by the quadratic mixing $\rho$. The size of the corresponding bispectrum is given by
    \begin{gather}
        f_{\text{NL}} \sim \frac{\rho}{\Lambda_2} \frac{\dot{\pi}_c^2\mathcal{D}^{-1}\dot{\pi}_c}{\mathcal{L}_2 \, \Delta_\zeta} \sim
        \begin{cases}
            c_s^{1/2} \Delta_\zeta^{-1} \rho/\Lambda_2 & \text{(low-speed collider)}\,,\\
            c_s^{-5/4} \Delta_\zeta^{-1}(H/\Lambda_2)(\rho/H)^{-3/4}  & \text{(modified dispersion relation)}\,.
            \raisetag{37pt}
    \end{cases}
    \end{gather}
    The perturbativity bounds on the couplings give $f_{\text{NL}} < c_s^{3/2}\, \Delta_\zeta^{-1}\, \rho/H < c_s^{1/2}\, \Delta_\zeta^{-1}$ in the low-speed collider regime, where the first inequality comes from the bound on $\Lambda_2$ and the second one is obtained by pushing $\rho/H$ to $1/c_s$, and $f_{\text{NL}} < \Delta_\zeta^{-1}$ at strong mixing.
    As opposed to the previous interaction, in the strong mixing regime, $f_{\text{NL}}$ decreases when the quadratic mixing increases. As we will see now, this is a common feature of all other cubic interactions.

    \item Next, we consider the interaction $\dot{\pi}_c\sigma^2$, governed by the dimensionless coupling $\lambda$. The non-Gaussian signal is
     \begin{gather}
        f_{\text{NL}} \sim \lambda\rho^2\,\frac{\dot{\pi}_c[\mathcal{D}^{-1}\dot{\pi}_c]^2}{\mathcal{L}_2 \, \Delta_\zeta} \sim
        \begin{cases}
            c_s^{5/2} \,\Delta_\zeta^{-1} \lambda(\rho/H)^2 & \text{(low-speed collider)}\,,\\
            c_s^{-1/4}\,\Delta_\zeta^{-1} \lambda \,(\rho/H)^{-3/4}  & \text{(modified dispersion relation)}\,.
            \raisetag{37pt}
    \end{cases}
    \end{gather}
    Considering perturbativity bounds on the associated couplings yields $f_{\text{NL}}<c_s^3 \Delta_\zeta^{-1}(\rho/H)^2< c_s \Delta_\zeta^{-1}$ in the low-speed collider regime, and $f_{\text{NL}} < \Delta_\zeta^{-1}$ at strong mixing.

    \item Finally, we consider non-linearities in the $\sigma$-sector which are fully characterised by the interaction $\sigma^3$. The amplitude of the non-Gaussian signal is
    \begin{gather}
        f_{\text{NL}} \sim \mu\rho^3\,\frac{[\mathcal{D}^{-1}\dot{\pi}_c]^3}{\mathcal{L}_2 \, \Delta_\zeta} \sim
        \begin{cases}
            c_s^{9/2} \,\Delta_\zeta^{-1} (\mu/H)(\rho/H)^3 & \text{(low-speed collider)}\,,\\
            c_s^{3/4} \,\Delta_\zeta^{-1}\,(\mu/H) (\rho/H)^{-3/4}  & \text{(modified dispersion relation)}\,,
            \raisetag{37pt}
    \end{cases}
    \end{gather}
    which gives $f_{\text{NL}}<c_s^{9/2} \Delta_\zeta^{-1} (\rho/H)^3< c_s^{3/2} \Delta_\zeta^{-1} $ in the low-speed collider regime and $f_{\text{NL}}<\Delta_\zeta^{-1}$ at strong mixing.
\end{itemize}

In the modified dispersion regime and for the interactions in $(\tilde{\partial}_i \pi_c)^2\sigma$ and $\sigma^3$, the defining property of the perturbativity bounds, $ f_\textrm{NL} \Delta_\zeta \sim {\cal L}_3/{\cal L}_2 < 1$, makes it automatic that saturating them gives $f_{\text{NL}}\sim \Delta_\zeta^{-1}$. However, this was not guaranteed for the other interactions $\dot{\pi}_c^2\sigma$ and $\dot{\pi}_c\sigma^2$, and it deserves an explanation. For definiteness, let us consider the first one, stemming from $\tilde{M}_3^3 \left(\delta g^{00}\right)^2 \sigma$ in the unitary gauge mixing action \eqref{eq:mixing-unitary-gauge}. The St{\"u}ckelberg transformation generates other interactions beyond the one in $\dot{\pi}_c^2\sigma$, and perturbative unitarity requires all of them to be suppressed by a scale greater than Hubble. However, these other interactions are necessarily suppressed by an energy scale larger than the one suppressing $\dot{\pi}_c^2\sigma$ once the theoretical upper bound on $\rho$ in Eq.~\eqref{eq: perturbativity strong mixing} is considered. Indeed, the latter ultimately comes precisely from ensuring that the quadratic terms are negligible for $\omega \sim H$ compared to the linear one in the combination $\delta g^{00} \rightarrow -2 \dot{\pi} - \dot{\pi}^2 + (\partial_i \pi)^2/a^2$, as the linear one generates the  dominant term $\rho \dot{\pi}_c  \sigma$ in the quadratic action, and the one in $(\partial_i \pi)^2$ generates the corresponding cubic interaction $(\tilde{\partial}_i \pi_c)^2\sigma$ (the other interaction being negligible in the modified dispersion relation regime). Therefore, among all the requirements $\left({\cal L}_n/{\cal L}_2\right)_{\omega=H} <1$ for the interactions generated by the St{\"u}ckelberg transformation, the one $n=3$ is the most stringent and provides the perturbativity bound, whose saturation then leads to $f_{\text{NL}} \sim \Delta_\zeta^{-1}$.

\vskip 4pt
In the low-speed collider regime, notice instead that the maximal sizes of $f_{\text{NL}}$ compatible with perturbative unitarity are suppressed by powers of $c_s$ compared to $\Delta_\zeta^{-1}$. This comes from the fact that the upper bound on $\rho$ set by the existence of the low-speed collider regime is more stringent than the corresponding perturbativity bound, or equivalently, that fluctuations in the strong mixing regime of the low-speed collider are automatically weakly coupled. Nonetheless, we stress that our results show that non-Gaussianities in this regime can be observationally large.
Additionally, in general, requiring the theory to be technically natural, i.e.~stable under radiative corrections, significantly lowers non-Gaussian signals (see e.g.~\cite{PhysRevD.85.103520, Lee:2016vti, CosmoFlow} for more details). To produce large non-Gaussianities, i.e.~$f_{\text{NL}}>1$, this theory requires new physics or fine-tuning, except for the interaction $\sigma^3$. However, as we will see in the next section, this interaction does not generate a visible low-speed collider resonance. Eventually, one should keep in mind that the scalings above provide estimates for the size of non-Gaussianities in the equilateral configuration, and do not take into account the enhancement of the signal at the low-speed collider resonance peak, which can all the more increase the size of the overall signal.

\subsection{Shapes of the Low-Speed Collider}

\begin{figure}[t!]
\centering
\begin{subfigure}{0.45\textwidth}
  \centering
  \hspace*{-1cm}
  \includegraphics[width=1\linewidth]{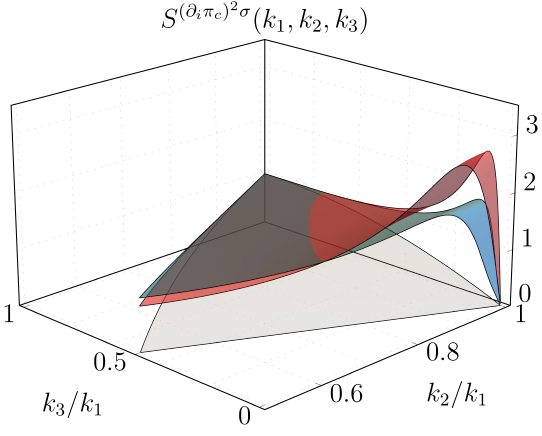}
\end{subfigure}
\begin{subfigure}{0.45\textwidth}
  \centering
  \includegraphics[width=1\linewidth]{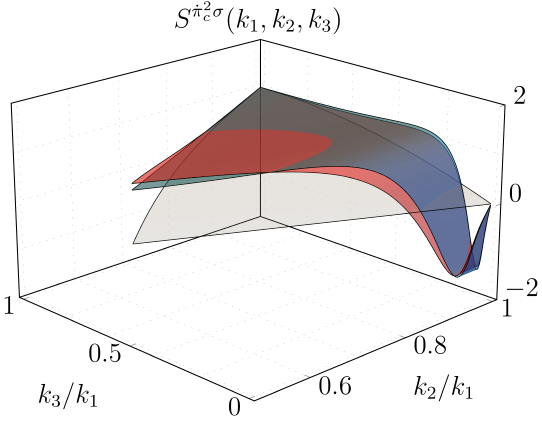}
\end{subfigure}
\begin{subfigure}{0.45\textwidth}
  \centering
  \hspace*{-1cm}
  \includegraphics[width=1\linewidth]{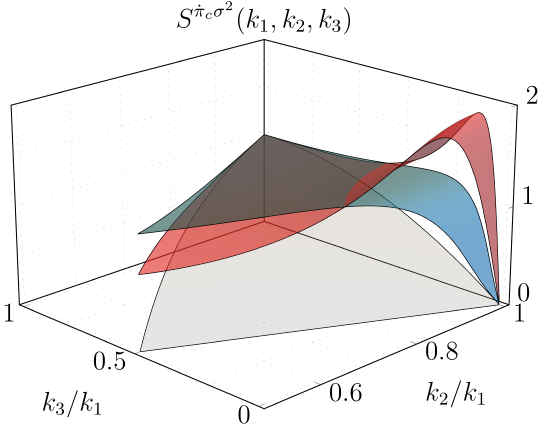}
\end{subfigure}
\begin{subfigure}{0.45\textwidth}
  \centering
  \includegraphics[width=1\linewidth]{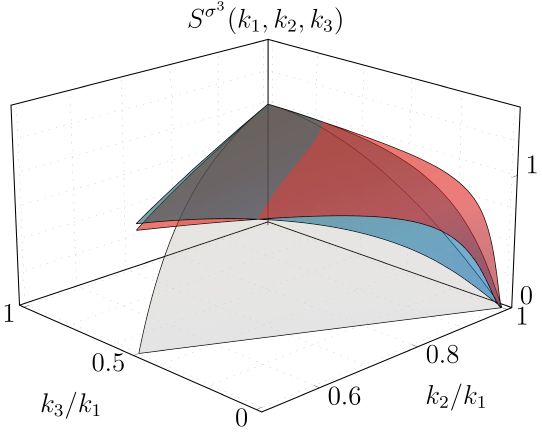}
\end{subfigure}
\caption{The dimensionless shape functions $S(k_1, k_2, k_3)$ normalised to unity in the equilateral configuration $k_1=k_2=k_3$ at \textcolor{pyblue}{weak mixing} in \textcolor{pyblue}{blue} for $\alpha=0.1$ and at \textcolor{pyred}{strong mixing} in \textcolor{pyred}{red} for $\alpha_{\text{eff}} = 0.1$. The different panels represent the shapes generated by the interactions $(\partial_i \pi_c)^2\sigma$, $\dot{\pi}_c^2\sigma$, $\dot{\pi}_c\sigma^2$ and $\sigma^3$. The shapes in the strong mixing regime have been computed with the cosmological flow approach described in Sec.~\ref{subsec: strong mixing} for $c_s=0.05, m/H=0.1, \rho/H=2$. We have represented in transparent \textcolor{gray}{gray} the standard equilateral shape for comparison.}
\label{fig: low-speed collider shapes alpha=0.1}
\end{figure}

\begin{figure}[t!]
\centering
\begin{subfigure}{0.45\textwidth}
  \centering
  \hspace*{-1cm}
  \includegraphics[width=1\linewidth]{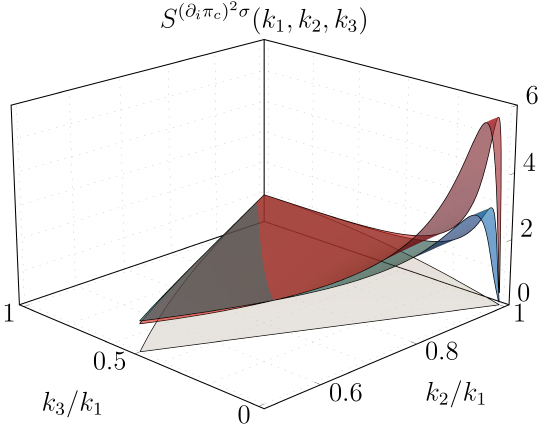}
\end{subfigure}
\begin{subfigure}{0.45\textwidth}
  \centering
  \includegraphics[width=1\linewidth]{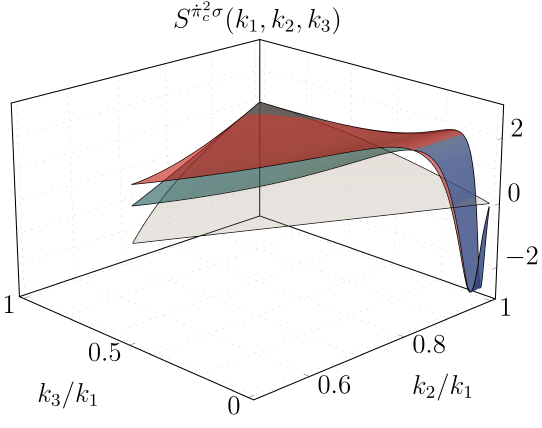}
\end{subfigure}
\begin{subfigure}{0.45\textwidth}
  \centering
  \hspace*{-1cm}
  \includegraphics[width=1\linewidth]{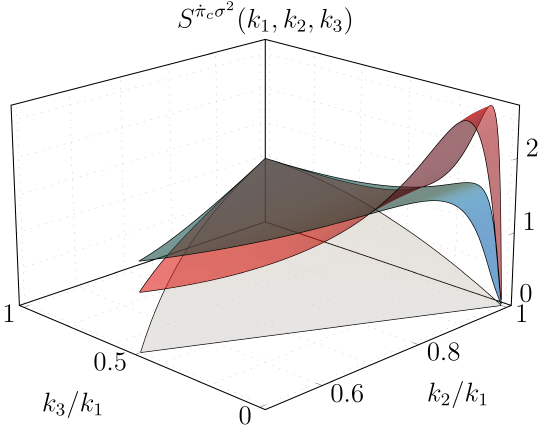}
\end{subfigure}
\begin{subfigure}{0.45\textwidth}
  \centering
  \includegraphics[width=1\linewidth]{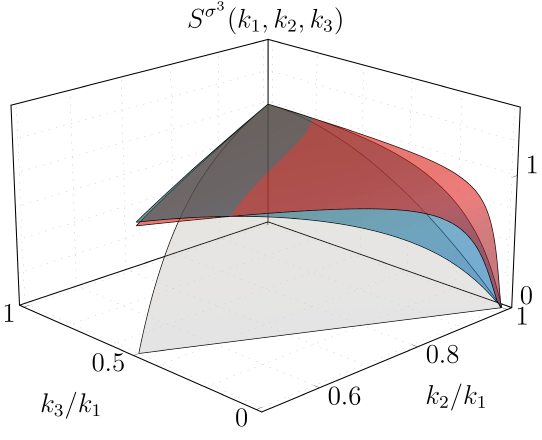}
\end{subfigure}
\caption{The dimensionless shape functions $S(k_1, k_2, k_3)$ normalised to unity in the equilateral configuration $k_1=k_2=k_3$ at \textcolor{pyblue}{weak mixing} in \textcolor{pyblue}{blue} for $\alpha=0.05$ and at \textcolor{pyred}{strong mixing} in \textcolor{pyred}{red} for $\alpha_{\text{eff}} = 0.05$, generated by the interactions $(\partial_i \pi_c)^2\sigma$, $\dot{\pi}_c^2\sigma$, $\dot{\pi}_c\sigma^2$ and $\sigma^3$. We have set $c_s=0.025, m/H=0.1, \rho/H=2$ in the strong mixing regime, and have represented in transparent \textcolor{gray}{gray} the standard equilateral shape.}
\label{fig: low-speed collider shapes alpha=0.05}
\end{figure}

In this section, we describe the low-speed collider shapes, as defined in Eq.~(\ref{eq: shape definition}). Without loss of generality, we order the momenta such that $k_3\leq k_2 \leq k_1$ and fix $k_1$ by virtue of scale invariance. Therefore, the shape information only depends on the two dimensionless ratios $k_2/k_1$ and $k_3/k_1$ that satisfy $0\leq k_3/k_1 \leq k_2/k_1 \leq 1$ and $k_1 \leq k_2 + k_3$. The second condition is the triangle inequality, enforcing the three momenta to close and form a triangle. We show the shape functions, using this parametrization, in Fig.~\ref{fig: low-speed collider shapes alpha=0.1} and \ref{fig: low-speed collider shapes alpha=0.05} for the four interactions $(\partial_i \pi_c)^2\sigma$, $\dot{\pi}_c^2\sigma$, $\dot{\pi}_c\sigma^2$ ad $\sigma^3$, both at weak and strong mixing, for $\alpha_{\text{eff}} = 0.1$ and $\alpha_{\text{eff}} = 0.05$, respectively. The shapes are normalised to unity in the equilateral configuration to ease comparison with the standard equilateral template, represented in gray.

\vskip 4pt
As we understood previously, the striking feature of the low-speed collider is the presence of a resonance in the squeezed limit. This resonance is readily visible in the shape generated by the interaction $(\partial_i \pi_c)^2\sigma$. At weak mixing, the location and the amplitude of the peak are governed by $\alpha$. For smaller $\alpha$, the resonance is visible in more squeezed configurations and has a larger amplitude, proportional to $\alpha^{-1}$. The shape generated by the interaction $\dot{\pi}_c^2 \sigma$---whose amplitude is not fixed by the quadratic coupling---presents a more complex resonance characterised by a double-peak structure, i.e.~a first small peak and a second one more noticeable. 
The shape generated by the interaction $\dot{\pi}_c\sigma^2$---which leads to double-exchange diagrams at weak mixing---have a resonance less pronounced than those generated by $(\partial_i \pi_c)^2\sigma$ and $\dot{\pi}_c^2 \sigma$. As for the interaction  $\sigma^3$, it does not lead to resonances, which can be traced back to the presence of additional propagators in the corresponding interactions once the heavy field $\sigma$ has been non-locally integrated out.
This is peculiar though to the situation considered here for simplicity of one additional heavy field, and it is straightforward to generalize our results to more generic setups.\footnote{More specifically, the natural generalisation of \eqref{eq: full theory} is obtained by introducing the operators $\sum_a \rho_a \sigma_a \dot{\pi}_c$, $\sum_a (\Lambda_{1,a})^{-1}\,\sigma_a  (\tilde{\partial}_i \pi_c)^2$, $\sum_a (\Lambda_{2,a})^{-1}\sigma_{a}\dot{\pi}_c^2$, $\sum_{a b} \lambda_{a b}\sigma_a \sigma_b\dot{\pi}_c$ and $\sum_{a,b,c}\mu_{abc}\sigma_a\sigma_b\sigma_c$, with $\sigma_{a=1,\dots, N}$ standing for $N$ heavy scalars with masses $m_a$, and $\rho_a, (\Lambda_{1,a})^{-1}, (\Lambda_{2,a})^{-1}, \lambda_{a b}, \mu_{abc}$ for flavour-indexed coupling constants. At weak mixing, these interactions induce single-, double- and triple-exchange diagrams, incorporating internal lines with different masses for the last two ones. The computation is then identical to the case with only one heavy field, mutatis mutandis. In fact, the expressions for the correlators become identical to Eqs.~\eqref{correction-power-spectrum}-\eqref{triplediag}, after a trivial adjustment for the coupling constants and accounting for the mass of each propagator inside the arguments of ${\cal I}_{1,2,3}$ in Eq.~\eqref{eq: seed Fs non-local EFT}.}
An interesting case then is the shape generated by the triple-exchange diagram of several heavy fields having different masses, say $m_1, m_2$ and $m_3$. When $m_2/H, m_3/H\geq 1/c_s$, the corresponding fields can be integrated out in a local manner, and the shape therefore 
resembles the one generated by $\dot{\pi}_c^2\sigma$, with an amplitude suppressed by $(H/m_2)^2 (H/m_3)^2$. In this case, the interaction only consists of the propagator associated with
the field of mass $m_1$ and the shape 
displays a resonance around $k_3/k_1\sim c_s m_1/H$. Similarly, if only one field is heavier than $1/c_s$, the resonance would be sensitive to the lightest field. In essence, the low-speed collider resonance is a natural discovery channel to detect heavy fields with masses not far from the Hubble scale when the Goldstone boson has a reduced speed of sound, in analogy with resonance peaks in ground-based colliders proving the existence of new particles.

\vskip 4pt
At strong mixing, in the ``effective mass regime" depicted in Fig.~\ref{fig: weak&strong low-speed collider pheno}, the low-speed collider physics leads to a similar phenomenology of shapes, though the resonances are more pronounced. We have understood that this regime can be qualitatively described by the weak mixing shapes after replacing $\alpha$ by $\alpha_{\text{eff}}$. Here, the exact computation using the cosmological flow approach gives a quantitative measure of the quality of this approximation.
Indeed, if this substitution was exactly accurate, the shapes at weak and strong mixing in Fig.~\ref{fig: low-speed collider shapes alpha=0.1} and \ref{fig: low-speed collider shapes alpha=0.05} would coincide. The slight mismatch signals that non-trivial physics occurs between sound-horizon crossing of $\pi_c$ and mass-shell crossing of $\sigma$ in the strong mixing regime that cannot be inferred from 
the dispersion relation analysis in Sec.~\ref{sec:physics-LSC}. As such, only an exact numerical calculation can in turn be decisive. Remarkably, substituting $\alpha \rightarrow \alpha_{\text{eff}}$ in the analytical shape generated by $\dot{\pi}_c^2\sigma$ at weak mixing gives an excellent match, and the ``effective mass regime" still provides an accurate description of the low-speed collider physics at strong mixing. In this respect, note that we have checked that this also applies to the amplitude of the signal, and not only to the shape information as shown in the figures.

\vskip 4pt
In the following, we give a quantitative analysis of the low-speed collider shapes by inspecting their overlap with the standard templates.

\subsection{Shape Correlations}

\begin{figure}[h!]
     \centering
     \hspace*{-0.7cm}
     \includegraphics[scale=0.8]{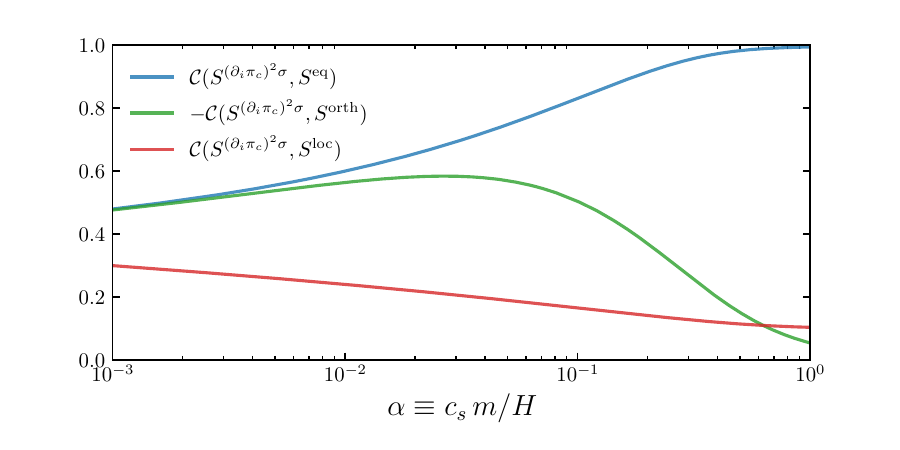}
     \caption{Correlations of the shape $S^{(\partial_i \pi_c)^2\sigma}$ with the standard \textcolor{pyblue}{equilateral} template $S^{\text{eq}}$ in \textcolor{pyblue}{blue}, \textcolor{pygreen}{orthogonal} template $S^{\text{orth}}$ in \textcolor{pygreen}{green}, and \textcolor{pyred}{local} template $S^{\text{loc}}$ in \textcolor{pyred}{red}, as functions of $\alpha \equiv c_s m/H$ in the weak mixing regime.}
     \label{fig: shape correlation}
 \end{figure}

To further characterise the low-speed collider shapes, we now compute their correlations with the standard local \cite{Gangui_1994, Wang:1999vf, Verde:1999ij, Komatsu:2001rj}, equilateral \cite{Babich:2004gb, Creminelli:2005hu} and orthogonal shapes \cite{Senatore:2009gt} used in data analysis, that are defined by 
\begin{equation}
\begin{aligned}
\label{eq: usual shape templates}
    S^{\text{loc}}(k_1, k_2, k_3) &= \frac{1}{3}\left(\frac{k_1^2}{k_2 k_3} + \text{2 perms}\right)\,, \\
    S^{\text{eq}}(k_1, k_2, k_3) &= \left(\frac{k_1}{k_2} + \text{5 perms}\right) - \left(\frac{k_1^2}{k_2 k_3} + \text{2 perms}\right) - 2\,, \\
    S^{\text{orth}}(k_1, k_2, k_3) &= 3\,S^{\text{eq}}(k_1, k_2, k_3) - 2 \,.
\end{aligned}
\end{equation}
Given two shapes $S^a$ and $S^b$, we define their correlation by \cite{Babich:2004gb, Fergusson:2008ra}

\begin{equation}
    \mathcal{C}(S^a, S^b) = \frac{\braket{S^a, S^b}}{\sqrt{\braket{S^a, S^a} \braket{S^b, S^b}}}\,,
\end{equation}
where the inner product is

\begin{equation}
    \braket{S^a, S^b} = \int_0^1\mathrm{d}x_2 \int_{1-x_2}^1 \mathrm{d}x_3\, S^a(1, x_2, x_3) \, S^b(1, x_2, x_3)\,.
\end{equation}
This correlation provides a quantitative measure of how distinguishable two shapes are, i.e.~two shapes are highly correlated if $|\mathcal{C}|$ is close to unity. For definiteness, we only focus on the shape \eqref{universal-weak-mixing}
generated by the interaction $(\partial_i \pi_c)^2\sigma$, that leads to the most sizeable low-speed collider resonance. 
We show in Fig.~\ref{fig: shape correlation} the correlations with the standard templates (\ref{eq: usual shape templates}) as functions of $\alpha$. As anticipated, for $\alpha \approx 1$, the shape is strongly correlated with the equilateral template. As the parameter $\alpha$ decreases, the low-speed collider resonance appears and becomes more pronounced. As a consequence, the shape correlation with $S^{\text{eq}}$ decreases while that with $S^{\text{loc}}$ increases. For sufficiently small $\alpha$, it is interesting to notice that we obtain a non-negligible correlation with $S^{\text{loc}}$, commonly attributed to the presence of multiple light fields during inflation. In contrast, the low-speed collider shape has been derived from an (effective) \textit{single}-field theory after integrating out a \textit{heavy} field. Overall though, in the bulk of the low-speed collider parameter space $0.01 \lesssim \alpha \lesssim 0.1$, there is a poor overlap between the representative shape $S^{(\partial_i \pi_c)^2\sigma}$ and the conventional templates.

\vskip 4pt
This analysis thus reveals that probing the bispectrum using only the standard templates $S^{\text{loc}}, S^{\text{eq}}$ and $S^{\text{orth}}$ could lead to a biased interpretation of data about the physics active in the primordial universe. Namely, conventional templates would not be blind to the low-speed collider signal, but a non-zero response to the equilateral/orthogonal and local templates could then be wrongly attributed to a scenario with multiple light fields endowed with a non-trivial sound speed, like in multifield DBI inflation \cite{RenauxPetel:2009sj}. Therefore, it appears essential to give a proper treatment to the low-speed collider shapes.

\subsection{Low-Speed Collider Template} 

For convenience of data analysis, we give a simple and universal template for the shape that captures the low-speed collider resonance. The derived analytical shapes are not convenient to deal with numerically because of fine cancellations appearing in Eq.~(\ref{eq: I1 EFT integral}). For this reason, we give a simpler template expressed in terms of elementary functions. To construct it, we realise that the low-speed collider shape behaves like the local one close to equilateral configurations and like the equilateral one in the squeezed limit, resulting in the low-speed collider resonance located in mildly-squeezed configurations. Naturally, the peak is shifted from the equilateral configuration to the squeezed limit as the parameter $\alpha$ decreases. A simple template for such a shape is given by  

\begin{figure}[t!]
\centering
\begin{subfigure}{0.45\textwidth}
  \centering
  \hspace*{-1cm}
  \includegraphics[width=1.2\linewidth]{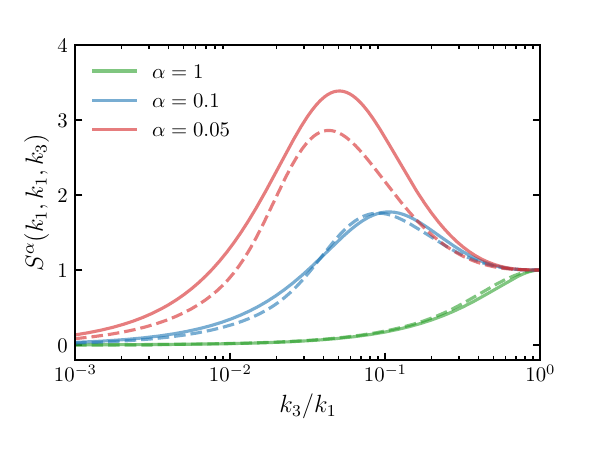}
\end{subfigure}
\begin{subfigure}{0.45\textwidth}
  \centering
  \includegraphics[width=1\linewidth]{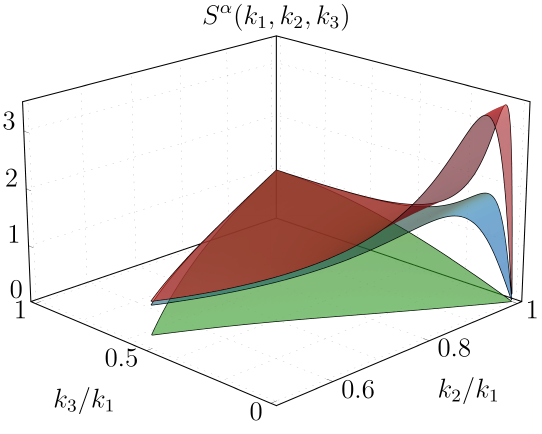}
\end{subfigure}
\caption{The low-speed collider shapes $S^\alpha(k_1, k_2, k_3)$ for $\alpha=1$ in \textcolor{pygreen}{green}, $\alpha=0.1$ in \textcolor{pyblue}{blue} and $\alpha=0.05$ in \textcolor{pyred}{red}. \textit{Left}: The template $S^\alpha$ (\textit{solid line}) and the shape $S^{(\partial_i \pi_c)^2\sigma}$ (\textit{dashed line}) in the isosceles-triangle configuration $k_1 = k_2$. \textit{Right}: The template $S^\alpha$ given in Eq.~(\ref{eq: low speed collider template}) in all triangular configurations. For illustration purposes, we have normalised the amplitudes of the signals to unity in the equilateral configuration.}
\label{fig: low speed collider template}
\end{figure}

\begin{tcolorbox}[colframe=white,arc=0pt,colback=greyish2]
\begin{equation}
\label{eq: low speed collider template}
    S^{\alpha}(k_1, k_2, k_3) = S^{\text{eq}}(k_1, k_2, k_3) + \frac{1}{3} \frac{k_1^2}{k_2 k_3} \left[1 + \left(\alpha \,\frac{k_1^2}{k_2 k_3}\right)^2\right]^{-1} + \text{ 2 perms}\,,
\end{equation}
\end{tcolorbox}
\noindent where permutations only apply to the second term and the equilateral shape $S^{\text{eq}}$ is defined in Eq.~(\ref{eq: usual shape templates}). By construction, the shape $S^\alpha$ interpolates between the usual equilateral and local shapes, and can be used for the observational search of primordial non-Gaussianity. As illustrated in Fig.~\ref{fig: low speed collider template}, it captures well the location and the amplitude of the low-speed collider resonance for the $\alpha$ range of interest.
Remarkably, this simple template also reproduces the qualitative properties of the shapes generated by the non-universal single- and double-exchange diagrams, with $\dot{\pi}_c^2\sigma$ and $\dot{\pi}_c \sigma^2$ interactions respectively, albeit replacing $\alpha$ by effective parameters $\alpha^{\dot{\pi}_c^2\sigma}_{\text{eff}} = 2\alpha \log_{10}(\alpha)$ and $\alpha^{\dot{\pi}_c\sigma^2}_{\text{eff}} = \alpha/2$.
Note that this template is different from the one obtained in the context of multiple light fields with different sound speeds where resonances were already noticed \cite{Renaux-Petel:2011rmu, Wang:2022eop}. 
In this case, when adjusting parameters of such a template to reproduce the location of the peak, the template misses the relative amplitude of the equilateral signal compared to the resonant one.
Contrariwise, the shape (\ref{eq: low speed collider template}) results from the exchange of a massive field. One can therefore observationally distinguish, despite similar features, the exchange of light fields from signatures of massive fields.

\vskip 4pt
It is well known that the computational complexity is drastically reduced when looking for a template that is \textit{product separable}, in the sense of a (sum of) product of functions of momentum variables. This is due to numerical challenges of integrating highly oscillating functions that arise when projecting cosmological observables onto two-dimensional redshift surfaces 
\cite{Creminelli:2005hu, Smith:2006ud,Fergusson:2006pr,Fergusson:2008ra}. One might rightly worry that the second term in Eq.~(\ref{eq: low speed collider template}) ruins factorizability. However, this apparent problem can be solved by introducing a Mellin parameter\footnote{A similar trick is used in \cite{Smith_2011} where ubiquitous
behaviours of the bispectrum related to what is now referred to as the total energy pole are made product separable by introducing a Schwinger parameter $(k_1 + k_2 + k_3)^{-n} = \Gamma(n)^{-1} \int_0^{\infty} \mathrm{d}x\, x^{n-1} e^{-(k_1+k_2+k_3)x}$.}

\begin{equation}
    \left[1 + \left(\alpha \,\frac{k_1^2}{k_2 k_3}\right)^2\right]^{-1} = \frac{1}{2i\pi} \int_{c-i \infty}^{c+i\infty}\d s\, \Gamma(s) \Gamma(1-s) \,\alpha^{-2s} \left(\frac{k_1^2}{k_2 k_3}\right)^{-2s}\,,
\end{equation}
where $0<c<1$ such that the contour separates poles of $\Gamma(s)$ from those of $\Gamma(1-s)$. The integrand is manifestly product separable in the momenta.\footnote{Note that it is not possible to write the term that ruins factorizability as a geometric series because it does not converge close to equilateral configurations for some permutations.} Along the contour, e.g.~by setting $s\rightarrow 1/2 + i \tilde{s}$, the Mellin-Barnes integral can be accurately computed numerically. More generally, a similar trick can also be used when the shape has a non-trivial pole structure.

\subsection{Self-interaction Contamination}
\label{sec:self-interaction}

In the presence of a reduced speed of sound, the non-linearly realised time diffeomorphism invariance unavoidably generates large Goldstone boson self-interactions which contaminate the low-speed collider signature. 
This is attributed to the cubic interaction $\dot{\pi}_c(\partial_i \pi_c)^2$ in Eq.~\eqref{eq: full theory}, whose size is fixed by $c_s$ and that generates an equilateral-type bispectrum of amplitude $\sim 1/c_s^2$ at weak mixing. Naturally, by tuning the Wilson coefficient setting the size of the other self-interaction $\dot{\pi}_c^3$, it is possible to lower the amplitude of the total self-interaction signal precisely at the location of the low-speed collider resonance, making it more visible.\footnote{In that respect, note that the ``multi-speed'' shapes put forth in \cite{Wang:2022eop} are not observable without a severe fine-tuning, as they are necessarily accompanied by a much larger equilateral-type shape that dwarfs it.} We do not allow this in the following: we concentrate on the signal fixed by symmetries, so that our assessment of the observability of the low-speed collider resonance is conservative.

\begin{figure}[h!]
    \centering
    \hspace*{-1.8cm}
    \subfloat{\includegraphics[width=1.2\textwidth]{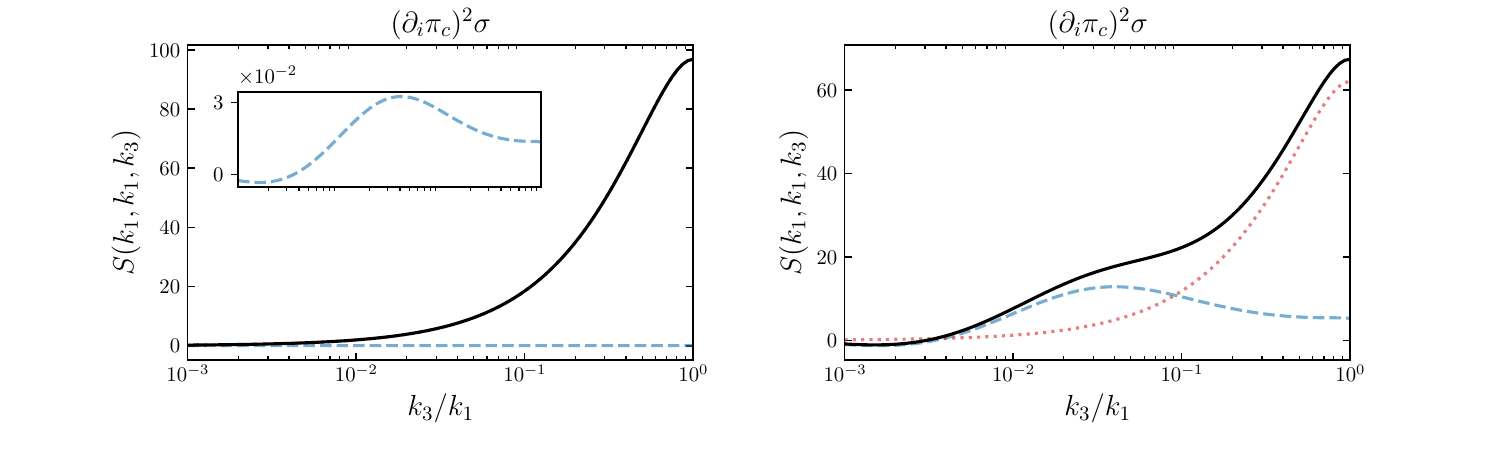}}
    \hfill
    \centering
    \hspace*{-1.8cm}
    \subfloat{\includegraphics[width=1.2\textwidth]{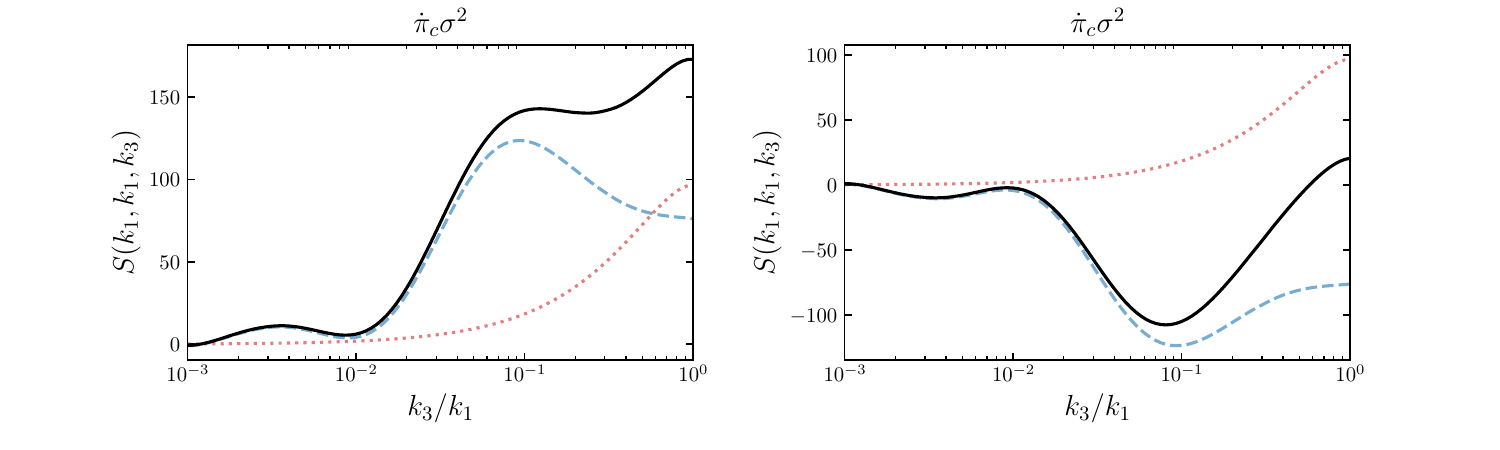}}
    \hfill
    \vspace*{-1cm}
    \caption{Shape functions generated by $(\partial_i \pi_c)^2\sigma$ (\textit{top}) and $\dot{\pi}_c\sigma^2$ (\textit{bottom}) in dashed \textcolor{pyblue}{blue} line and self-interactions of the Goldstone boson $\dot{\pi}_c(\partial_i \pi_c)^2$ fixed by the speed of sound $c_s$ in dotted \textcolor{pyred}{red} line, and the sum of both signals in solid black line, in the isosceles-triangle configuration $k_1=k_2$. \textit{Top panel}: In the weak mixing regime (\textit{left}), we have set $m/H=2, \rho/H=0.1$ and $c_s=0.05$ so that $\alpha = 0.1$. At strong mixing in the ``effective mass regime" (\textit{right}), we have set $m/H=0.1, \rho/H=2$ and $c_s=0.05$ so that $\alpha_{\text{eff}} = 0.1$. \textit{Bottom panel}: We have set $\rho/H=m/H=2, c_s=0.05$ with $\lambda=0.01$ (\textit{left}) and $\lambda=-0.01$ (\textit{right}).}
    \label{fig: equilateral contaminations}
\end{figure}

\vskip 4pt
Let us first estimate the relative size of the low-speed collider resonance with respect to the self-interaction contamination at weak mixing. 
Focusing on the interaction $(\partial_i \pi_c)^2 \sigma$, the shape in the squeezed limit around $k_3/k_1\sim \alpha$ scales as $S^{(\partial_i \pi_c)^2 \sigma} \sim \alpha^{-1}(\rho/H)^2$, see Eq.~(\ref{eq: fNL size of S11}), where the $\alpha^{-1}$ is attributed to the resonance. The standard equilateral-type shape decreases like $k_3/k_1$ so that $S^{\dot{\pi}_c(\partial_i \pi_c)^2} \sim c_s^{-2} \alpha$. Therefore, the ratio of both signals is $S^{(\partial_i \pi_c)^2 \sigma}/S^{\dot{\pi}_c(\partial_i \pi_c)^2} \sim (\rho/m)^2 \lesssim 1$.
In the strict weak mixing regime $\rho \ll m$, the low-speed collider resonance is dwarfed by the ever-present self-interaction signal. 
In Fig.~\ref{fig: equilateral contaminations}, we show the exact low-speed collider signal and the corresponding self-interaction contamination, as well as the total shape, for different scenarios. At weak mixing (top left), the low-speed collider is indeed not visible in the total signal one would measure. As a consequence, it can only be detected after properly subtracting the equilateral-type contamination, considered as a ``background noise" in this case.
Although the scalings above miss $\mathcal{O}(1)$ coefficients,
the two contributions become comparable for $\rho \sim m$, which motivates to consider strong mixing where the situation is more interesting. 
In the ``effective mass regime", as discussed previously, replacing $m$ by the effective mass $m_{\text{eff}}$ gives an accurate description of the low-speed collider physics at strong mixing.
The low-speed collider resonance therefore scales as $S^{(\partial_i \pi_c)^2 \sigma} \sim \alpha_{\text{eff}}^{-1}(\rho/H)^2$, where $\alpha_{\text{eff}}\approx c_s \rho/H$, while the equilateral-type shape is $S^{\dot{\pi}_c(\partial_i \pi_c)^2} \sim c_s^{-2} \alpha_{\text{eff}}$. The ratio leads to $S^{(\partial_i \pi_c)^2 \sigma}/S^{\dot{\pi}_c(\partial_i \pi_c)^2} \sim 1$. It means that the resonance is not suppressed, and is well visible in the total shape, as seen in the top right panel of Fig.~\ref{fig: equilateral contaminations}.
Note that the estimates of the self-interaction signal parametrically hold, as the propagation of $\pi_c$ is not much affected by its mixing with $\sigma$ for the regime of interest $\alpha_{\text{eff}} \lesssim 1$. However, the full dynamics is taken into account in the numerical results of Fig.~\ref{fig: equilateral contaminations}.

\vskip 4pt
It is straightforward to generalise these estimates to the other interactions, which can lead to low-speed collider signals with even larger amplitude as their corresponding cubic couplings are not tied to $\rho/H$.
For definiteness, the case of the interaction $\dot{\pi}_c \sigma^2$ is considered on the bottom panels of Fig.~\ref{fig: equilateral contaminations}, for two different values $\lambda=\pm 0.01$ of the coupling constant in Eq.~\eqref{eq: full theory} ensuring perturbative control. 
The low-speed collider resonance is well visible, notably due to the enhancement of the signal by $\Delta_\zeta^{-1}$.
Additionally, note that when the signal generated by the mixing cubic interaction and the one generated by self-interactions have opposite signs, the low-speed collider signature is well distinct in the total shape.

\vskip 4pt
Measuring equilateral-type non-Gaussianities would constrain the speed of sound $c_s$, and the position of the low-speed collider resonance would then pinpoint the (effective) mass of the additional heavy field. While this picture is correct at weak mixing, it is in general more subtle at strong mixing. Indeed, the total signal near equilateral configurations is not entirely fixed by $c_s$, as the low-speed collider and self-interaction signals can have a comparable amplitude there, see for instance the situations in the bottom panels in Fig.~\ref{fig: equilateral contaminations}.

\section{Conclusions}
\label{sec: conclusions}

In this paper, we have studied the imprint of massive fields on cosmological correlators in the presence of a reduced speed of sound for the Goldstone boson of broken time translations. This minimal framework leads to a whole new class of signatures---that we call \textit{cosmological low-speed collider signals}---which are characterised by a distinctive resonance in the mildly-squeezed limit of the bispectrum. This work generalises \cite{Jazayeri:2022kjy} to include all leading cubic interactions, which at weak mixing lead to single-, double- and triple-exchange three-point diagrams, and the corresponding strong mixing regime in which the unavoidable quadratic mixing cannot be treated perturbatively. Let us summarise our results.

\begin{itemize}
    \item Using the framework of EFT of inflationary fluctuations, we have coupled the Goldstone boson of broken time translations $\pi$ to a massive scalar field $\sigma$, identifying allowed leading cubic interactions giving sizeable non-Gaussianities. When the speed of sound of $\pi$ is reduced $c_s \lesssim 1$, the heavy field can be integrated out in an unusual manner, yielding a \textit{non-local} single-field theory, see Eq.~(\ref{eq: non-local theory}). This theory emerges as the leading-order term in a time-derivative expansion. Although the heavy field becomes effectively non-dynamical, this simple theory accurately describes the full dynamics of the Goldstone boson and captures all multi-field physical effects---except for the non-perturbative particle production leading to the conventional cosmological collider signal---in the almost entire parameter space. 
    
    \item We have characterised this non-local EFT by analysing the dispersion relation, and have shown that this theory encapsulates several known regimes like the effective reduced sound speed and the modified dispersion relation regimes. 
    This theory provides a systematic and tractable single-field EFT treatment of the imprint of massive fields on cosmological correlators. We have also determined the precise regimes of validity of this theory and shown how to systematically incorporate higher-order corrections in the derivative expansion.

    \item At weak mixing, we have used bootstrap techniques to analytically compute the corresponding bispectra. We have demonstrated that all three-point correlators originate from a \textit{unique simple seed correlator}, which in turn is mapped to correlators of $\pi$ by well-chosen \textit{weight-shifting operators}. At strong mixing, we have used the cosmological flow approach to obtain exact numerical results for the correlators of interest. Remarkably, we have also shown that the analytical understanding at weak mixing extends to the strong mixing regime of the low-speed collider. In all cases, our approaches allow for a complete characterisation of correlators in all kinematic configurations.
    
    \item The phenomenology of the low-speed collider physics is summarised in Figs.~\ref{fig: intro} and \ref{fig: weak&strong low-speed collider pheno}. We have determined the size of non-Gaussianities and have studied the corresponding shape dependence of the bispectra, whose distinctive signature is governed by a single parameter, see Eq.~(\ref{eq: alpha-eff parameter}). For observational relevance, we give a simple product-separable template for the low-speed collider shape in Eq.~(\ref{eq: low speed collider template}). Our results are cautiously optimistic: within the regime of validity of the EFT, we can accommodate observationally large non-Gaussianity and sizeable low-speed collider signatures, in particular in the strong mixing regime.
\end{itemize}

\vskip 4pt
The promising and rich low-speed collider phenomenology exposed in this paper is an invitation to extend and generalise these signatures to spinning intermediate fields and mixed correlators involving gravitons.
Finally, it would be interesting from a theoretical perspective to investigate the possibility of systematically constructing a set of boost breaking effective field theories that are non-local in space, yet still consistent with a local UV completion.

\vspace*{0.5cm}
\paragraph{Acknowledgements.} We thank 
Giovanni Cabass, 
Sebastian Cespedes,
Thomas Colas,
Paolo Creminelli,
Gregory Kaplanek,
Scott Melville,
Enrico Pajer,
Andrew Tolley,
Xi Tong and Yuhang Zhu 
for helpful discussions.
SJ, SRP and DW are supported by the European Research Council under the European Union's Horizon 2020 research and innovation programme (grant agreement No 758792, Starting Grant project GEODESI). This article is distributed under the Creative Commons Attribution International Licence (\href{https://creativecommons.org/licenses/by/4.0/}{CC-BY 4.0}).

\newpage
\appendix
\section{Weight-Shifting Operator Derivation} 
\label{app: WS Operators}
It was pointed out in \cite{Arkani-Hamed:2018kmz} that de Sitter invariant, four-point diagrams consisting of external massless fields can be related to those of the external conformally coupled field through a set of weight-shifting operators. In \cite{Jazayeri:2022kjy}, it was shown that even without de Sitter boost symmetry similar relationships continue to hold between the correlators of the Goldstone boson and the ones of $\vpi$ (defined to be a massive field with $m_\vpi^2=2H^2$ that propagates with the same speed of sound as $\pi_c$).\footnote{In \cite{Jazayeri:2022kjy}, $\vpi$ was taken to be the conformally coupled field, hence having a relativistic dispersion relation. The speed of sound for the Goldstone boson was then adjusted by rescaling the energies of the external $\vpi$ legs, i.e.~$k\to c_s k$. In contrast, here we start with a theory for $\vpi$ that has a reduced sound speed. The ending results of both approaches are identical up to easily identifiable factors of $c_s$.}
In this appendix, we generalise the result in the latter paper to the double- and triple-exchange diagrams that contribute to the bispectrum of $\pi_c$. We achieve this by linking these graphs to similar ones of $\vpi$'s four- and six-point functions, defined in \eqref{vpicordef}. For the simplicity of the presentation, here we concentrate on the non-local EFT limit where $\sigma$ can be integrated out. 
However, \textit{the final weight-shifting operators also apply to the same diagrams in the full two-field theory, where $\sigma$ is a dynamical degree of freedom}. 

\vskip 4pt
In the non-local theory, the heavy field is non-dynamical, therefore its associated bulk-to-bulk propagators at leading order in derivative is proportional to a Dirac delta function, i.e.

\begin{equation}
\raisebox{-\height/2}{
\begin{tikzpicture}[line width=1. pt, scale=2]
\node at (0.75,-0.2) {\footnotesize{$\bm{s}$}};
\draw[pyblue] (0.4,0) -- (1.1, 0);
\end{tikzpicture}}
=
    {\cal D}^{-1}(\bm{s},\tau) \dfrac{1}{a(\tau)}\delta (\tau-\tau')\,.
\end{equation}
Therefore, we only need to identify operators that map the vertices of the diagrams in the theory containing $\vpi$ onto those in the theory for the Goldstone boson $\pi_c$. Specifically, we want to find the map between the following building block 

\begin{equation}
    \mathbb{B}_{\vpi^2}(k_1,k_2,\tau) = 
    \raisebox{-\height/2}{
\begin{tikzpicture}[line width=1. pt, scale=2.5]
\draw[black] (0.2, 0) -- (0.4, -0.6);
\draw[black] (0.6, 0) -- (0.4, -0.6);
\draw[fill=black] (0.4, -0.6) circle (.03cm);
\draw[lightgray2, line width=0.8mm] (0, 0) -- (0.8, 0);
\node at (0.4, -0.8) {\footnotesize{$\vpi^2$}};
\node at (0.2,-0.4) {\footnotesize{$\bm{k}_1$}};
\node at (+0.6,-0.4) {\footnotesize{$\bm{k}_2$}};
\end{tikzpicture} 
} 
    = a^4(\tau)\vpi_{k_1}(\tau)\vpi_{k_2}(\tau)\vpi_{k_1}^*(\tau_0)\vpi_{k_2}^*(\tau_0)\,,
\end{equation}
and the building blocks below, associated with graphs with external $\pi$ lines:

\begin{equation}
   \mathbb{B}_{\dot{\pi}_c}(k,\tau) = 
   \raisebox{-\height/2}{
\begin{tikzpicture}[line width=1. pt, scale=2.5]
\draw[pyred] (0.2, 0) -- (0.2, -0.6);
\draw[fill=black] (0.2, -0.6) circle (.03cm);
\draw[lightgray2, line width=0.8mm] (0, 0) -- (0.4, 0);
\node at (0.2, -0.8) {\footnotesize{$\dot{\pi}_c$}};
\node at (0,-0.4) {\footnotesize{$\bm{k}$}};
\end{tikzpicture} 
} 
    = a^3(\tau) \pi'_{c,k}(\tau)\pi_{c,k}^*(\tau_0)\,,
\end{equation}

\begin{equation}    
    \mathbb{B}_{\dot{\pi}_c^2}(k_1,k_2,\tau) =
    \raisebox{-\height/2}{
\begin{tikzpicture}[line width=1. pt, scale=2.5]
\draw[pyred] (0.2, 0) -- (0.4, -0.6);
\draw[pyred] (0.6, 0) -- (0.4, -0.6);
\draw[fill=black] (0.4, -0.6) circle (.03cm);
\draw[lightgray2, line width=0.8mm] (0, 0) -- (0.8, 0);
\node at (0.4, -0.8) {\footnotesize{$\dot{\pi}_c^2$}};
\node at (0.2, -0.4) {\footnotesize{$\bm{k}_1$}};
\node at (0.6, -0.4) {\footnotesize{$\bm{k}_2$}};
\end{tikzpicture} 
} 
    =a^2(\tau)\pi'_{c,k_1}(\tau)\pi'_{c,k_2}(\tau)\pi_{c,k_1}^*(\tau_0)\pi_{c,k_2}^*(\tau_0)\,,
\end{equation}

\begin{equation}
    \mathbb{B}_{(\partial_i\pi_c)^2}(k_1,k_2,\tau) = 
    \raisebox{-\height/2}{
\begin{tikzpicture}[line width=1. pt, scale=2.5]
\draw[pyred] (0.2, 0) -- (0.4, -0.6);
\draw[pyred] (0.6, 0) -- (0.4, -0.6);
\draw[fill=black] (0.4, -0.6) circle (.03cm);
\draw[lightgray2, line width=0.8mm] (0, 0) -- (0.8, 0);
\node at (0.4, -0.8) {\footnotesize{$(\partial_i \pi_c)^2$}};
\node at (0.2, -0.4) {\footnotesize{$\bfk_1$}};
\node at (0.6, -0.4) {\footnotesize{$\bfk_2$}};
\end{tikzpicture} 
}
    =-a^2(\tau) \,\bm{k}_1\cdot\bm{k}_2\,\pi_{c,k_1}(\tau)\pi_{c,k_2}(\tau)\pi_{c,k_1}^*(\tau_0)\pi_{c,k_2}^*(\tau_0)\,.
\end{equation}
It can be easily verified that all the above components can be mapped onto $\mathbb{B}_{\vpi^2}$ via the following boundary operators

\begin{gather}
    \begin{aligned}
\label{eq: weight-shifting operators}
    \mathbb{B}_{\dot{\pi}_c}(k,\tau)&=-\dfrac{2 c_s k_{\text{soft}}}{H\tau_0^2}\lim_{k_{\text{soft}}\to 0}\mathbb{B}_{\vpi^2}(k,k_{\text{soft}},\tau)\,,\\
    \mathbb{B}_{\dot{\pi}_c^2}(k_1,k_2,\tau)&=\dfrac{H^2}{2c_s^2 \tau_0^2 k_1 k_2}\dfrac{\partial^2}{\partial k_{12}^2}\,\left[(k_1 k_2)\,\mathbb{B}_{\vpi^2}(k_1,k_2,\tau)\right]\,, \\
     \mathbb{B}_{(\partial_i\pi_c)^2}(k_1,k_2,\tau)&=\dfrac{H^2}{c_s^4\tau_0^2 k_1^2 k_2^2}\dfrac{\bm{k}_1\cdot \bm{k}_2}{k_1 k_2}\left(1-k_{12}\dfrac{\partial}{\partial k_{12}}+k_1 k_2 \dfrac{\partial^2}{\partial k_{12}^2}\right)\left[(k_1 k_2)\,\mathbb{B}_{\vpi^2}(k_1,k_2,\tau)\right]\,, 
     \end{aligned}
\raisetag{57pt}
\end{gather}
where we have defined $k_{12}=k_1+k_2$. Notice that the combination $(k_1 k_2)\,\mathbb{B}_{\vpi^2}(k_1,k_2,\tau)$ only depends on $k_{12}$, therefore acting with $\partial_{k_{12}}$ is unambiguous. It should be noted that, following the same reasoning presented above, other building blocks associated with higher derivative vertices in $\pi_c$ can be straightforwardly related to $\mathbb{B}_{\vpi^2}$ via similar (albeit higher-derivative) boundary operators. 

\vskip 4pt
In the non-local EFT, all exchange contributions reduce to simple contact terms. Therefore, the \textit{in-in} expressions for diagrams are given by the image of the product of their corresponding building blocks, followed by an integration over time and multiplied by appropriate coupling constants and symmetry factors.
For example, for the double-exchange diagram contributing to the $\vpi$ six-point function, we have

\begin{equation}
    \begin{aligned}
    \raisebox{-\height/3}{
\begin{tikzpicture}[line width=1. pt, scale=2.5]
\node at (0.2, 0.15) {\footnotesize{$\bfk_1$}};
\node at (0.6, 0.15) {\footnotesize{$\bfk_2$}};
\node at (0.85, 0.15) {\footnotesize{$\bfk_3$}};
\node at (1.2, 0.15) {\footnotesize{$\bfk_4$}};
\node at (1.42, 0.15) {\footnotesize{$\bfk_5$}};
\node at (1.85, 0.15) {\footnotesize{$\bfk_6$}};
\draw[black] (0.2, 0) -- (0.4, -0.6);
\draw[black] (0.6, 0) -- (0.4, -0.6);
\draw[black] (0.8, 0) -- (1, -0.6);
\draw[black] (1.2, 0) -- (1, -0.6);
\draw[black] (1.4, 0) -- (1.6, -0.6);
\draw[black] (1.8, 0) -- (1.6, -0.6);
\draw[pyblue] (0.4, -0.6) -- (1, -0.6);
\draw[pyblue] (1, -0.6) -- (1.6, -0.6);
\draw[fill=black] (0.4, -0.6) circle (.03cm);
\draw[fill=black] (1, -0.6) circle (.03cm);
\draw[fill=black] (1.6, -0.6) circle (.03cm);
\draw[lightgray2, line width=0.8mm] (0, 0) -- (2, 0);
\end{tikzpicture} 
}
    &=-2\times 2^4\,g_1^2 g_2\, \text{Im}\,\int_{-\infty}^0\, d\tau\,\mathbb{B}_{\vpi^2}(k_1,k_2,\tau){\cal D}^{-1}(\bm{k}_1+\bm{k}_2,\tau) \\ \nn &
    \qquad\qquad\times \mathbb{B}_{\vpi^2}(k_3,k_4,\tau) {\cal D}^{-1}(\bm{k}_5+\bm{k}_6,\tau) \mathbb{B}_{\vpi^2}(k_5,k_6,\tau)\,. 
    \end{aligned}
\end{equation}
The double-exchange diagram of $\pi$, on the other hand, reads 
\begin{equation}
\begin{aligned}
\raisebox{-\height/3}{
\begin{tikzpicture}[line width=1. pt, scale=2.5]
\draw[pyred] (0.2, 0) -- (0.4, -0.6);
\draw[pyred] (1.8, 0) -- (1.6, -0.6);
\draw[pyred] (1, 0) -- (1, -0.6);
\draw[pyblue] (0.4, -0.6) -- (1, -0.6);
\draw[pyblue] (1, -0.6) -- (1.6, -0.6);
\draw[fill=black] (0.4, -0.6) circle (.03cm);
\draw[fill=black] (1, -0.6) circle (.03cm);
\draw[fill=black] (1.6, -0.6) circle (.03cm);
\draw[lightgray2, line width=0.8mm] (0, 0) -- (2, 0);
\node at (0.4, -0.8) {\footnotesize{$\dot{\pi}_c\sigma$}};
\node at (1.6, -0.8) {\footnotesize{$\dot{\pi}_c\sigma$}};
\node at (1, -0.8) {\footnotesize{$\dot{\pi}_c\sigma^2$}};
\node at (0.2, 0.15) {\footnotesize{$\bfk_1$}};
\node at (1, 0.15) {\footnotesize{$\bfk_2$}};
\node at (1.8, 0.15) {\footnotesize{$\bfk_3$}};
\end{tikzpicture} 
} 
&= \text{Im}\,\int_{-\infty}^0\, d\tau\,\mathbb{B}_{\dot{\pi}_c}(k_1,\tau){\cal D}^{-1}(\bm{k}_1,\tau) \\ \nn 
    &\qquad\qquad\times \mathbb{B}_{\dot{\pi}_c}(k_1,\tau) {\cal D}^{-1}(\bm{k}_2,\tau) \mathbb{B}_{\dot{\pi}_c}(k_3,\tau) \,. 
    \end{aligned}
\end{equation}
With the two expressions above and using the relationships between the building blocks of the two diagrams in \eqref{eq: weight-shifting operators}, one arrives at \eqref{wsform2}. Equations \eqref{wsform} and \eqref{wsform3} follow from a similar procedure. We should emphasise again that identical relationships apply to the full two-field theory, and all the more so to the sub-leading contributions arising from integrating out $\sigma$ at higher time-derivative orders. 
\section{Higher-order Non-local EFT Corrections} \label{subleadingnonlocal}

We have explicitly computed the power spectrum and the bispectra for all the interactions in the non-local EFT~(\ref{eq: non-local theory}). However, let us recall that this theory has been derived by considering only the leading-order term in the time derivative expansion~(\ref{eq: time derivative expansion}). In this section, we show how one can systematically correct correlators of $\pi_c$ by going to higher order in the derivative expansion~(\ref{eq: time derivative expansion}). While adding more terms to the non-local EFT improves the accuracy to which correlators are determined, and can be used to diagnose failures of the EFT, see \cite{Jazayeri:2022kjy}, it does not account for particle production effects, resulting in some level of unavoidable, but exponentially suppressed errors. 

\vskip 4pt
The seed correlators of the field $\varphi$ attributed to the single-, double- and triple-exchange diagrams, introduced in Eq.~\eqref{vpicordef}, at arbitrary order in the time-derivative expansion, are given by
\begin{align}
\nn
    &\braket{\varphi_{\bm{k}_1} \ldots \varphi_{\bm{k}_4}}_{(1)}' = -16g_1^2 \left(\prod_{i=1}^4 \varphi_{k_i}^*(\tau_0)\right) \\ 
    &\hspace*{0.5cm}\times \text{Im} \int_{-\infty}^0 \d \tau a^4 \varphi_{k_1} \varphi_{k_2} \left[\sum_{n=0}^\infty\textcolor{pyblue}{\hat{\mathcal{O}}_n(\Ms_{34})} \varphi_{k_3} \varphi_{k_4}\right] + \text{ 2 perms}\,,\\ \nn
    &\braket{\varphi_{\bm{k}_1} \ldots \varphi_{\bm{k}_6}}_{(2)}' = -96g_1^2 g_2 \left(\prod_{i=1}^6 \varphi_{k_i}^*(\tau_0)\right) \\
    &\hspace*{0.5cm}\times \text{Im} \int_{-\infty}^0 \d \tau a^4 \varphi_{k_1} \varphi_{k_2} \left[\sum_{n=0}^\infty\textcolor{pyblue}{\hat{\mathcal{O}}_n(\Ms_{34})} \varphi_{k_3} \varphi_{k_4}\right] \left[\sum_{n=0}^\infty\textcolor{pyblue}{\hat{\mathcal{O}}_n(\Ms_{56})} \varphi_{k_5} \varphi_{k_6}\right] + \text{ 15 perms}\,,\\ 
    \nn
    &\braket{\varphi_{\bm{k}_1} \ldots \varphi_{\bm{k}_6}}_{(3)}' = -96g_1^3 g_3 \left(\prod_{i=1}^6 \varphi_{k_i}^*(\tau_0)\right) \\
    &\hspace*{0.5cm}\times \text{Im} \int_{-\infty}^0 \d \tau a^4 \left[\sum_{n=0}^\infty\textcolor{pyblue}{\hat{\mathcal{O}}_n(\Ms_{12})} \varphi_{k_1} \varphi_{k_2}\right]\left[\sum_{n=0}^\infty\textcolor{pyblue}{\hat{\mathcal{O}}_n(\Ms_{34})} \varphi_{k_3} \varphi_{k_4}\right] \left[\sum_{n=0}^\infty\textcolor{pyblue}{\hat{\mathcal{O}}_n(\Ms_{56})} \varphi_{k_5} \varphi_{k_6}\right] + \text{ 15 perms}\,,
\end{align}
To avoid clutter, above we have omitted the (conformal) time dependence of $a(\tau)$, $\varphi_k(\tau)$, as well as the differential operators $\textcolor{pyblue}{\hat{\mathcal{O}}_n}$ that is defined by

\begin{equation}
\label{eq: non-local differential operator n}
    \textcolor{pyblue}{\hat{\mathcal{O}}_n(k)} = \frac{(-1)^n}{(k\tau)^2 + (m/H)^2} \left\{[\tau^2 \partial_\tau^2 - 2\tau \partial_\tau][(k\tau)^2 + (m/H)^2]^{-1}\right\}^n\,.
\end{equation}
The above expressions for the correlators of $\vpi$ systematically correct the leading-order results given in \eqref{eq: seed Fs non-local EFT}. Up to any given number of (time) derivatives, one can compute the time-integrals given above and straightforwardly extract the quantities $F_{1,2,3}$ that were defined in \eqref{vpicordef}. Subsequently, in order to compute with the same precision e.g. the power spectrum and the bispectrum of $\zeta$ (induced by the single-, double-, and triple-exchange graphs), one should act with the bespoke weight-shifting operators dictated by Eqs.~\eqref{wsform}-\eqref{wsform3}.

\section{Non-local EFT Resummation} 
\label{app: EFT Resummation}

In this appendix, as an interesting curiosity, we show how one can fully resum the non-local EFT in flat space. We will see that the effective calculation reproduces the one computed in the two-field theory in some particular limit. Additionally, we will comment on the singularity structure of these correlators.

\paragraph{Flat-space propagators.} In flat space, after setting $H=0$, the correlators are computed at a fixed time $t_0=0$, which is analogous to the end of inflation surface $\tau_0$. Let us consider a theory composed of a massless field $\phi$ propagating at speed $c_s$ and a relativistic field $\sigma$ with mass $m$. We assume that both fields are coupled at cubic order through the interaction $\mathcal{L} \supset -\frac{g}{2}\phi^2\sigma$ where $g$ is some coupling constant. 

\vskip 4pt
Following the diagrammatic rules developed in \cite{Giddings:2010ui, chen2017schwinger} albeit adapted to flat space, the bulk-to-boundary propagator of $\phi$ (also called the Wightman propagator) and the bulk-to-bulk propagator of $\sigma$ (also called the Feynman propagator) read

\begin{equation}
    \begin{aligned}
    W_k(t, t') &= \frac{1}{2c_s k}e^{-ic_s k (t - t')}\,,\\
    G_k(t, t') &= \frac{1}{2\omega} \left(e^{-i\omega(t-t')}\theta(t-t') + e^{i\omega(t-t')}\theta(t'-t)\right)\,,
    \end{aligned}
\end{equation}
with $\omega^2 = k^2+m^2$. In the following, we will consider the $s$-channel four-point function $\braket{\phi_{\bm{k}_1} \phi_{\bm{k}_2} \phi_{\bm{k}_3} \phi_{\bm{k}_4}}$. The $t$- and $u$- channels can be easily found by suitable permutations of the external momenta.

\paragraph{Exchange diagram.} In the two-field theory, the correlator is composed of four different sub-diagrams, corresponding to the four different time ordering for the two vertices of the exchange diagram. Using the Feynman rules, the first contribution is

\begin{equation}
\label{eq: I++ diagram}
    \begin{aligned}
    I_{++} &= 
    \raisebox{-\height/3}{
\begin{tikzpicture}[line width=1. pt, scale=2.5]
\draw[pyred] (0.2, 0) -- (0.4, -0.6);
\draw[pyred] (0.6, 0) -- (0.4, -0.6);
\draw[pyred] (1, 0) -- (1.2, -0.6);
\draw[pyred] (1.4, 0) -- (1.2, -0.6);
\draw[pyblue] (0.4, -0.6) -- (1.2, -0.6);
\draw[fill=black] (0.4, -0.6) circle (.03cm);
\draw[fill=black] (1.2, -0.6) circle (.03cm);
\draw[lightgray2, line width=0.8mm] (0, 0) -- (1.6, 0);
\end{tikzpicture} 
} \\
&= -g^2 \int_{-\infty}^0 \d t' \d t''\, W_{k_1}(0, t') W_{k_2}(0, t') G_{s_{12}}(t', t'') W_{k_3}(0, t'') W_{k_4}(0, t'') \\
&=  \frac{g^2}{16c_s^4 k_1 k_2 k_3 k_4} \left(\frac{1}{E E_L E_R} + \frac{1}{2\omega E_L E_R}\right)\,,
    \end{aligned}
\end{equation}
where we have defined $s_{ij} = |\bm{k}_i + \bm{k}_j|$ and $\omega^2 = s_{12}^2 + m^2$ being the exchanged energy of the massive field. We also have introduced the total energy $E \equiv c_s(k_{12} + k_{34})$ with $k_{ij} = k_i + k_j$ and the energy flowing into the left (resp. right) vertex $E_L \equiv c_s k_{12} + \omega$ (resp. $E_R \equiv c_s k_{34} + \omega$). After inspecting the properties of the propagators, it is easy to show that $I_{--} = I_{++}^*$, so that one has $I_{++}+I_{--} = 2\text{Re}(I_{++})$. However in our case, $I_{++}$ is real so that the sum of both diagrams is just twice the result obtained in (\ref{eq: I++ diagram}). The $I_{+-}$ contribution is

\begin{equation}
\label{eq: I+- diagram}
    \begin{aligned}
    I_{+-} &= 
\raisebox{-\height/2}{
\begin{tikzpicture}[line width=1. pt, scale=2.5]
\draw[pyred] (0.2, 0) -- (0.4, -0.6);
\draw[pyred] (0.6, 0) -- (0.4, -0.6);
\draw[pyred] (1, 0) -- (1.2, 0.6);
\draw[pyred] (1.4, 0) -- (1.2, 0.6);
\draw[fill=black] (0.4, -0.6) circle (.03cm);
\draw[fill=black] (1.2, 0.6) circle (.03cm);
\draw[lightgray2, line width=0.8mm] (0, 0) -- (1.6, 0);
\draw[pyblue] (0.4, -0.6) -- (1.2, 0.6);
\end{tikzpicture} 
}\\
    &= g^2 \int_{-\infty}^0 \d t' \d t''\, W_{k_1}(0, t') W_{k_2}(0, t') W_{\omega}(t'', t') W_{k_3}(t'', 0) W_{k_4}(t'', 0)\\
    &= \frac{g^2}{16c_s^4 k_1 k_2 k_3 k_4} \frac{1}{2\omega E_L E_R}\,.
    \end{aligned}
\end{equation}
Similarly to the previous case, we have $I_{-+}=I_{+-}^*$ so that one just needs to multiply by a factor two the result in (\ref{eq: I+- diagram}) to obtain $I_{+-} + I_{-+}$. Summing all the four contributions, the final correlator is

\begin{equation}
\label{eq: flatspace full theory correlator}
    \braket{\phi_{\bm{k}_1} \phi_{\bm{k}_2} \phi_{\bm{k}_3} \phi_{\bm{k}_4}}' = \frac{g^2}{8c_s^4 k_1 k_2 k_3 k_4} \left(\frac{1}{E E_L E_R} + \frac{1}{\omega E_L E_R}\right)\,.
\end{equation}
The correlator has a singularity when the total energy vanishes $E\rightarrow 0$, and when partial energies flowing into the left or right vertices add up to zero $E_{L, R}\rightarrow 0$. Note that these singularities are never reached for physical momentum configurations. In fact, the residue of the total energy singularity is the corresponding scattering amplitude

\begin{equation}
    \lim\limits_{E \to 0} \braket{\phi_{\bm{k}_1} \phi_{\bm{k}_2} \phi_{\bm{k}_3} \phi_{\bm{k}_4}}' = \frac{1}{8c_s^4 k_1 k_2 k_3 k_4} \times \frac{\mathcal{A}_4}{E}\,,
\end{equation}
where $\mathcal{A}_4 = -g^2/\mathcal{S}$ with $\mathcal{S} = c_s^2k_{12}^2 - \omega^2$ being the (non-relativistic) Mandelstam variable. When the sound speed of $\phi$ is sufficiently reduced, an interesting limit is that of $c_s k_{12, 34} \ll \omega$, corresponding to the massive field $\sigma$ being much more energetic than the external field $\phi$. In this case, we have $E_{L, R} = \omega + \mathcal{O}(c_sk_{12, 34}/\omega)$ and the correlator can be written as an expansion in $c_sk_{12, 34}/\omega$ whose leading order term reads

\begin{equation}
\label{eq: flat-space correlator limit}
    \braket{\phi_{\bm{k}_1} \phi_{\bm{k}_2} \phi_{\bm{k}_3} \phi_{\bm{k}_4}}' = \frac{g^2}{8c_s^4 k_1 k_2 k_3 k_4} \left(\frac{1}{E\omega^2} + \mathcal{O}\left(\frac{c_sk_{12, 34}}{\omega}\right)\right)\,.
\end{equation}
An interesting feature of this limit is that the partial energy singularities effectively disappear. Indeed in this limit, the massive field can be integrated out and the exchange diagram becomes an effective contact one. 

\paragraph{Non-local contact diagrams.} Following the development of Sec.~\ref{subsection: integrate out}, let us now explicitly integrate out the massive field at the level of the Lagrangian. This procedure generates an infinite tower of non-local contact operators

\begin{equation}
    \mathcal{L} \supset -\frac{g^2}{4}\phi^2\sum_{n=0}^\infty \mathcal{O}_n \phi^2\,, \hspace*{0.5cm}\text{with}\hspace*{0.5cm} \mathcal{O}_n = \frac{1}{-\partial_i^2 + m^2}\frac{(-\partial_t^2)^n}{(-\partial_i^2 + m^2)^n}\,.
\end{equation}
A considerable simplification, compared to the cosmological case, is that the physical momentum does not redshift and the operator ordering does not matter. This Lagrangian leads to an infinite number of contact diagrams to evaluate. However, interestingly enough, one can easily compute them all. The contribution $I_+$ reads

\begin{equation}
\label{eq: flat-space non-local diagrams computation}
    \begin{aligned}
    I_+ &= 
\raisebox{-\height/2}{
\begin{tikzpicture}[line width=1. pt, scale=2.5]
\draw[pyred] (0.2, 0) -- (0.5, -0.6);
\draw[pyred] (0.4, 0) -- (0.5, -0.6);
\draw[pyred] (0.6, 0) -- (0.5, -0.6);
\draw[pyred] (0.8, 0) -- (0.5, -0.6);
\draw[fill=black] (0.5, -0.6) circle (.03cm);
\draw[lightgray2, line width=0.8mm] (0.1, 0) -- (0.9, 0);
\end{tikzpicture} 
} \\
    &= \frac{ig^2}{2} \int_{-\infty}^0 \d t\, W_{k_1}(0, t) W_{k_2}(0, t) \sum_{n=0}^{\infty} \frac{(-1)^n}{\omega^2} \frac{\partial_t^{2n}}{\omega^{2n}} W_{k_3}(0, t) W_{k_4}(0, t) + (k_{12} \leftrightarrow k_{34}) \\
    &= \frac{ig^2}{32c_s^4 k_1 k_2 k_3 k_4}\frac{1}{\omega^2} \int_{-\infty}^0 \d t\, e^{ic_s k_{12}t} \sum_{n=0}^{\infty} \frac{(-1)^n}{\omega^{2n}} \partial_t^{2n} e^{ic_s k_{34}t} + (k_{12} \leftrightarrow k_{34}) \\
    &= \frac{ig^2}{32c_s^4 k_1 k_2 k_3 k_4}\frac{1}{\omega^2} \left(\sum_{n=0}^\infty \left(\frac{c_s k_{34}}{\omega}\right)^{2n}\right) \int_{-\infty}^0 \d t\, e^{ic_s(k_{12} + k_{34})t} + (k_{12} \leftrightarrow k_{34})
    \\
    &= \frac{g^2}{32c_s^4 k_1 k_2 k_3 k_4}\frac{1}{E}\left(\frac{1}{\omega^2 - c_s^2 k_{12}^2} + \frac{1}{\omega^2 - c_s^2 k_{34}^2}\right)\,,
    \end{aligned}
\end{equation}
where in the third line we have used $\partial_t^{2n} e^{ic_s k_{ij}t} = (-1)^n (c_s k_{ij})^{2n} e^{ic_s k_{ij}t}$. The second contribution is $I_- = I_+$.
In the end, the infinite sum of contact diagrams has the following simple and compact form

\begin{equation}
    \braket{\phi_{\bm{k}_1} \phi_{\bm{k}_2} \phi_{\bm{k}_3} \phi_{\bm{k}_4}}' = \frac{g^2}{16c_s^4 k_1 k_2 k_3 k_4}\frac{1}{E}\left(\frac{1}{\omega^2 - c_s^2 k_{12}^2} + \frac{1}{\omega^2 - c_s^2 k_{34}^2}\right)\,,
\end{equation}
where we note that the same result is also found if the massive field is integrated out in a local manner, in that case resumming an infinite number of local contact operators.
This explicit correlator has a few interesting features. First of all, we do not fully recover the correlator computed in the two-field theory (\ref{eq: flatspace full theory correlator}). This means that the effective single-field theory used here, integrating out the field $\sigma$ \textit{\`a la} amplitude, does not  faithfully capture the effects present in the original theory, see below. Second, we note that the correlator diverges when $c_s k_{12, 34} \rightarrow \omega$ which is equivalent to the limit $E_{L, R} \rightarrow 2\omega$. This is expected because in this configuration the geometric series in (\ref{eq: flat-space non-local diagrams computation}) does not converge. Physically, it corresponds to the case when the energy of two external field $\phi$ coincides with the exchanged energy of the intermediate particle. In this configuration, the massive field could not have been integrated out in the first place. The most interesting configuration is when we place ourselves deep inside the radius of convergence of the geometric series $c_s k_{12, 34} \ll \omega$, which corresponds to the massive field carrying way more energy than the external fields. In this limit, the correlator simplifies to 

\begin{equation}
    \braket{\phi_{\bm{k}_1} \phi_{\bm{k}_2} \phi_{\bm{k}_3} \phi_{\bm{k}_4}}' = \frac{g^2}{8c_s^4 k_1 k_2 k_3 k_4} \left(\frac{1}{E\omega^2} + \mathcal{O}\left(\frac{c_sk_{12, 34}}{\omega}\right)\right)\,.
\end{equation}
This form perfectly matches the correct limit of the four-point correlator computed in the original theory (\ref{eq: flat-space correlator limit}) and reproduces the residue at the total energy singularity. However, higher-order corrections do not coincide. This apparent mismatch, as well as the others mentioned above, have been recently solved by adding the proper boundary term accounting for the homogeneous solution to the equation of motion for the heavy field, that we have discarded \cite{Salcedo:2022aal}.

\clearpage
\phantomsection
\addcontentsline{toc}{section}{References}
\bibliographystyle{utphysEFT}
{\linespread{1.075}
\bibliography{references.bib}
}

\end{document}